\documentclass[a4paper,11pt]{article}
\pdfoutput=1 

\usepackage[dvipsnames,table]{xcolor}
\definecolor{linkcol}{rgb}{0,0,.2}
\usepackage{jheppub} 

\usepackage[T1]{fontenc} 

\usepackage{multirow}

\usepackage{framed}
\definecolor{shadecolor}{rgb}{0.94,0.94,0.94} 
\usepackage{pbox}
\usepackage{mathtools}

\usepackage{bbm}
\usepackage[colorinlistoftodos,prependcaption,textsize=tiny,backgroundcolor=blue!20]{todonotes}

\usepackage{hhline}
\usepackage{lscape}
\newlength{\distQTabShift}
\newlength{\distQTabRowHeight}
\newlength{\exLin}
\newlength{\exLinX}
\newlength{\exSpace}
\newcommand{\grline}[1]{\arrayrulecolor{orange}\cline{#1}\arrayrulecolor{gray}}
\newcommand{\grcell}[1]{\multicolumn{1}{c!{\color{orange}\vrule}}{#1}}
\newcommand{\grcellL}[1]{\multicolumn{1}{|c!{\color{orange}\vrule}}{#1}}
\newcommand{\grcellFirstL}[1]{\multicolumn{1}{!{\color{orange}\vrule}c|}{#1}}
\newcommand{\grcellFirstLR}[1]{\multicolumn{1}{!{\color{orange}\vrule}c!{\color{orange}\vrule}}{#1}}
\newcommand{\arrc}{\arrayrulecolor{gray}}
\newcommand{\arrgr}{\arrayrulecolor{orange}}
\newcommand{\grzero}{\textcolor{gray}{~0~}}
\newcommand{\grone}{\textcolor{gray}{~1~}}
\newcommand{\sparg}{\textcolor{gray}{u}}
\newcommand{\trivAnz}[1]{\textcolor{gray}{#1}}
\newcommand{\rhigh}{\rule{0pt}{4.6ex}}
\newcommand{\grc}{\cellcolor{lightorange}}
\newcommand{\koSqrt}{{\setlength\fboxsep{0.6pt}\sqrt{\boxed{\tx}}}}
\definecolor{lightorange}{rgb}{.98,.96,.93}
\colorlet{gradlin}{orange}
\definecolor{momcell}{rgb}{.96,.98,1}
\definecolor{xcol}{RGB}{8,91,19}
\definecolor{betacol}{RGB}{8,91,19}
\definecolor{spcol}{RGB}{0,0,103}
\definecolor{zhucol}{RGB}{0,0,103}
\definecolor{zcol}{RGB}{185,73,21}
\definecolor{percol}{RGB}{10,80,88}
\definecolor{etacol}{RGB}{90,0,77}
\definecolor{gcol}{RGB}{120,24,6}
\definecolor{adocol}{RGB}{79,21,11}
\definecolor{pfcol}{RGB}{74,68,0}
\definecolor{step1}{RGB}{42,158,182}
\definecolor{step2}{RGB}{32,119,137}
\definecolor{step3}{RGB}{232,114,51}
\definecolor{step4}{RGB}{20,79,92}
\definecolor{step5}{RGB}{148,84,49}
\definecolor{nongradrootcol}{RGB}{158,1,136}
\definecolor{gradrootcol}{RGB}{0,1,166}
\colorlet{twisthighlightcol}{betacol!30}
\colorlet{rootcol1}{gradrootcol}
\colorlet{rootcol2}{nongradrootcol}

\newcommand{\algu}{\mathfrak{u}}
\newcommand{\su}{\mathfrak{su}}
\renewcommand{\sl}{\mathfrak{sl}}
\newcommand{\psu}{\mathfrak{psu}}

\newcommand{\gl}{\mathfrak{gl}}

\def\a{\alpha}
\def\b{\beta}

\DeclareMathOperator{\Tr}{Tr}
\DeclareMathOperator{\Pf}{Pf}
\newcommand{\asympt}{\sim}

\def\be{\begin{eqnarray}}
\def\ee{\end{eqnarray}}
\def\no{\nonumber}

\def\CO{{\cal O}}

\newcommand{\bP}{{\bf P}}
\def\Pt{{\tilde \bP}}
\def\bQ{{\bf Q}}

\newcommand{\hl}{\hat{\lambda}}
\newcommand{\hn}{\hat{\nu}}

\newcommand{\emp}{\emptyset}

\newcommand{\ZZ}{\mathcal{Z}}
\newcommand{\ZZb}{\bar{\mathcal{Z}}}
\newcommand{\XX}{\mathcal{X}}
\newcommand{\XXb}{\bar{\mathcal{X}}}
\newcommand{\YY}{\mathcal{Y}}
\newcommand{\YYb}{\bar{\mathcal{Y}}}
\newcommand{\DD}{\mathcal{D}}
\newcommand{\FF}{\mathcal{F}}

\newcommand{\Psib}{\bar{\Psi}}

\newcommand{\dQ}{\mathbb{Q}}

\newcommand{\ns}{\mathbf{n}}
\newcommand{\aaa}{\mathbf{a}}
\newcommand{\bbb}{\mathbf{b}}
\newcommand{\fff}{\mathbf{f}}

\newcommand{\adim}{\textcolor{adocol}{\gamma}}
\newcommand{\ado}[1]{\textcolor{adocol}{\adim^{(#1)}}}

\def\cA{{\textsf{A}}}
\def\cB{{\textsf{B}}}
\def\cc{{\textsf{c}}}
\def\cd{{\textsf{d}}}

\def\tz{{\textsf{z}}}

\newcommand{\subeq}{\stackrel{\text{\tiny sub.}}{=}}

\def\ii{\mathbbm{i}\mspace{2mu}}

\newcommand\grpath[4]{{\color{Orange}\bf#1#2#3#4}}

\newcommand{\spar}{\textcolor{spcol}{u}}
\newcommand{\zdim}{\Delta}
\newcommand{\zv}[1]{\textcolor{zcol}{\zeta_{#1}}} 
\newcommand{\Zv}{\textcolor{zcol}{Z_{11}^{(2)}}} 
\newcommand{\etaF}[1]{\textcolor{etacol}{\eta_{#1}}}
\newcommand{\Per}[1]{\textcolor{percol}{\mathcal{P}_{#1}}}
\newcommand{\perphi}[3]{\textcolor{percol}{\phi_{#1,#2}^{(#3)}}}
\newcommand{\Lanz}{L^\star}
\newcommand{\zh}{\textcolor{zhucol}{x}}
\newcommand{\tc}[1]{\textcolor{xcol}{\mathrm{c}_{#1}}} 

\newcommand{\g}{\textcolor{gcol}{g}}

\def\tb{\textcolor{betacol}{\beta}}
\def\tx{{\textcolor{xcol}{\textsf{x}}}}

\newcommand{\deltaL}{\delta_\lambda}

\definecolor{redmellow}{rgb}{.9, .2, 0.2}
\definecolor{greenmellow}{rgb}{0, .5, 0.2}
\definecolor{bluemellow}{rgb}{0.1,0.1,0.7}

\title{\boldmath The fate of the Konishi multiplet in the $\beta$-deformed Quantum Spectral Curve} 

\author[a]{Christian Marboe}
\author[a,b]{\;\; Erik Wid\'en}

\affiliation[a]{Nordita, Stockholm University \& KTH Royal Institute of Technology,\\ Roslagstullsbacken 23, SE-106 91 Stockholm, Sweden}
\affiliation[b]{Department of Physics and Astronomy, Uppsala University\\
SE-751 08 Uppsala, Sweden}

\emailAdd{christian.marboe@su.se}
\emailAdd{chrmarboe@gmail.com}
\emailAdd{erik.widen@nordita.org}
\emailAdd{e.a.widen@gmail.com}

\abstract{We investigate the solution space of the $\beta$-deformed Quantum Spectral Curve by studying a sample of solutions corresponding to single-trace operators that in the undeformed theory belong to the Konishi multiplet. We discuss how to set the precise boundary conditions for the leading Q-system for a given state, how to solve it, and how to build perturbative corrections to the $\bP\mu$-system. We confirm and add several loop orders to known results in the literature.}

\begin{document} 
\maketitle
\flushbottom

\section{Introduction}\label{sec:intro}

The Quantum Spectral Curve (QSC) \cite{Gromov:2013pga,Gromov:2014caa} is an incredibly efficient and elegent framework for computing the spectrum of planar $\mathcal{N}=4$ Super Yang-Mills theory (SYM). Its power has been demonstrated in many applications \cite{Gromov:2014bva,Alfimov:2014bwa,Marboe:2014gma,Marboe:2014sya,Gromov:2015wca,Gromov:2015vua,Hegedus:2016eop,Marboe:2016igj,Gromov:2016rrp,Marboe:2017dmb,Marboe:2018ugv,Alfimov:2018cms,Cavaglia:2018lxi,Harmark:2018red,Gromov:2017cja}, along with similar developments in ABJM theory \cite{Cavaglia:2014exa,Bombardelli:2017vhk,Gromov:2014eha,Anselmetti:2015mda,Cavaglia:2016ide,Bombardelli:2018bqz,Lee:2017mhh,Lee:2018jvn}. However, the twisted version of the QSC \cite{Kazakov:2015efa,Gromov:2015dfa} has not yet been used to investigate the spectrum of anomalous dimensions of single-trace operators, except for the study of the $\gamma$-deformed BMN vacuum in \cite{Kazakov:2015efa} and the study \cite{Gromov:2017cja} of the spectrum of the strongly $\gamma$-deformed fishnet theory \cite{Gurdogan:2015csr}. In this work, we set out to put the twisted QSC to work by studying its solution space for the $\tb$-deformation \cite{Leigh:1995ep,Lunin:2005jy,Roiban:2003dw}. Put very shortly, what we do in this paper is to generalize the methods \cite{Marboe:2017dmb,Marboe:2018ugv} to the twisted QSC \cite{Kazakov:2015efa}, and these three papers are essential reading to follow this paper. 

The spectrum of single-trace operators in $\tb$-deformed planar $\mathcal{N}=4$ SYM has previously been studied, both using conventional quantum field theory methods and integrability. 
The anomalous dimensions of one-magnon of two-magnon $\su(2)$ states were found from QFT-calculations to four loops in \cite{Fiamberti:2008sm}, and for one-magnon states to the first wrapping order in \cite{Fiamberti:2008sn}. The complete one-loop dilatation operator was studied in \cite{Fokken:2013mza}, see also \cite{Wilhelm:2016izi,Fokken:2017qpu}. 
Twist-2 and twist-3 operators in the $\sl(2)$ sector were treated up to four loops in \cite{deLeeuw:2010ed} by using the asymptotic Bethe ansatz and L\"{u}scher corrections. The $\su(2)$ Konishi operator was studied up to four loops using L\"{u}scher corrections in \cite{Ahn:2010yv}, in agreement with \cite{Fiamberti:2008sm}. In this paper we will reproduce some of these results and demonstrate the power of the QSC by going well beyond in loop order.

Section \ref{sec:sym} is an informal discussion of the (broken) symmetries of the $\tb$-deformed theory and the resulting splitting of the symmetry multiplets of single-trace operators in the undeformed theory. Section \ref{sec:QSC} contains a short recap of the QSC and the features that are relevant for our purposes. In section \ref{sec:lead}, we explain how to set the precise boundary conditions for the leading solution of the Q-system and a strategy for how to solve it. Section \ref{sec:pert} gives a summary of the algorithm used to construct perturbative corrections to the leading solutions. Section \ref{sec:ex} presents a sample of solutions for different parts of the broken Konishi multiplet.

\section{Symmetry and \texorpdfstring{$\beta$}{beta}-deformation}\label{sec:sym}

The field content and the multiplet structure of single-trace operators in $\mathcal{N}=4$ SYM is dictated by the global $\psu(2,2|4)$ superconformal symmetry. By multiplet, we refer to an irreducible representation formed by a vector space of operators that are connected by the generators of the symmetry. 
In the deformations of the theory, the field content remains the same, though the multiplets split into smaller pieces due to the breaking of some of the symmetries.

In this section we briefly recall the basics of the full $\mathcal{N}=4$ superconformal symmetry and discuss the splitting of the Konishi multiplet in the $\tb$-deformation.

\subsection{Symmetry of \texorpdfstring{$\mathcal{N}=4$}{N=4} SYM}
Similarly to \cite{Marboe:2017dmb}, we use the oscillator language used to describe the symmetry and its representations. We start with a short recap of the basic concepts.

\paragraph{Oscillator construction for \texorpdfstring{$\psu(2,2|4)$}{psu(2,2|4)}.}
At zero coupling, oscillators provide a convenient way to parametrize the $\psu(2,2|4)$ generators $E_{mn}$:
\be\label{psugen}
E_{mn}=\chi_m^\dagger\chi_n 
\,,\quad
\chi^\dagger=\{-\bbb_1,-\bbb_2,\fff_1^\dagger,\fff_2^\dagger,\fff_3^\dagger,\fff_4^\dagger,\aaa_1^\dagger,\aaa_2^\dagger\}
\,,\quad
\chi=\{\bbb_1^\dagger,\bbb_2^\dagger,\fff_1,\fff_2,\fff_3,\fff_4,\aaa_1,\aaa_2\}. \no\\
\ee
The supersymmetry generators are of the form $\fff_a^\dagger\bbb^\dagger_{\dot\a}$ and $\fff_a^\dagger\aaa_{\a}$, the $\su(4)$ $R$-symmetry is generated by $\fff_a^\dagger\fff_b$, while the non-compact $\su(2,2)$ conformal symmetry is generated by combinations of $\aaa$'s and $\bbb$'s.

\paragraph{Field content.} The field content of the theory can be constructed according to
\begin{table}[h]
\centering
\begin{tabular}{|c|c|c|c|} \hline
scalar&fermion&field strength& $\begin{matrix}\text{covariant}\\ \text{derivative}\end{matrix}$\\\hline
$\begin{matrix}\Phi_{ab}\equiv \fff_a^\dagger\fff_b^\dagger|0\rangle\\
\scriptscriptstyle \Phi_{12}\equiv\ZZ,\; \Phi_{13}\equiv\XX,\; \Phi_{14}\equiv\YY, \\\scriptscriptstyle \Phi_{23}\equiv\YYb,\; \Phi_{24}\equiv\XXb,\; \Phi_{34}\equiv\ZZb
\end{matrix}$
&\!
$\begin{matrix} \Psi_{a\alpha}\equiv\fff_a^\dagger\aaa_\alpha^\dagger|0\rangle \\[2mm] \bar{\Psi}_{a\dot\a}\equiv \epsilon_{abcd}\fff_b^\dagger\fff_c^\dagger\fff_d^\dagger \bbb_{\dot\a}^\dagger |0\rangle \end{matrix}$
\!&\!
$\begin{matrix}\mathcal{F}_{\a\b}\equiv\aaa_\a^\dagger\aaa_\b^\dagger|0\rangle \\[2mm] \mathcal{F}_{\dot\a\dot\b}\equiv\bbb_{\dot\a}^\dagger\bbb_{\dot\b}^\dagger\fff_1^\dagger\fff_2^\dagger\fff_3^\dagger\fff_4^\dagger|0\rangle \end{matrix}$
\!&
$\DD_{\alpha\dot\alpha}\equiv\aaa_\alpha^\dagger\bbb_{\dot\alpha}^\dagger$\\\hline
\end{tabular}
\end{table}

\paragraph{Quantum numbers.}
We use the conventions of \cite{Marboe:2017dmb}, and describe single-trace operators by the oscillator content needed to construct them, i.e.
\be
\ns=[n_{\bbb_1},n_{\bbb_2}|n_{\fff_1},n_{\fff_2},n_{\fff_3},n_{\fff_4}|n_{\aaa_1},n_{\aaa_2}]
\,,
\ee
where $n_{\bullet}$ are number operators, e.g. $n_{\aaa_2}\equiv \aaa_2^\dagger \aaa_2$.
We will also use the $\su(4)$ and $\su(2,2)$ weights $\lambda_a$ and $\nu_i$ given by
\be \label{weights}
\lambda_a\equiv n_{\fff_a}\,,\quad\quad \nu_i\equiv \left\{-L-n_{\bbb_{\dot\a}}-\frac{\gamma}{2},n_{\aaa_\a}+\frac{\gamma}{2} \right\}_i\,,
\ee
where $\gamma$ is the anomalous dimension and $L$ is the length, i.e. the number of fields in the operator. Note that only six quantum numbers are needed to characterize a $\psu(2,2|4)$ representation, e.g. the differences $\lambda_a-\lambda_{a+1}$ and $\nu_i-\nu_{i+1}$.

\paragraph{Grading.}
To denote a $\psu(2,2|4)$ grading, i.e. an ordering of the oscillators in \eqref{psugen}, 
we can use a sequence of $2\times 4$ numbers $1,2,3,4,\hat1,\hat2,\hat3,\hat4$ that correspond to the ordering of the oscillators. 
More often, we will simply use a shorthand notation with four numbers that corresponds to the positions of the fermionic oscillators $\fff_a$ in the grading.
For example, the grading $\chi=\{\fff_1,\bbb_1^\dagger,\bbb_2^\dagger,\fff_2,\aaa_1,\fff_3,\fff_4,\aaa_2\}$ is denoted by $\hat{1}12\hat{2}3\hat{3}\hat{4}4$ or \grpath{0}{2}{3}{3}.

\paragraph{Highest weight state.}
For a given grading, we define a highest weight state (HWS) as the operator within a multiplet that is annihilated by all $E_{mn}$ for which $m<n$.

\paragraph{Young diagram.} Following the practice of \cite{Gunaydin:2017lhg,Marboe:2017dmb}, one can use non-compact Young diagrams to characterize multiplets at $\g=0$ in the undeformed theory. We refer to figure 4 in \cite{Marboe:2017dmb} for the definition and to figure \ref{fig:koYD} below for the Young diagrams corresponding to the Konishi multiplet.

\subsection{Leftover symmetry in the \texorpdfstring{$\tb$}{beta}-deformation}

The $\tb$-deformation breaks the off-diagonal part of the $R$-symmetry and 12 out of the 16 supercharges. An overview of the $\psu(2,2|4)$ generators \eqref{psugen} that correspond to (un)broken symmetries is given in figure \ref{brokengen}. The leftover continuous symmetry is $\su(2,2|1)\oplus\algu(1)^3$. 
As the oscillators $\fff_1$, $\fff_2$ and $\fff_3$ are treated on an equal footing, there is an additional discrete $S_3$ symmetry that permutes these three oscillators.

\colorlet{brokencol}{xcol!50}

\begin{figure}[h]
\be
\begin{array}{rrrr|rrrr}
{\color{black}-\bbb_1\bbb_1^\dagger} &
{\color{black}-\bbb_1\bbb_2^\dagger} & 
{\color{black}-\bbb_1\aaa_1} & 
{\color{black}-\bbb_1\aaa_2} &
{\color{brokencol}-\bbb_1\fff_1} & 
{\color{brokencol}-\bbb_1\fff_2} & 
{\color{brokencol}-\bbb_1\fff_3} & 
{\color{black}-\bbb_1\fff_4} \\

{\color{black}-\bbb_2\bbb_1^\dagger} & 
{\color{black}-\bbb_2\bbb_2^\dagger} & 
{\color{black}-\bbb_2\aaa_1} & 
{\color{black}-\bbb_2\aaa_2} &
{\color{brokencol}-\bbb_2\fff_1} & 
{\color{brokencol}-\bbb_2\fff_2} & 
{\color{brokencol}-\bbb_2\fff_3} & 
{\color{black}-\bbb_2\fff_4} \\

{\color{black}\aaa_1^\dagger \bbb_1^\dagger} & 
{\color{black}\aaa_1^\dagger \bbb_2^\dagger} & 
{\color{black}\aaa_1^\dagger \aaa_1} & 
{\color{black}\aaa_1^\dagger \aaa_2} &
{\color{brokencol}\aaa_1^\dagger \fff_1} & 
{\color{brokencol}\aaa_1^\dagger \fff_2} & 
{\color{brokencol}\aaa_1^\dagger \fff_3} & 
{\color{black}\aaa_1^\dagger \fff_4} \\

{\color{black}\aaa_2^\dagger \bbb_1^\dagger} & 
{\color{black}\aaa_2^\dagger \bbb_2^\dagger} & 
{\color{black}\aaa_2^\dagger \aaa_1} & 
{\color{black}\aaa_2^\dagger \aaa_2} &
{\color{brokencol}\aaa_2^\dagger \fff_1} & 
{\color{brokencol}\aaa_2^\dagger \fff_2} & 
{\color{brokencol}\aaa_2^\dagger \fff_3} &
{\color{black}\aaa_2^\dagger \fff_4} 
 \\ \hline

{\color{brokencol}\fff_1^\dagger \bbb_1^\dagger} & 
{\color{brokencol}\fff_1^\dagger \bbb_2^\dagger} & 
{\color{brokencol}\fff_1^\dagger \aaa_1} & 
{\color{brokencol}\fff_1^\dagger \aaa_2} &
{\color{black}\fff_1^\dagger \fff_1} & 
{\color{brokencol}\fff_1^\dagger \fff_2} & 
{\color{brokencol}\fff_1^\dagger \fff_3} & 
{\color{brokencol}\fff_1^\dagger \fff_4} \\

{\color{brokencol}\fff_2^\dagger \bbb_1^\dagger} & 
{\color{brokencol}\fff_2^\dagger \bbb_2^\dagger} & 
{\color{brokencol}\fff_2^\dagger \aaa_1} & 
{\color{brokencol}\fff_2^\dagger \aaa_2} &
{\color{brokencol}\fff_2^\dagger \fff_1} & 
{\color{black}\fff_2^\dagger \fff_2} & 
{\color{brokencol}\fff_2^\dagger \fff_3} & 
{\color{brokencol}\fff_2^\dagger \fff_4} \\

{\color{brokencol}\fff_3^\dagger \bbb_1^\dagger} & 
{\color{brokencol}\fff_3^\dagger \bbb_2^\dagger} & 
{\color{brokencol}\fff_3^\dagger \aaa_1} & 
{\color{brokencol}\fff_3^\dagger \aaa_2} &
{\color{brokencol}\fff_3^\dagger \fff_1} & 
{\color{brokencol}\fff_3^\dagger \fff_2} & 
{\color{black}\fff_3^\dagger \fff_3} & 
{\color{brokencol}\fff_3^\dagger \fff_4} \\

{\color{black}\fff_4^\dagger \bbb_1^\dagger} & 
{\color{black}\fff_4^\dagger \bbb_2^\dagger} & 
{\color{black}\fff_4^\dagger \aaa_1} & 
{\color{black}\fff_4^\dagger \aaa_2} &
{\color{brokencol}\fff_4^\dagger \fff_1} & 
{\color{brokencol}\fff_4^\dagger \fff_2} & 
{\color{brokencol}\fff_4^\dagger \fff_3} & 
{\color{black}\fff_4^\dagger \fff_4}

\end{array}\no
\ee
\caption{$\psu(2,2|4)$ generators. The generators marked in {\color{xcol}green} correspond to broken symmetries in the $\tb$-deformed theory.}
\label{brokengen}
\end{figure}

\paragraph{Diagonal twist of \texorpdfstring{$\su(4)$}{su(4)}.} The $\tb$-deformation twists the $\su(4)$-symmetry with twist parameters \cite{Kazakov:2015efa}
\be \label{betatw}
\tx_a=\left\{
e^{\ii \tb(n_{\fff_2}-n_{\fff_3})},
e^{\ii \tb(n_{\fff_3}-n_{\fff_1})},
e^{\ii \tb(n_{\fff_1}-n_{\fff_2})},
1 \right\}\,.
\ee
Notice that the twist depends on the quantum numbers, i.e. it depends on the operator in question. Throughout the paper, we will use the shorthand notation $\tx\equiv e^{\ii\tb}$.

\paragraph{Shifted weights.} The concept of shifted weights, $\hl$ and $\hn$, is important, because they govern the asymptotics of the QSC. They are given by \cite{Kazakov:2015efa}
\begin{subequations} \label{shws}
\be
\hl_a &=& \lambda_a -\sum_{b\prec a}\delta_{\tx_a,\tx_b} + \sum_{i\prec a}\delta_{\tx_a,1} + \Lambda \\
\hn_i &=& \nu_i -i+1
+ \sum_{a\prec i}\delta_{\tx_a,1} - \Lambda
\ee
\end{subequations}
where $\prec$ means that the oscillator corresponding to the left index is placed before the one corresponding to the one on the right side in the grading for which the operator is a HWS. $\Lambda$ is an arbitrary integer shift that we will return to.

\subsection{The Konishi multiplet}
The Konishi multiplet is the archetypical example in the study of $\mathcal{N}=4$ SYM. We here review some facts about this multiplet in the undeformed theory and look at how it splits up due to the $\tb$-deformation.

\subsubsection{Undeformed theory}

In the undeformed theory, the Konishi multiplet contains the simplest operators not protected from quantum corrections. The operator of lowest dimension ($\zdim_0=2$) within the multiplet is the two-scalar state 
\be\label{su4ko}
\Tr[\ZZ\bar\ZZ + \XX\bar\XX + \YY\bar\YY]\,,
\ee 
often referred to as the "$\su(4)$ Konishi". It is the highest weight state in the grading $12\hat1\hat2\hat3\hat434$/\grpath{2}{2}{2}{2}. We can act on the state \eqref{su4ko} with the symmetry generators \eqref{psugen} to build an infinite tower of states. Throughout the paper, we will leave out the symbol "$\Tr[...]$", so the reader should keep in mind that a trace is always implicit when discussing operators. Also, we will loosely refer to the states by a representative of its field content, e.g. we will refer to \eqref{su4ko} simply as $\ZZ\ZZb$.

\paragraph{Supercharges and gradings.} 
Acting once on \eqref{su4ko} with the supercharges $\aaa^\dagger\fff$ or $\bbb^\dagger\fff^\dagger$ produces operators containing a scalar and a fermion, i.e. of the type $\Phi\Psi$, with $\Delta_0=\frac{5}{2}$. These states are of highest weight in different gradings. For example, acting with $\aaa_1^\dagger\fff_4$ results in a state with content of the type $\ZZ\Psi_{31}$ and takes us to the HWS grading $12\hat1\hat2\hat33\hat44$, i.e. simply the replacement $\hat43\to3\hat4$. Acting with $\aaa_2^\dagger\fff_2$, we get a state of the kind $\bar\ZZ\Psi_{12}$, which is a HWS in $12\hat1\hat3\hat44\hat23$, where we also needed to make rearrangements within the fermionic and bosonic oscillators, respectively.

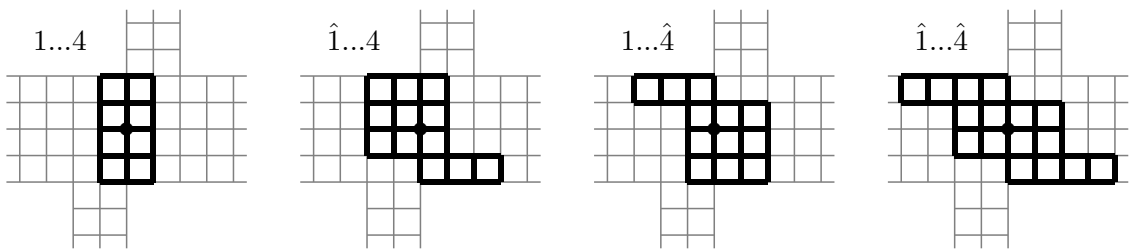
\begin{figure}[b]
\centering
\begin{picture}(400,90)
\put(5,70){$1...4$}
\put(115,70){$\hat1...4$}
\put(225,70){$1...\hat4$}
\put(335,70){$\hat1...\hat4$}

\color{gray}
\linethickness{.2mm}

\multiput(-5,20)(0,10){5}{\line(1,0){90}}
\put(20,0){\line(1,0){20}}
\put(20,10){\line(1,0){20}}
\put(40,70){\line(1,0){20}}
\put(40,80){\line(1,0){20}}
\put(0,20){\line(0,1){40}}
\put(10,20){\line(0,1){40}}
\put(20,-5){\line(0,1){65}}
\put(30,-5){\line(0,1){65}}
\put(40,-5){\line(0,1){90}}
\put(50,20){\line(0,1){65}}
\put(60,20){\line(0,1){65}}
\put(70,20){\line(0,1){40}}
\put(80,20){\line(0,1){40}}

\multiput(105,20)(0,10){5}{\line(1,0){90}}
\put(130,0){\line(1,0){20}}
\put(130,10){\line(1,0){20}}
\put(150,70){\line(1,0){20}}
\put(150,80){\line(1,0){20}}
\put(110,20){\line(0,1){40}}
\put(120,20){\line(0,1){40}}
\put(130,-5){\line(0,1){65}}
\put(140,-5){\line(0,1){65}}
\put(150,-5){\line(0,1){90}}
\put(160,20){\line(0,1){65}}
\put(170,20){\line(0,1){65}}
\put(180,20){\line(0,1){40}}
\put(190,20){\line(0,1){40}}

\multiput(215,20)(0,10){5}{\line(1,0){90}}
\put(240,0){\line(1,0){20}}
\put(240,10){\line(1,0){20}}
\put(260,70){\line(1,0){20}}
\put(260,80){\line(1,0){20}}
\put(220,20){\line(0,1){40}}
\put(230,20){\line(0,1){40}}
\put(240,-5){\line(0,1){65}}
\put(250,-5){\line(0,1){65}}
\put(260,-5){\line(0,1){90}}
\put(270,20){\line(0,1){65}}
\put(280,20){\line(0,1){65}}
\put(290,20){\line(0,1){40}}
\put(300,20){\line(0,1){40}}

\multiput(325,20)(0,10){5}{\line(1,0){90}}
\put(350,0){\line(1,0){20}}
\put(350,10){\line(1,0){20}}
\put(370,70){\line(1,0){20}}
\put(370,80){\line(1,0){20}}
\put(330,20){\line(0,1){40}}
\put(340,20){\line(0,1){40}}
\put(350,-5){\line(0,1){65}}
\put(360,-5){\line(0,1){65}}
\put(370,-5){\line(0,1){90}}
\put(380,20){\line(0,1){65}}
\put(390,20){\line(0,1){65}}
\put(400,20){\line(0,1){40}}
\put(410,20){\line(0,1){40}}

\color{black}
\linethickness{0.7mm}

\put(30,20){\line(1,0){20}}
\put(30,30){\line(1,0){20}}
\put(30,40){\line(1,0){20}}
\put(30,50){\line(1,0){20}}
\put(30,60){\line(1,0){20}}

\put(30,20){\line(0,1){40}}
\put(40,20){\line(0,1){40}}
\put(50,20){\line(0,1){40}}

\put(130,60){\line(1,0){30}}
\put(130,50){\line(1,0){30}}
\put(130,40){\line(1,0){30}}
\put(130,30){\line(1,0){50}}
\put(150,20){\line(1,0){30}}

\put(130,30){\line(0,1){30}}
\put(140,30){\line(0,1){30}}
\put(150,20){\line(0,1){40}}
\put(160,20){\line(0,1){40}}
\put(170,20){\line(0,1){10}}
\put(180,20){\line(0,1){10}}

\put(230,60){\line(1,0){30}}
\put(230,50){\line(1,0){50}}
\put(250,40){\line(1,0){30}}
\put(250,30){\line(1,0){30}}
\put(250,20){\line(1,0){30}}

\put(230,50){\line(0,1){10}}
\put(240,50){\line(0,1){10}}
\put(250,20){\line(0,1){40}}
\put(260,20){\line(0,1){40}}
\put(270,20){\line(0,1){30}}
\put(280,20){\line(0,1){30}}

\put(330,60){\line(1,0){40}}
\put(330,50){\line(1,0){60}}
\put(350,40){\line(1,0){40}}
\put(350,30){\line(1,0){60}}
\put(370,20){\line(1,0){40}}

\put(330,50){\line(0,1){10}}
\put(340,50){\line(0,1){10}}
\put(350,30){\line(0,1){30}}
\put(360,30){\line(0,1){30}}
\put(370,20){\line(0,1){40}}
\put(380,20){\line(0,1){30}}
\put(390,20){\line(0,1){30}}
\put(400,20){\line(0,1){10}}
\put(410,20){\line(0,1){10}}

\put(40,40){\circle*{5}}
\put(150,40){\circle*{5}}
\put(260,40){\circle*{5}}
\put(370,40){\circle*{5}}
\end{picture}
\caption{Young diagrams for the four short multiplets that constitute the Konishi multiplet. The given gradings are those where the operators in the short multiplets remain highest weight states at finite coupling.}
\label{fig:koYD}
\end{figure}

\paragraph{Shortening.} An important feature of the Konishi multiplet is that it is composed of operators of different lengths. The superconformal algebra at $\g=0$ does not connect the full Konishi multiplet. This effect is known as shortening. Only when quantum corrections to the superconformal algebra are taken into account is it possible to connect the complete Konishi multiplet. At zero coupling the multiplet splits up into four short multiplets, one with $L=2$, two with $L=3$, and one with $L=4$. The Young diagrams for each of these four short multiplets are given in figure \ref{fig:koYD}. For example, if we act on the state \eqref{su4ko} with first $\aaa_1^\dagger\fff_4$ and then $\aaa_2^\dagger\fff_4$, we annihilate the state. But the quantum corrections to the generators would in fact produce a state of length three with field content $\ZZ\XX\bar\YY$.

Figure \ref{fig:ko} gives a sample of states in the Konishi multipet. Besides the "$\su(4)$ Konishi" \eqref{su4ko}, other popular members of the Konishi multiplet are the "$\su(2)$ Konishi" with field content $\ZZ^2\XX^2$, being the HWS in \grpath{0}{2}{2}{4}, and the "$\sl(2)$ Konishi" with content $\DD_{12}^2\ZZ^2$, being the HWS in \grpath{1}{1}{3}{3}.

\begin{figure}[t]
\centering
\begin{picture}(430,200)

\color{gray!88}
\put(0,165){\line(1,0){430}}
\put(0,125){\line(1,0){430}}
\put(0,85){\line(1,0){430}}
\put(0,45){\line(1,0){430}}
\put(0,5){\line(1,0){430}}

\put(290,-5){\line(0,1){205}}
\put(365,-5){\line(0,1){205}}

\put(0,182){\footnotesize$\Delta_0\!=\!2$}
\put(0,142){\footnotesize$\Delta_0\!=\!\frac{5}{2}$}
\put(0,102){\footnotesize$\Delta_0\!=\!3$}
\put(0,62){\footnotesize$\Delta_0\!=\!\frac{7}{2}$}
\put(0,22){\footnotesize$\Delta_0\!=\!4$}

\put(140,200){\footnotesize$L\!=\!2$}
\put(317,200){\footnotesize$L\!=\!3$}
\put(392,200){\footnotesize$L\!=\!4$}

\color{black}

\put(150,180){$\boxed{\ZZ\bar\ZZ}$}
\put(150.4,180){$\colorbox{blue!10}{$\ZZ\bar\ZZ$}$}

\put(130,140){$\boxed{\ZZ\Psi_{31}}$}

\put(210,140){$\boxed{\ZZ\bar\Psi_{22}}$}
\put(210.4,140){$\colorbox{blue!10}{$\ZZ\bar\Psi_{22}$}$}

\put(40,100){$\boxed{\DD(\ZZ\bar\ZZ)}$}

\put(126,100){$\boxed{\Psi_{11}\Psi_{21}}$}

\put(215,100){$\boxed{\DD_{12}\ZZ\XX}$}

\put(310,100){$\boxed{\ZZ\XX\YY}$}
\put(310.4,100){$\colorbox{blue!10}{$\ZZ\XX\YY$}$}

\put(126,60){$\boxed{\Psi_{11}\FF_{11}}$}
\put(126.4,60){$\colorbox{blue!10}{$\Psi_{11}\FF_{11}$}$}

\put(210,60){$\boxed{\DD_{12}\ZZ\Psi_{11}}$}


\put(305,60){$\boxed{\ZZ\XX\Psi_{11}}$}

\put(40,20){$\boxed{\DD^2(\ZZ\bar\ZZ)}$}

\put(231,20){$\boxed{\DD_{12}^2\ZZ^2}$}
\put(231.4,20.3){$\colorbox{blue!10}{$\DD_{12}^2\ZZ^2$}$}
\put(168,20){$\boxed{\DD_{12}^2\ZZ\XX}$}
\put(168.4,20.3){$\colorbox{blue!10}{$\DD_{12}^2\ZZ\XX$}$}
\put(110,20){$\boxed{\DD_{12}^2\XX^2}$}

\put(310,20){$\boxed{\ZZ\Psi_{11}^2}$}
\put(310.4,20.3){$\colorbox{blue!10}{$\ZZ\Psi_{11}^2$}$}

\put(380,20){$\boxed{\ZZ^2\XX^2}$}
\put(380.4,20.3){$\colorbox{blue!10}{$\ZZ^2\XX^2$}$}

\thicklines

\colorlet{arrowcol}{xcol!50}

\put(150,177){\vector(-4,-3){84}}
\put(95,150){\tiny$\aaa^\dagger\bbb^\dagger$}

\put(60,93){\vector(0,-1){59}}
\put(63,60){\tiny$\aaa^\dagger\bbb^\dagger$}

\put(160,177){\vector(-2,-3){17}}
\put(153,160){\tiny$\aaa_1^\dagger\fff_4$}

\put(145,134){\color{arrowcol}\vector(0,-1){22}}
\put(148,122){\tiny$\aaa_1^\dagger\fff_3$}

\put(145,94){\color{arrowcol}\vector(0,-1){22}}
\put(148,82){\tiny$\aaa_1^\dagger\fff_2$}

\put(171,177){\color{arrowcol}\vector(2,-1){47}}
\put(190,170){\tiny$\bbb_2^\dagger\fff_1^\dagger$}

\put(167,137){\color{arrowcol}\vector(2,-1){47}}
\put(183,130){\tiny$\bbb_2^\dagger\fff_1^\dagger$}

\put(232,134){\vector(0,-1){22}}
\put(235,122){\tiny$\aaa_1^\dagger\fff_4$}

\put(232,94){\color{arrowcol}\vector(0,-1){22}}
\put(235,82){\tiny$\aaa_1^\dagger\fff_3$}

\put(242,54){\color{arrowcol}\vector(0,-1){20}}
\put(245,42){\tiny$\bbb_2^\dagger\fff_2^\dagger$}

\put(246,135){\color{arrowcol}\vector(3,-1){63}}
\put(270,128){\tiny\color{gcol}$\bbb_1^\dagger\fff_1^\dagger$}

\put(325,94){\vector(0,-1){22}}
\put(328,82){\tiny$\aaa_1^\dagger\fff_4$}

\put(325,54){\color{arrowcol}\vector(0,-1){20}}
\put(328,42){\tiny$\aaa_1^\dagger\fff_3$}

\put(229,23){\color{arrowcol}\vector(-1,0){18}}
\put(215,28){\tiny$\fff_3^\dagger\fff_2$}

\put(167,23){\color{arrowcol}\vector(-1,0){18}}
\put(152,28){\tiny$\fff_3^\dagger\fff_2$}

\put(262,57){\color{arrowcol}\vector(2,-1){47}}
\put(273,52){\tiny\color{gcol}$\bbb_1^\dagger\fff_1^\dagger$}

\put(350,57){\vector(2,-1){44}}
\put(371,50){\tiny\color{gcol}$\aaa_2^\dagger\fff_4$}

\end{picture}
\caption{Sample operators from the Konishi multiplet. Black arrows correspond to symmetries that remain in the $\tb$-deformed theory, while the {\color{xcol}green} arrows correspond to broken symmetries. Oscillators marked in {\color{gcol}brown} indicate that it is the quantum corrected version of the corresponding generator that connects the two operators. Note that the given states are not the exact eigenstates, but just a sample of the field content that they can contain. The $\colorbox{blue!10}{highlighted}$ operators correspond to the examples that we treat in this paper, see table \ref{tab:examples}. 
}
\label{fig:ko}
\end{figure}

\paragraph{\texorpdfstring{$R$}{R}-symmetry structure.}
The $\su(4)$ Konishi \eqref{su4ko} is a singlet under the $\su(4)$ $R$-symmetry. However, e.g. the $\su(2)$ and $\sl(2)$ Konishi are not. We can act on the $\sl(2)$ Konishi, $\DD_{12}^2\ZZ^2$, with $\fff_3^\dagger\fff_2$ once to produce states with field content $\DD_{12}^2\ZZ\XX$ and twice to get $\DD_{12}^2\XX^2$, and similarly for the other $R$-symmetry generators.

\paragraph{Conformal generators.} The conformal generators are composed by $\aaa$ and $\bbb$ oscillators. Those of type $\aaa^\dagger\aaa$ and $\bbb^\dagger\bbb$ act similarly to the $R$-symmetry generators, and their action can lead to highest weight states in different gradings, corresponding to the permutations $12\leftrightarrow21$ and $34\leftrightarrow43$. The generators of type $\aaa^\dagger\bbb^\dagger$, corresponding to derivatives, are different. You can act with these generators infinitely many times, and they only produce descendants.

\subsubsection{Deformed theory}
The $\tb$-deformation breaks the Konishi multiplet into a large number of smaller symmetry multiplets, which we will refer to as {\it submultiplets}. 
The operators belonging to a submultiplet are related by the unbroken supercharges ($\aaa_\alpha^\dagger\fff_4$ and $\bbb_{\dot\alpha}\fff_4$ at zero coupling) and the conformal generators. Due to the non-compact conformal symmetry, the submultiplets are all infinite-dimensional.

For example, the state \eqref{su4ko} will remain in the same multiplet as the operator that is generated by acting with $\aaa_1^\dagger\fff_4$ (of the type $\ZZ\Psi_{31}$), but not with the ones generated by acting with $\aaa_1\fff_{a<4}$ (e.g. of type $\ZZ\Psi_{41}$) as these are no longer symmetries.

In this paper, we refrain from classifying all submultiplets of the Konishi multiplet and instead simply consider a sample of submultiplets that together illustrate some of the features of the solution space of the $\tb$-deformed QSC. The eight examples that we discuss are listed in table \ref{tab:examples}. Most notably, we will consider the submultiplets containing the $\su(4)$, $\su(2)$ and $\sl(2)$ Konishi operators (first, eighth and fifth entry in table \ref{tab:examples}, respectively), which can be compared to known results in the literature.

In fact, two of the eight operators in table \ref{tab:examples} are in the same submultiplet: the third and eighth operator, $\ZZ\XX\YY$ and the $\su(2)$ state $\ZZ^2\XX^2$, are related by the action of the unbroken generators $\aaa_1^\dagger\fff_4$ and $\aaa_2^\dagger\fff_4$ (to be more precise by the quantum corrected versions of these generators). Using the freedom to choose any $\Lambda$ in the shifted weights \eqref{shws}, we see that they have identical shifted weights $\hl$ and $\hn$, and the same twists $\tx_a$. Notice furthermore that the operator in the second row, $\ZZ\bar\Psi_{22}$, also has the same $\hl$ and $\hn$, but twists differing by the replacement $\tx\to\tx^{\frac{1}{2}}$.

\begin{table}[t!]
\centering
\begin{tabular}{|c|c|c|c|c|c|c|} \hline
\begin{tabular}{c} HWS \\ grading \end{tabular} & $\!\Delta_0\!$ & $L$ & $\ns$& \begin{tabular}{c}possible \\ field \\ content\end{tabular} & \begin{tabular}{c} $\hl_a-\Lambda$ \\ $\hn_i|_{g=0}+\Lambda$ \end{tabular} & $\tx_a$  \\ \hline\hline
\begin{tabular}{c}
\hspace{1.1mm}\begin{picture}(39,40)
\linethickness{0.2mm}\color{gray}
\put(12,36){\line(1,0){12}}
\put(12,30){\line(1,0){12}}
\put(12,12){\line(1,0){12}}
\put(12,18){\line(1,0){12}}
\put(12,24){\line(1,0){12}}
\put(12,12){\line(0,1){24}}
\put(18,12){\line(0,1){24}}
\put(24,12){\line(0,1){24}}
\linethickness{0.6mm}\color{orange}
\put(6,12){\line(1,0){12}}
\put(18,12){\line(0,1){24}}
\put(18,36){\line(1,0){12}}
\end{picture}\vspace{-5mm}
\\
\grpath{2}{2}{2}{2}
\end{tabular}
& 2 & 2&
$[0,\!0|1,\!1,\!1,\!1|0,\!0]$ & \begin{tabular}{c} \colorbox{blue!10}{$\ZZ\bar\ZZ$}\\ $\XX\bar\XX$ \\ $\YY\bar\YY$ \end{tabular} & 
\begin{tabular}{c} \{3,2,1,0\} \\ \{-2,-3,2,1\} \end{tabular}
& $\{1,1,1,1\}$
 \\ \hline
\begin{tabular}{c}
\hspace{1.1mm}\begin{picture}(39,40)
\linethickness{0.2mm}\color{gray}
\put(12,36){\line(1,0){12}}
\put(12,30){\line(1,0){12}}
\put(12,12){\line(1,0){12}}
\put(12,18){\line(1,0){12}}
\put(12,24){\line(1,0){12}}
\put(12,12){\line(0,1){24}}
\put(18,12){\line(0,1){24}}
\put(24,12){\line(0,1){24}}
\linethickness{0.6mm}\color{orange}
\put(6,12){\line(1,0){6}}
\put(12,12){\line(0,1){6}}
\put(12,18){\line(1,0){6}}
\put(18,18){\line(0,1){18}}
\put(18,36){\line(1,0){12}}
\end{picture}\vspace{-5mm}
\\
\grpath{1}{2}{2}{2}
\end{tabular}
&
$\frac{5}{2}$ & 2
&
$[0,\!1|2,\!1,\!1,\!1|0,\!0]$ &
 \begin{tabular}{c}\colorbox{blue!10}{$\ZZ\bar\Psi_{22}$} \\ $\XX\bar\Psi_{32}$ \\ $\YY\bar\Psi_{42}$
  \end{tabular} 
&
\begin{tabular}{c} \{3,1,1,2\} \\ \{-2,-3,0,-1\} \end{tabular}
& 
$\{1,\tx^{-1},\tx,1\}$
 \\ \hline
\begin{tabular}{c}
\hspace{1.1mm}\begin{picture}(39,40)
\linethickness{0.2mm}\color{gray}
\put(18,12){\line(1,0){18}}
\put(6,18){\line(1,0){30}}
\put(6,24){\line(1,0){18}}
\put(6,30){\line(1,0){18}}
\put(6,36){\line(1,0){18}}
\put(6,18){\line(0,1){18}}
\put(12,18){\line(0,1){18}}
\put(18,12){\line(0,1){24}}
\put(24,12){\line(0,1){24}}
\put(30,12){\line(0,1){6}}
\put(36,12){\line(0,1){6}}
\linethickness{0.6mm}\color{orange}
\put(6,12){\line(0,1){6}}
\put(6,18){\line(1,0){12}}
\put(18,18){\line(0,1){18}}
\put(18,36){\line(1,0){12}}
\end{picture}\vspace{-5mm}
\\
\grpath{0}{2}{2}{2}
\end{tabular}
&
3 & 3 
& 
$[0,\!0|3,\!1,\!1,\!1|0,\!0]$ & \colorbox{blue!10}{$\ZZ\XX\YY$}  
& 
\begin{tabular}{c} \{3,1,1,2\} \\ \{-2,-3,0,-1\} \end{tabular}
& 
$\{1,\tx^{-2},\tx^2,1\}$
 \\ \hline
\begin{tabular}{c}
\hspace{1.1mm}\begin{picture}(39,40)
\linethickness{0.2mm}\color{gray}
\put(12,36){\line(1,0){12}}
\put(12,30){\line(1,0){12}}
\put(12,12){\line(1,0){12}}
\put(12,18){\line(1,0){12}}
\put(12,24){\line(1,0){12}}
\put(12,12){\line(0,1){24}}
\put(18,12){\line(0,1){24}}
\put(24,12){\line(0,1){24}}
\linethickness{0.6mm}\color{orange}
\put(6,12){\line(1,0){12}}
\put(18,12){\line(0,1){6}}
\put(18,18){\line(1,0){6}}
\put(24,18){\line(0,1){18}}
\put(24,36){\line(1,0){6}}
\end{picture}\vspace{-5mm}
\\
\grpath{2}{3}{3}{3}
\end{tabular}
& $\frac{7}{2}$ & 2 
&
$[0,\!0|1,\!0,\!0,\!0|3,\!0]$ & \colorbox{blue!10}{$\Psi_{11}\FF_{11}$} 
& 
\begin{tabular}{c} \{3,0,0,2\} \\ \{-2,-3,2,-1\} \end{tabular}
& 
$\{1,\tx^{-1},\tx,1\}$
\\ \hline
\begin{tabular}{c}
\hspace{1.1mm}\begin{picture}(39,40)
\linethickness{0.2mm}\color{gray}
\put(12,36){\line(1,0){12}}
\put(12,30){\line(1,0){12}}
\put(12,12){\line(1,0){12}}
\put(12,18){\line(1,0){12}}
\put(12,24){\line(1,0){12}}
\put(12,12){\line(0,1){24}}
\put(18,12){\line(0,1){24}}
\put(24,12){\line(0,1){24}}
\linethickness{0.6mm}\color{orange}
\put(6,12){\line(1,0){6}}
\put(12,12){\line(0,1){12}}
\put(12,24){\line(1,0){12}}
\put(24,24){\line(0,1){12}}
\put(24,36){\line(1,0){6}}
\end{picture}\vspace{-5mm}
\\
\grpath{1}{1}{3}{3}
\end{tabular}
& 4 & 2 &
$[0,\!2|2,\!2,\!0,\!0|2,\!0]$ & \colorbox{blue!10}{$\DD_{12}^2\ZZ^2$}
&
\begin{tabular}{c} \{2,2,3,2\} \\ \{-2,-5,0,-1\} \end{tabular} 
& 
$\{\tx^2,\tx^{-2},1,1\}$
\\ \hline
\begin{tabular}{c}
\hspace{1.1mm}\begin{picture}(39,40)
\linethickness{0.2mm}\color{gray}
\put(12,12){\line(1,0){12}}
\put(6,18){\line(1,0){18}}
\put(6,24){\line(1,0){24}}
\put(12,30){\line(1,0){18}}
\put(12,36){\line(1,0){12}}
\put(6,18){\line(0,1){6}}
\put(12,12){\line(0,1){12}}\put(12,30){\line(0,1){6}}
\put(18,12){\line(0,1){24}}
\put(24,12){\line(0,1){6}}\put(24,24){\line(0,1){12}}
\put(30,24){\line(0,1){6}}
\linethickness{0.6mm}\color{orange}
\put(6,12){\line(1,0){6}}
\put(12,12){\line(0,1){12}}
\put(12,24){\line(1,0){12}}
\put(24,24){\line(0,1){12}}
\put(24,36){\line(1,0){6}}
\end{picture}\vspace{-5mm}
\\
\grpath{1}{1}{3}{3}
\end{tabular}
& 4 & 2 
&
$[0,\!2|2,\!1,\!1,\!0|2,\!0]$ & \begin{tabular}{c} \colorbox{blue!10}{$\DD_{12}^2\ZZ\XX$} \\ $\DD_{12}\Psi_{11}\bar\Psi_{42}$ \end{tabular}  & 
\begin{tabular}{c} \{3,1,1,2\} \\ \{-2,-4,1,-1\} \end{tabular} 
& 
$\{1,\tx^{-1},\tx,1\}$
\\ \hline
\begin{tabular}{c}
\hspace{1.1mm}\begin{picture}(39,40)
\linethickness{0.2mm}\color{gray}
\put(18,12){\line(1,0){18}}
\put(6,18){\line(1,0){30}}
\put(6,24){\line(1,0){18}}
\put(6,30){\line(1,0){18}}
\put(6,36){\line(1,0){18}}
\put(6,18){\line(0,1){18}}
\put(12,18){\line(0,1){18}}
\put(18,12){\line(0,1){24}}
\put(24,12){\line(0,1){24}}
\put(30,12){\line(0,1){6}}
\put(36,12){\line(0,1){6}}
\linethickness{0.6mm}\color{orange}
\put(6,12){\line(0,1){6}}
\put(6,18){\line(1,0){12}}
\put(18,18){\line(0,1){6}}
\put(18,24){\line(1,0){6}}
\put(24,24){\line(0,1){12}}
\put(24,36){\line(1,0){6}}
\end{picture}\vspace{-5mm}
\\
\grpath{0}{2}{3}{3}
\end{tabular}
& 4 & 3 
&
$[0,\!0|3,\!1,\!0,\!0|2,\!0]$ & \colorbox{blue!10}{$\ZZ\Psi_{11}^2$}  & 
\begin{tabular}{c} \{3,1,0,3\} \\ \{-3,-4,0,-2\} \end{tabular} 
& 
$\{\tx,\tx^{-3},\tx^2,1\}$
\\ \hline
\begin{tabular}{c}
\hspace{1.1mm}\begin{picture}(39,40)
\linethickness{0.2mm}\color{gray}
\put(18,12){\line(1,0){24}}
\put(6,18){\line(1,0){36}}
\put(6,24){\line(1,0){24}}
\put(-6,30){\line(1,0){36}}
\put(-6,36){\line(1,0){24}}
\put(-6,30){\line(0,1){6}}
\put(0,30){\line(0,1){6}}
\put(6,18){\line(0,1){18}}
\put(12,18){\line(0,1){18}}
\put(18,12){\line(0,1){24}}
\put(24,12){\line(0,1){18}}
\put(30,12){\line(0,1){18}}
\put(36,12){\line(0,1){6}}
\put(42,12){\line(0,1){6}}
\linethickness{0.6mm}\color{orange}
\put(6,12){\line(0,1){6}}
\put(6,18){\line(1,0){12}}
\put(18,18){\line(0,1){12}}
\put(18,30){\line(1,0){12}}
\put(30,30){\line(0,1){6}}
\end{picture}\vspace{-5mm}
\\
\grpath{0}{2}{2}{4}
\end{tabular}
& 4 & 4 &
$[0,\!0|4,\!2,\!2,\!0|0,\!0]$ & \colorbox{blue!10}{$\ZZ^2\XX^2$}  
& 
\begin{tabular}{c} \{4,2,2,3\} \\ \{-3,-4,-1,-2\} \end{tabular}
& 
$\{1,\tx^{-2},\tx^2,1\}$
\\ \hline
%
\end{tabular}
\caption{Representative operators from the submultiplets that we consider in this paper. We will refer to the operators by the $\colorbox{blue!10}{highlighted}$ examples listed in the \textit{possible field content} column. Note that this does not refer to the precise structure of the operator.}
\label{tab:examples}
\end{table}

\begin{shaded}
\noindent 
{\bf Example: $\Psi_{11}\FF_{11}$ - the HWS in \grpath{2}{3}{3}{3}}

\noindent 
Throughout the paper, we will exemplify our approach by considering the submultiplet containing the HWS of the undeformed Konishi multiplet in the $12\hat13\hat2\hat3\hat44$ (\grpath{2}{3}{3}{3}) grading, with oscillator numbers $\ns=[0,0|1,0,0,0|3,0]$ and consequently field content 
\be
\Psi_{11}\FF_{11}&=& \aaa_1^\dagger \fff_1^\dagger |0\rangle \otimes (\aaa_1^{\dagger})^2 |0\rangle\,. \no
\ee 
This is the fourth example in table \ref{tab:examples}. The grading path and the Young diagram corresponding to this operator in the undeformed theory are

\begin{center}
\begin{picture}(100,85)

\color{black}
\linethickness{0.7mm}

\put(20,0){\line(1,0){40}}
\put(20,20){\line(1,0){40}}
\put(20,40){\line(1,0){40}}
\put(20,60){\line(1,0){40}}
\put(20,80){\line(1,0){40}}

\put(20,0){\line(0,1){80}}
\put(40,0){\line(0,1){80}}
\put(60,0){\line(0,1){80}}

\color{orange}
\linethickness{1.3mm}

\put(0,0){\line(1,0){40}}
\put(40,0){\line(0,1){20}}
\put(40,20){\line(1,0){20}}
\put(60,20){\line(0,1){60}}
\put(60,80){\line(1,0){20}}

\end{picture}
\end{center}

\noindent For this operator (and the submultiplet that it belongs to) the twist \eqref{betatw} is
\be
\tx_a = \{1,e^{-\ii \tb},e^{+\ii \tb},1\} \equiv  \{1,\tx^{-1},\tx,1\}
\ee
while the shifted weights \eqref{shws} are
\be
\hl_a = \{3,0,0,2\}+\Lambda\,,\quad\quad \hn_i = \{-2,-3,2,-1\}-\Lambda\,.
\ee

\end{shaded}


\section{QSC essentials}\label{sec:QSC}
The Quantum Spectral Curve is a Riemann-Hilbert problem whose solutions, among other things, capture the spectrum of anomalous dimensions of single-trace operators. The generalization of the QSC to the twisted case does not change its algebraic structure, only the boundary conditions. In the following discussion of the twisted QSC, we closely follow the results and conventions of \cite{Kazakov:2015efa}.

\subsection{Q-system}
A very elegant aspect of the QSC is the $\gl(4|4)$ Q-system \cite{Gromov:2014caa}. We will consider it when finding the leading solutions of the QSC in section \ref{sec:lead}. It consists of a set of Q-functions, $Q_{ab...|ij...}$, with up to four asymmetric indices of each of two types taking values between 1 and 4. The Q-functions satisfy three types of QQ-relations:
\begin{subequations}
\label{QQ}
\be
Q_{A|I}Q_{Aab|I}&=&Q^+_{Aa|I}Q^-_{Ab|I}-Q^-_{Aa|I}Q^+_{Ab|I}\label{QQ1}\\
Q_{A|I}Q_{A|Iij}&=&Q^+_{A|Ii}Q^-_{A|Ij}-Q^-_{A|Ii}Q^+_{A|Ij}\label{QQ2}\\
Q_{Aa|I}Q_{A|Ii}&=&Q^+_{Aa|Ii}Q^-_{A|I}-Q^-_{Aa|Ii}Q^+_{A|I} \label{QQ3} \;.
\ee
\end{subequations}
We require that $Q_{\emp|\emp}=Q_{1234|1234}=1$.

\paragraph{Distinguished Q-functions.} We will call a Q-function {\it distinguished} if its indices take the lowest possible values, i.e.
\be \label{distQ}
\dQ_{a,s}\equiv Q_{12...a|12...s}\,.
\ee
The set of distinguished Q-functions $\dQ_{a,s}$ ($a,s=0, \dots, 4$) are related only by the QQ-relation \eqref{QQ3}.

\paragraph{Asymptotics.} 
The large $\spar$ asymptotics of a general Q-function is governed by \cite{Kazakov:2015efa}
\be\label{Qasymp}
Q_{A|I}\simeq 
\left(\prod_{a\in A} \frac{\cA_a \tx_a^{\ii \spar}\spar^{-\hl_a}}{\tx_a^{\frac{|A|-|I|-1}{2}}}\right)
\left(\prod_{i\in I} \cB_i \spar^{-\hn_i}\right)
\left( \frac{\prod_{a<b\in A}z_{a,b} \, \spar^{-\delta_{\tx_a,\tx_b}}\prod_{i<j\in I}\ii\frac{\hn_j-\hn_i}{u}}{\prod_{a\in A}\prod_{i\in I} z_{a,i}\,\spar^{-\delta_{\tx_a,1}}} \right)
\ee
where
\begin{subequations}
\be
z_{a,b}&=& \left\{ 
\begin{matrix} 
\tx_b - \tx_a  & \tx_a\neq \tx_b \\ 
\ii \tx_a (\hl_b-\hl_a)  & \tx_a=\tx_b
\end{matrix}
\right. \\
z_{a,i}&=& \left\{ 
\begin{matrix} 
1 - \tx_a  & \tx_a\neq 1 \\ 
\ii (1-\hl_a-\hn_i)  & \tx_a=1
\end{matrix}
\right. \,.
\ee
\end{subequations}
Of particular importance are the functions
\begin{subequations} \label{PQas}
\be
Q_{a|\emp}\equiv \bP_a &\simeq& \cA_a \tx_a^{\ii \spar} u^{-\hl_a} \\
 -\epsilon^{abcd} Q_{bcd|1234} \equiv \bP^a &\simeq& \cA^a \tx_a^{-\ii u} \spar^{
 \hl_a-4\delta_{\tx_a,1}+\sum_{b\neq a}\delta_{\tx_a,\tx_b}
 } \equiv  \cA^a \tx_a^{-\ii \spar} \spar^{\hl_a^\star} \label{hlstar} \\
Q_{\emp|i}\equiv \bQ_i &\simeq& \cB_i u^{-\hn_i} \\
 -\epsilon^{ijkl} Q_{1234|jkl} \equiv \bQ^i &\simeq& \cB^i 
\spar^{
\hn_i
-\sum_{a} \delta_{\tx_a,1} +3
 } \equiv \cB^i \spar^{\hn_i^\star}\,,
\ee
\end{subequations}
where the normalizations $\cA$ and $\cB$ satisfy (no sums over $a$ or $i$) \cite{Kazakov:2015efa}
\begin{subequations}\label{AABB}
\be
\cA_a\cA^a &=& \frac{1}{\tx_a}\frac{\prod_i z_{a,i}}{\prod_{b\neq a}z_{b,a}}\\
\cB_i\cB^i &=& \frac{\prod_a z_{a,i}}{\prod_{j\neq i} \ii (\hn_i-\hn_j)}\,.
\ee
\end{subequations}
We have the freedom to choose $\cA_a$ and $\cB_i$ freely as long as the products \eqref{AABB} are satisfied, but to maintain $Q_{1234|1234}=1$, the choice should satisfy 
\be
\prod_{a=1}^4 \cA_a \prod_{i=1}^4 \cB_i =  \frac{\prod_{a,i}z_{a,i}}{\prod_{1\le a<b\le 4} z_{a,b} \prod_{1\le i < j \le 4}\ii(\hn_j - \hn_i)}\,.
\ee

\subsection{\texorpdfstring{$\bP\mu$}{Pmu}-system}\label{sec:PmuSys}
The full Q-system carries a lot of redundant information which can be reduced into a much more compact formulation, called the $\bP\mu$-system. This consists of the $2\times 4$ functions $\bP_a$ and $\bP^a$ introduced in \eqref{PQas}, and six additional functions of the spectral parameter arranged in the anti-symmetric symbol $\mu_{ab}$. These, in turn, build the upper-index functions
\begin{equation}
  \label{eq:muUpper}
  \mu^{ab} = - \frac{1}{2} \epsilon^{abcd} \mu_{cd} \; \frac{1}{\Pf(\mu)}
  ,
\end{equation}
where the Pfaffian of $\mu$,
\begin{equation}
  \Pf(\mu) =  \frac{1}{8} \epsilon^{abcd} \mu_{ab} \mu_{cd}
  ,
\end{equation}
is in fact a constant determined by the normalization of the Q-system. The upper- and lower-indexed functions satisfy the relations
\begin{align}
  \label{eq:PmuRels}
  \bP_a \bP^a &= 0
  , &
  \mu_{ab} \mu^{bc} &= \delta_a^c
  .
\end{align}
They are all multivalued functions of the spectral parameter $\spar$ and have a very precise analytic structure, as we will see below. It contains an infinite number of branch cuts, that are all of square root type, while the QSC functions are required to be analytic everywhere else. Let us state the analytic properities of the functions individually.

\begin{figure}
  \centering
  \includegraphics{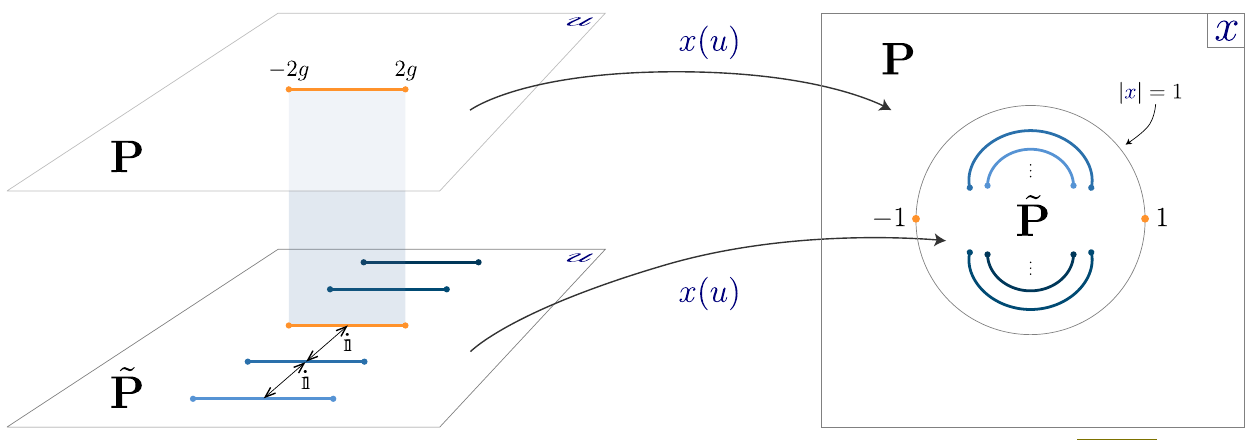}
  \caption{Analytic structure of $\bP$ and the effect of the Zhukowsky map (defined below in section \ref{sec:anzP}). $\bP$ has only one cut in between the branch points $\pm 2 \g$ on its first sheet. It connects to the second sheet where $\Pt$ has an infinite number of cuts at $\pm 2\g + \ii \mathbb{Z}$. The Zhukowsky map resolves the cut at $\pm 2\g$ while mapping the first sheet onto the exterior of the unit circle and the second into the interior. The plot uses the example value $\g=2$ and indicates the infinite number of cuts by ellipses in the $\zh$-plane.}
  \label{fig:analStructP}
\end{figure}

\paragraph{Analytic structure of \texorpdfstring{$\bP$}{P}.}
The multivalued functions $\bP$ have one Riemann sheet with only a single branch cut\footnote{In this paper we exclusively choose \emph{short} cuts. One could of course choose different branch cuts, e.g. \emph{long} cuts that connect the branch points through infinity, but the short cuts are the natural choice in the weak coupling limit, $\g\to0$.}, in between the points $\pm  2\g$. We denote the function values on this sheet by $\bP(\spar)$.  The analytical continuation into the second sheet is denoted $\Pt(\spar)$, and on this sheet there is an infinite number of cuts at $\pm 2\g + \ii \mathbb{Z}$. This is illustrated to the left in figure \ref{fig:analStructP}.

\paragraph{Analytic structure of \texorpdfstring{$\mu$}{mu}.} The functions $\mu$ have an infinite number of cuts at $\pm 2\g + \ii \mathbb{Z}$ on all sheets but with the very special property
\begin{equation}
  \label{eq:muTildeShift}
  \tilde{\mu} = \mu^{[2]}
  \,,
\end{equation}
i.e. the analytic continuation through the cut at $\pm 2\g$ is the same as the values on the first sheet, only shifted by $\ii$. An important consequence of this it that both the following expressions are regular on the real axis:
\begin{equation}
  \label{eq:regConds}
      \mu + \mu^{[2]} \,,\quad\quad    \dfrac{\mu - \mu^{[2]}}{\sqrt{\spar^2 - 4 \g^2}}\,,
\end{equation}
which is exploited in the perturbative algorithm below.

\paragraph{Solutions corresponding to single-trace operators.} 
An important constraint is that for the solutions to the QSC that correspond to single-trace operators, $\mu$ need to have power-like large $\spar$ asymptotics, up to an overall exponential twist factor, i.e.
\be
\mu_{ab}\sim (\tx_a\tx_b)^{\ii \spar} \cdot \spar^{M_{ab}}
\ee
where $M_{ab}$ is some integer. Furthermore, $\mu$ should satisfy the zero-momentum condition
\be
\lim_{\spar\to0} \frac{\mu_{ab}(u)}{\mu_{ab}(u+\ii)}=1\,.
\ee

\paragraph{Relations.} The analytic continuation $\Pt$ is given through $\mu$ and $\bP$ by
\begin{align}
  \label{eq:PmuPt}
  \Pt_a &= \mu_{ab} \bP^b ,	&	\Pt^a &= \mu^{ab} \bP_{b}
  \,,
\end{align}
and these functions are further related by the equation
\begin{equation}
  \mu_{ab} - \tilde{\mu}_{ab} = - \bP_a \Pt_b + \bP_b \Pt_a
  \label{eq:muPPt}
  .
\end{equation}
Using equations \eqref{eq:muTildeShift} and \eqref{eq:PmuPt}, it can be rewritten as
\begin{equation}
  \label{eq:muDiffEq}
 \nabla \mu_{ab} \equiv \mu_{ab} - \mu_{ab}^{[2]} = - \bP_a \bP^c \mu_{bc}^{[1 \pm 1]} + \bP_b \bP^c \mu_{ac}^{[1 \pm 1]}
  .
\end{equation}
These are in fact the same difference equations that are satisfied by the central Q-functions $Q_{ab|ij}^-$. Thus each $\mu_{ab}$ can be written as a linear combination of the six corresponding $Q_{ab|ij}^-$.

\paragraph{Symmetries.}
The $\bP\mu$-system is subject to two important symmetries \cite{Gromov:2014caa}. First of all, the \emph{gauge} transformations
\be \label{gauge}
\bP_a\to \zh^\Lambda \bP_a \quad\quad \bP^a \to \zh^{-\Lambda}\bP^a
\ee
related to the freedom in the shifted weights \eqref{shws}; $\zh$ is defined below in equatoin \eqref{eq:zhukowsky}. Second, one can use the $H$-symmetry
\be
\bP_a\to H_a^{\;b} \bP_b \quad\quad \mu_{ab}\to H_a^{\;c} H_b^{\;d}\mu_{cd}\,,
\ee
where $H$ is a constant matrix, to rotate in the basis of $\bP$'s and $\mu$'s.

\subsection{All-loop ansatz for \texorpdfstring{$\bP$}{P}}\label{sec:anzP}
It is possible to construct an ansatz for $\bP$ thanks to their simple analytic structure, which is central for the perturbative algorithm described in section \ref{sec:pert}. This procedure was described thoroughly in \cite{Marboe:2018ugv}, and we simply make the natural generalization to the twisted scenario. The crucial idea is to use the Zhukowsky map to write an expansion that converges everywhere on the first sheet $\bP(u)$, and also in a finite region on the second sheet $\Pt(u)$.

\paragraph{The Zhukowsky map.} The two first sheets of $\bP$ can be brought together into one by introducing the Zhukowsky variable
\begin{equation}
  \zh + \frac{1}{\zh} = \frac{\spar}{\g} 
  .
  \label{eq:zhukowsky}
\end{equation}
The single cut on the first $\spar$-sheet is mapped to the unit circle in the $\zh$-plane. It is hence dissolved as the first sheet is mapped to the region $|\zh|>1$ while the second sheet is mapped into the interior of the unit circle. %
As $\zh$ is a double-valued function of $\spar$, we always choose the branch $|\zh|>1$ and substitute $\zh \to \frac{1}{\zh}$ for values on the second $\spar$-sheet. The map is illustrated in figure \ref{fig:analStructP}. Note that expanding the Zhukowsky variable in $\g$ gives
\begin{align}
  \zh = \frac{\spar}{\g} - \frac{\g}{\spar} - \frac{\g^3}{\spar^3} - \frac{2\g^5}{\spar^5} - \frac{5\g^7}{\spar^7} + \CO(\g^9)
\end{align}
and that the large $\spar$-asymptotics is $\zh \asympt \frac{\spar}{\g}$.

\paragraph{Explicit ansatz for \texorpdfstring{$\bP$}{P}.} 
By generalizing the ansatz of \cite{Marboe:2018ugv} and by testing it in explicit calculations, we propose the following ansatz for the functions $\bP$:
\begin{subequations}
  \label{eq:PAnz}
  \begin{align}
    \bP_a &= \tx_a^{\ii \spar}  \left( \g \zh \right)^{-\Lanz + \Lambda - \deltaL}
    \left(
      \sum_{k=0}^{\Lanz - \hl_a+\Lambda+\deltaL} \cd_{a,k} (\g \zh)^k
      + \sum_{k=1}^{\infty} \cc_{a,k} \left( \frac{\g}{\zh} \right)^k
    \right)
    \,,
    \\
    \bP^a &= \tx_a^{-\ii \spar}  \left( \g \zh \right)^{-\Lambda+\deltaL}
    \left(
      \sum_{k=0}^{\hl_a^{\star}-\Lambda-\deltaL} \cd^{a,k} (\g \zh)^k
      + \sum_{k=1}^{\infty} \cc^{a,k} \left( \frac{\g}{\zh} \right)^k
    \right)
    \,.
  \end{align}
\end{subequations}
Note that the combinations $\hl_a-\Lambda$ and $\hl_a^\star-\Lambda$ are independent of the choice of $\Lambda$, cf. the definitions \eqref{shws} and \eqref{hlstar}. We here introduced two new numbers, $L^\star$ and $\deltaL$. The former is a modified version of the operator length. The pattern that we find is that $\Lanz$ corresponds to the lowest operator length with which the quantum numbers $\hl$ and $\hn$ can be achieved. The number $\deltaL$ is an offset that we, for the states from the Konishi multiplet, find to be 0 for all gradings except those that end with $...\hat4$, where it takes the value 1. 
\begin{shaded}
\noindent {\bf $\Lanz$ and $\deltaL$ for the examples in table \ref{tab:examples}.}

\noindent 
For five of the eight examples in table \ref{tab:examples} the length is $L=2$, which is the lowest length of any operator in the Konishi multiplet, so naturally we have $\Lanz=2$. For the \grpath{0}{2}{2}{2} HWS $\ZZ\XX\YY$ with $L=3$ and the \grpath{0}{2}{2}{4} HWS $\ZZ^2\XX^2$ with $L=4$, we also have $\Lanz=2$ as they have the same $\hl$ and $\hn$ as the \grpath{1}{2}{2}{2} HWS $\ZZ\bar\Psi_{22}$ with $L=2$. However, the \grpath{0}{2}{3}{3} HWS $\ZZ\Psi_{11}^2$ with $L=3$ has quantum numbers that cannot be achieved for a length-2 state, and consequently it has $\Lanz=3$.

The \grpath{0}{2}{2}{4} HWS $\ZZ^2\XX^2$ has $\deltaL=1$ due to its grading ending as {\color{Orange} \bf ...4}, but all others have $\deltaL=0$. 

\end{shaded}
\noindent
Furthermore, for the states from the Konishi multiplet we assume\footnote{Some QSC-solutions may have $\g$-expansions that also contain odd powers, as discussed in \cite{Marboe:2018ugv}. For our scope, the current assumption suffices.} that the coefficients $\cc$ and $\cd$ in the ansatz \eqref{eq:PAnz} have regular expansions in $\g^2$: 
\begin{align}
  \cc &= \sum_{j=0}^{\infty} \cc^{(j)} \g^{2j}
  \,, &
  \cd &= \sum_{j=0}^{\infty} \cd^{(j)} \g^{2j}
  .
\end{align}
Naturally, the coefficient of the highest power in $\spar$ is fixed by the chosen normalizations $\cA$. 

The beauty of the expansion in $\zh$ is that it converges for all $|\zh|>1$ but can be extended through the resolved cut into a finite region inside $|\zh|<1$ (until the first singularity at $\zh(\spar \pm \ii)$). As such, it does also cover a region on the second $\spar$-sheet, where the ansatz for $\Pt$ close to $|\zh| = 1$ ($\spar = 0$ for small $\g$) can be obtained by replacing $\zh \to \frac{1}{\zh}$ in \eqref{eq:PAnz}.

For the purpose of perturbation theory, the ansatz \eqref{eq:PAnz} should be expressed in terms of $\spar$ and expanded in $\g$, giving us an expansion of $\bP$ on the form
\be
\bP=\bP^{(0)}+\g^2\bP^{(1)} + g^4\bP^{(2)}+...
\ee
A crucial feature is that each perturbative contribution $\bP^{(n)}$ only contain a finite number of unknown coefficients $\cc$ and $\cd$. As we will see in section \ref{sec:pert}, this gives us a starting point for doing perturbation theory.

\begin{shaded}
  \noindent {\bf Example: $\Psi_{11} \FF_{11}$ - \grpath{2}{3}{3}{3}}

  \noindent 
  For this example, the asymptotics of $\bP$ and $\bQ$ are dictated by the weights
  \begin{align}
  \hl &= \{3, 0, 0, 2\} +\Lambda \,, & \no
  \hn\big|_{\g=0} &= \{-2, -3, 2, -1\}+\Lambda \,, 
  \\
  \hl^\star &= \{0, 0, 0, -1\}-\Lambda\,, & 
  \hn^\star\big|_{\g=0} &= \{-1, -2, 3, 0\}-\Lambda \,.
\end{align}
  The products \eqref{AABB} are
  \begin{align}
 \cA_1\cA^1  &= \frac{\ii \tx \adim  (\adim +2)^2 (\adim +8)}{16 (\tx-1)^2} \,, & \cB_1 \cB^1 &=  \frac{\ii (\tx-1)^2 \adim  (\adim +2)}{4 \tx (\adim +1) (\adim +4)} \no
\,,\\
 \cA_2\cA^2  &= \frac{2}{\tx+1}-1 \,, & \cB_2 \cB^2 &=  -\frac{\ii (\tx-1)^2 (\adim +4)}{4 \tx (\adim +5)} \no
\,,\\
 \cA_3\cA^3  &= \frac{\tx-1}{\tx+1} \,, & \cB_3 \cB^3 &=  \frac{\ii (\tx-1)^2 (\adim +6) (\adim +8)}{12 \tx (\adim +4) (\adim +5)} \no
\,,\\
 \cA_4\cA^4  &= -\frac{\ii \tx \adim  (\adim +2) (\adim +4) (\adim +6)}{16 (\tx-1)^2} \,, & \cB_4 \cB^4 &=  -\frac{\ii (\tx-1)^2 \adim }{12 \tx (\adim +1)} 
\,,
  \end{align}
where we make the choices
\begin{align}
 \cA_1 &= -\frac{\ii \tx \adim  (\adim +2)}{4 (\tx-1)^2} \,, & \cB_1 &=  \frac{\ii}{\adim ^2+5 \adim +4} \,,\no
\\
 \cA_2 &= \frac{\tx-1}{\tx (\tx+1)} \,, & \cB_2 &=  -\frac{1}{\adim ^2+7 \adim +10} \,,\no
\\
 \cA_3 &= -(\tx-1)^2 \,, & \cB_3 &=  \frac{1}{6} (-\adim -8) \,,\no
\\
\cA_4 &= \frac{1}{8} \ii (\adim +2) (\adim +4) (\adim +6) \,, & \cB_4 &=  \frac{(\tx-1)^2 \adim  (\adim +2)}{4 \tx} \,.
\end{align}
For this state, we have $\Lanz = 2$ and $\deltaL = 0$, and setting $\Lambda = 0$ the ansatz \eqref{eq:PAnz} for $\bP$ looks like
\begin{subequations}
  \begin{align}
    \bP_a &= \tx_a^{\ii \spar}  \left( \g \zh \right)^{2}
    \left(
      \sum_{k=0}^{ \{-1, 2, 2, 0\}_a} \cd_{a,k} (\g \zh)^k
      + \sum_{k=1}^{\infty} \cc_{a,k} \left( \frac{\g}{\zh} \right)^k
    \right)
    \,,
    \\
    \bP^a &= \tx_a^{-\ii \spar}  
    \left(
      \sum_{k=0}^{ \{0, 0, 0, -1 \}_a} \cd^{a,k} (\g \zh)^k
      + \sum_{k=1}^{\infty} \cc^{a,k} \left( \frac{\g}{\zh} \right)^k
    \right)
    \,.
  \end{align}
\end{subequations}
Two examples of the explicit $\g$-expansion to second order are
\begin{align}
  \bP_2 & = \tx^{-\ii \spar} \left(\frac{\cd_{2,0}^{(0)}}{\spar^2}+\frac{\cd_{2,1}^{(0)}}{\spar}+\frac{\tx-1}{\tx (\tx+1)}\right) + \g^2 \tx^{-\ii \spar} \left(\frac{\cc_{2,1}^{(0)}+\cd_{2,1}^{(0)}}{\spar^3}+\frac{2 \cd_{2,0}^{(0)}}{\spar^4}+\frac{\cd_{2,0}^{(1)}}{\spar^2}+\frac{\cd_{2,1}^{(1)}}{\spar}\right)  + \CO(\g^4)
 \,,
\no \\
 \Pt_2 &= -\tx^{-\ii \spar} \left( \spar^2 \cd_{2,0}^{(0)} + \spar^2 \sum_{k=1}^{\infty} \spar^k \cc_{2,k}^{(0)} \right) 
 + \g^2 \tx^{-\ii \spar} \Bigg( -2 \cd_{2,0}^{(0)}  + \spar \Big( -3 \cc_{2,1}^{(0)} + \cd_{2,1}^{(0)}\Big)\no \\
  &\quad\, 
  + \spar^2\Big( -4 \cc_{2,2}^{(0)} + \cd_{2,0}^{(1)} \Big) 
+ \spar^2 \sum_{k=1}^{\infty} \spar^k \Big( \cc_{2,k}^{(1)} - (k+2) \cc_{2,k+2}^{(0)}\Big) \Bigg) + \CO(\g^{4})
\,,
\end{align}
where we have substituted the asymptotic coefficients $\cA_2$ and $\cA^2$.
\end{shaded}

\section{The leading Q-system}\label{sec:lead}

For each symmetry multiplet, there is a distinct solution to the QSC. In the undeformed theory, the infinite set of operators in the Konishi multiplet correspond to just a single solution. In the $\tb$-deformed theory, this solution splits up into several solutions corresponding to the submultiplets. These solutions should all reduce to the undeformed solution in the limit $\tb\to0$.

To find the solution, we generalize the strategy \cite{Marboe:2017dmb} to find the leading Q-system. The idea of the method is to first find the subset of distinguished Q-functions \eqref{distQ}, which are related only by one type of QQ-relation \eqref{QQ3}, by imposing so-called {\it zero-remainder conditions} on them. To do this, we first have to understand the boundary conditions of the problem, i.e. the precise structure of the distinguished Q-functions, given the quantum numbers and HWS grading of the operator in question.

\subsection{Explicit boundary conditions}
For the undeformed theory, the precise boundary conditions for the leading Q-system were discussed in \cite{Marboe:2017dmb,Marboe:2017zdv}. In that work, the concept of a larger Young diagram Q-system was introduced, from which the $\gl(4|4)$ Q-system could be picked out as a subset. In the deformed theory, the concept of extended Young diagrams is not immediately applicable, so in our context we will work directly with the $\gl(4|4)$ Q-system.

The boundary conditions only change slightly for the twisted case. First of all, exponential twist-dependent factors appear. Furthermore, it is well-known that the number Bethe roots, i.e. the number of zeros in the Q-functions, is affected by twisting. The number of roots that appear in the Bethe equations is unchanged, but the number of roots in other Q-functions\footnote{Note that one can in principle also write down Bethe equations for these functions by performing so-called duality transformations.} will be altered. Before describing the boundary conditions for the twisted $\psu(2,2|4)$ spin chain, we take a look at the $\su(2)$ Q-system to see these features.

\begin{shaded}
\noindent {\bf Example: boundary conditions of twisted $\su(2)$ Q-system}

\noindent For example, eigenstates of the $\su(2)$ spin chain of the form $\ZZ^{L-M}\XX^{M}$ correspond to polynomial solutions to the single QQ-relation
\be\label{su2QQ}
Q_1^+Q_2^- - Q_1^-Q_2^+ = \spar^L
\ee
where $Q_1$ and $Q_2$ have the form
\be
Q_1=\prod^{M}_{k=1} (\spar-u_k)\,,\quad \quad\quad Q_2 \propto \prod^{{\color{rootcol2}L-M+1}}_{k=1} (\spar-v_k)\,.
\ee
In the twisted case, the QQ-relation remains the same%
, but the Q-functions change structure to
\be
Q_1= 
\tz^{\ii u}
\prod^{M}_{k=1} (\spar-u_k)\,,\quad \quad\quad 
Q_2 \propto \tz^{-\ii u} \prod^{{\color{rootcol2}L-M}}_{k=1} (\spar-v_k)\,,
\ee
where $\tz$ is some twist.  %
The degree of $Q_2$ is lowered by one, as a consequence of the fact that the leading powers in $u$ of the two terms on the left hand side of \eqref{su2QQ} do not cancel. 

The important point is that the change in the number of roots happens only in the Q-functions that do not appear in the Bethe equations, which are
\be
-\frac{Q_1(u_k+\ii)}{Q_1(u_k-\ii)} = \left(\frac{u_k+\frac{\ii}{2}}{u_k-\frac{\ii}{2}}\right)^L\,.
\ee
\end{shaded}

\noindent In analogy with this example, we can make an educated guess for a concrete ansatz for the larger $\psu(2,2|4)$ Q-system. 
Our main requirement is that the number of Bethe roots in the Q-functions on the HWS grading path of the operator remains unchanged from the twisted to the untwisted case.
Importantly, this requirement is in agreement with the asymptotic structure of the Q-functions \eqref{Qasymp}.

\paragraph{Structure of distinguished Q-functions.} The distinguished Q-functions \eqref{distQ} have the overall structure
\be
\dQ_{a,s}=\left(\prod_{b=1}^a \tx_b\right)^{\ii u} \, f_{a,s}(\spar) \, {q}_{a,s}(\spar) 
\ee
where $f_{a,s}(\spar)$ is trivial "fusion factor" containing shifted powers of $\spar^{\pm L}$ and where 
$q_{a,s}(\spar)$ 
is a polynomial factor carrying the Bethe roots.

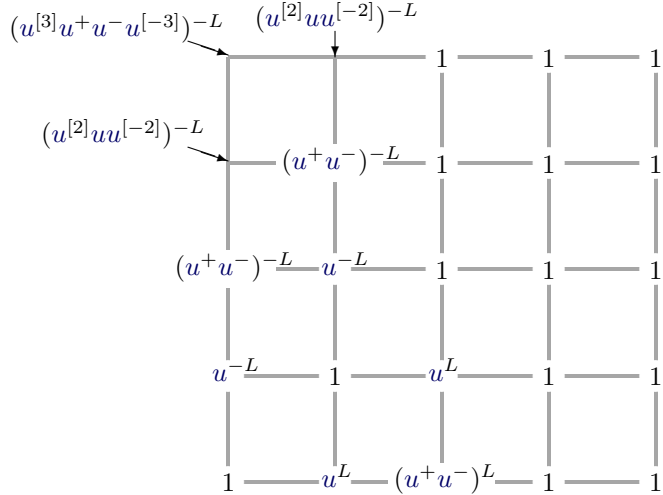
\begin{figure}[h!]
\centering
\begin{picture}(160,195)

{\color{gray!70}
\linethickness{0.5mm}

\put(10,10){\line(1,0){160}}
\put(10,50){\line(1,0){160}}
\put(10,90){\line(1,0){160}}
\put(10,130){\line(1,0){160}}
\put(10,170){\line(1,0){160}}

\put(10,10){\line(0,1){160}}
\put(50,10){\line(0,1){160}}
\put(90,10){\line(0,1){160}}
\put(130,10){\line(0,1){160}}
\put(170,10){\line(0,1){160}}
}

\small
\linethickness{5mm}

\put(-12,90){\color{white}\line(1,0){40}} \put(-10,88){$(\spar^+\spar^-)^{-L}$} 
\put(50,90){\color{white}\circle*{12}} \put(45,88){$\spar^{-L}$}
\put(90,90){\color{white}\circle*{12}} \put(87.6,86.5){$1$}
\put(130,90){\color{white}\circle*{12}} \put(127.6,86.5){$1$}
\put(170,90){\color{white}\circle*{12}} \put(167.6,86.5){$1$}

\put(10,50){\color{white}\circle*{12}} \put(4,48){$\spar^{-L}$}
\put(50,50){\color{white}\circle*{12}} \put(47.6,46.5){$1$}
\put(90,50){\color{white}\circle*{12}} \put(85,48){$\spar^{L}$}
\put(130,50){\color{white}\circle*{12}} \put(127.6,46.5){$1$}
\put(170,50){\color{white}\circle*{12}} \put(167.6,46.5){$1$}

\put(10,10){\color{white}\circle*{12}} \put(7.6,6.5){$1$}
\put(50,10){\color{white}\circle*{12}} \put(45,8){$\spar^{L}$}
\put(70,10){\color{white}\line(1,0){40}} \put(72,8){$(\spar^+\spar^-)^{L}$} 
\put(130,10){\color{white}\circle*{12}} \put(127.6,6.5){$1$}
\put(170,10){\color{white}\circle*{12}} \put(167.6,6.5){$1$}

\put(-60,138){$(\spar^{[2]}\spar\spar^{[-2]})^{-L}$} \put(-5,136){\vector(3,-1){15}}
\put(28,130){\color{white}\line(1,0){40}} \put(30,128){$(\spar^+\spar^-)^{-L}$} 
\put(90,130){\color{white}\circle*{12}} \put(87.6,126.5){$1$}
\put(130,130){\color{white}\circle*{12}} \put(127.6,126.5){$1$}
\put(170,130){\color{white}\circle*{12}} \put(167.6,126.5){$1$}

\thinlines
\put(-72,178){$(\spar^{[3]}\spar^+\spar^-\spar^{[-3]})^{-L}$} \put(-5,176){\vector(3,-1){15}}
\put(20,182){$(\spar^{[2]}\spar\spar^{[-2]})^{-L}$} \put(50,180){\vector(0,-1){10}}
\put(90,170){\color{white}\circle*{12}} \put(87.6,166.5){$1$}
\put(130,170){\color{white}\circle*{12}} \put(127.6,166.5){$1$}
\put(170,170){\color{white}\circle*{12}} \put(167.6,166.5){$1$}
\end{picture}
\vspace{-3mm}
\caption{The value of $f_{a,s}$ in the $\psu(2,2|4)$ Q-system.}
\label{fig:fus}
\end{figure}

\paragraph{Fusion factor \texorpdfstring{$f_{a,s}$}{f_a,s}.} The fusion factors have the form
\be \label{fpow}
f_{a,s}(\spar)=\left\{ \begin{matrix} \prod_{k=-|s-a|}^{|s-a|} \left( \spar-\frac{\ii\,k}{2} \right)^{\text{sign}(s-a)\,L} & \quad\text{if} \quad s < 0\vee (s=0\wedge a<0)  \\ 1 & \quad\text{otherwise} \end{matrix} \right.\,.
\ee
Note that this corresponds to a particular fixing of the gauge symmetry \eqref{gauge}. %
The values of $f_{a,s}$ for the $\psu(2,2|4)$ Q-system are illustrated in figure \ref{fig:fus}.

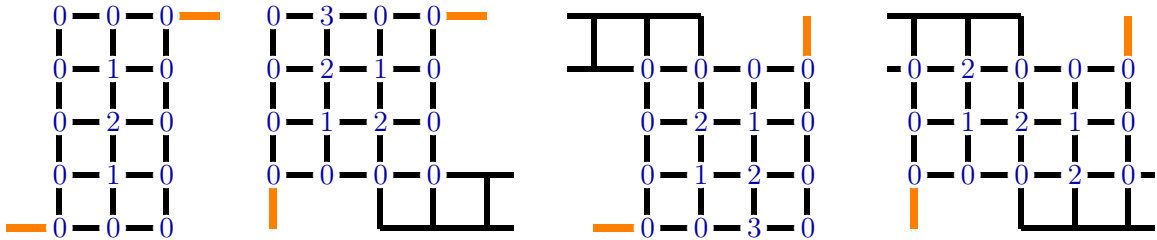
\begin{figure}[h]%
\centering
\begin{picture}(420,85)

\color{orange}
\linethickness{1mm}

\put(0,0){\line(1,0){20}}
\put(60,80){\line(1,0){20}}

\put(100,0){\line(0,1){20}}
\put(160,80){\line(1,0){20}}

\put(220,0){\line(1,0){20}}
\put(300,60){\line(0,1){20}}

\put(340,0){\line(0,1){20}}
\put(420,60){\line(0,1){20}}

\color{black}
\linethickness{0.7mm}

\put(20,0){\line(1,0){40}}
\put(20,20){\line(1,0){40}}
\put(20,40){\line(1,0){40}}
\put(20,60){\line(1,0){40}}
\put(20,80){\line(1,0){40}}

\put(20,0){\line(0,1){80}}
\put(40,0){\line(0,1){80}}
\put(60,0){\line(0,1){80}}

\put(100,80){\line(1,0){60}}
\put(100,60){\line(1,0){60}}
\put(100,40){\line(1,0){60}}
\put(100,20){\line(1,0){90}}
\put(140,0){\line(1,0){50}}

\put(100,20){\line(0,1){60}}
\put(120,20){\line(0,1){60}}
\put(140,0){\line(0,1){80}}
\put(160,0){\line(0,1){80}}
\put(180,0){\line(0,1){20}}

\put(210,80){\line(1,0){50}}
\put(210,60){\line(1,0){90}}
\put(240,40){\line(1,0){60}}
\put(240,20){\line(1,0){60}}
\put(240,0){\line(1,0){60}}

\put(220,60){\line(0,1){20}}
\put(240,0){\line(0,1){80}}
\put(260,0){\line(0,1){80}}
\put(280,0){\line(0,1){60}}
\put(300,0){\line(0,1){60}}

\put(330,80){\line(1,0){50}}
\put(330,60){\line(1,0){90}}
\put(340,40){\line(1,0){80}}
\put(340,20){\line(1,0){90}}
\put(380,0){\line(1,0){50}}

\put(340,20){\line(0,1){60}}
\put(360,20){\line(0,1){60}}
\put(380,0){\line(0,1){80}}
\put(400,0){\line(0,1){60}}
\put(420,0){\line(0,1){60}}


\color{rootcol1}

\put(20,0){\color{white}\circle*{10}} \put(17.5,-3.4){0}
\put(40,0){\color{white}\circle*{10}} \put(37.5,-3.4){0}
\put(60,0){\color{white}\circle*{10}} \put(57.5,-3.4){0}

\put(20,20){\color{white}\circle*{10}} \put(17.5,16.6){0}
\put(40,20){\color{white}\circle*{10}} \put(37.5,16.6){1}
\put(60,20){\color{white}\circle*{10}} \put(57.5,16.6){0}

\put(20,40){\color{white}\circle*{10}} \put(17.5,36.6){0}
\put(40,40){\color{white}\circle*{10}} \put(37.5,36.6){2}
\put(60,40){\color{white}\circle*{10}} \put(57.5,36.6){0}

\put(20,60){\color{white}\circle*{10}} \put(17.5,56.6){0}
\put(40,60){\color{white}\circle*{10}} \put(37.5,56.6){1}
\put(60,60){\color{white}\circle*{10}} \put(57.5,56.6){0}

\put(20,80){\color{white}\circle*{10}} \put(17.5,76.6){0}
\put(40,80){\color{white}\circle*{10}} \put(37.5,76.6){0}
\put(60,80){\color{white}\circle*{10}} \put(57.5,76.6){0}

\put(100,20){\color{white}\circle*{10}} \put(97.5,16.6){0}
\put(120,20){\color{white}\circle*{10}} \put(117.5,16.6){0}
\put(140,20){\color{white}\circle*{10}} \put(137.5,16.6){0}
\put(160,20){\color{white}\circle*{10}} \put(157.5,16.6){0}

\put(100,40){\color{white}\circle*{10}} \put(97.5,36.6){0}
\put(120,40){\color{white}\circle*{10}} \put(117.5,36.6){1}
\put(140,40){\color{white}\circle*{10}} \put(137.5,36.6){2}
\put(160,40){\color{white}\circle*{10}} \put(157.5,36.6){0}

\put(100,60){\color{white}\circle*{10}} \put(97.5,56.6){0}
\put(120,60){\color{white}\circle*{10}} \put(117.5,56.6){2}
\put(140,60){\color{white}\circle*{10}} \put(137.5,56.6){1}
\put(160,60){\color{white}\circle*{10}} \put(157.5,56.6){0}

\put(100,80){\color{white}\circle*{10}} \put(97.5,76.6){0}
\put(120,80){\color{white}\circle*{10}} \put(117.5,76.6){3}
\put(140,80){\color{white}\circle*{10}} \put(137.5,76.6){0}
\put(160,80){\color{white}\circle*{10}} \put(157.5,76.6){0}

\put(240,0){\color{white}\circle*{10}} \put(237.5,-3.4){0}
\put(260,0){\color{white}\circle*{10}} \put(257.5,-3.4){0}
\put(280,0){\color{white}\circle*{10}} \put(277.5,-3.4){3}
\put(300,0){\color{white}\circle*{10}} \put(297.5,-3.4){0}

\put(240,20){\color{white}\circle*{10}} \put(237.5,16.6){0}
\put(260,20){\color{white}\circle*{10}} \put(257.5,16.6){1}
\put(280,20){\color{white}\circle*{10}} \put(277.5,16.6){2}
\put(300,20){\color{white}\circle*{10}} \put(297.5,16.6){0}

\put(240,40){\color{white}\circle*{10}} \put(237.5,36.6){0}
\put(260,40){\color{white}\circle*{10}} \put(257.5,36.6){2}
\put(280,40){\color{white}\circle*{10}} \put(277.5,36.6){1}
\put(300,40){\color{white}\circle*{10}} \put(297.5,36.6){0}

\put(240,60){\color{white}\circle*{10}} \put(237.5,56.6){0}
\put(260,60){\color{white}\circle*{10}} \put(257.5,56.6){0}
\put(280,60){\color{white}\circle*{10}} \put(277.5,56.6){0}
\put(300,60){\color{white}\circle*{10}} \put(297.5,56.6){0}

\put(340,20){\color{white}\circle*{10}} \put(337.5,16.6){0}
\put(360,20){\color{white}\circle*{10}} \put(357.5,16.6){0}
\put(380,20){\color{white}\circle*{10}} \put(377.5,16.6){0}
\put(400,20){\color{white}\circle*{10}} \put(397.5,16.6){2}
\put(420,20){\color{white}\circle*{10}} \put(417.5,16.6){0}

\put(340,40){\color{white}\circle*{10}} \put(337.5,36.6){0}
\put(360,40){\color{white}\circle*{10}} \put(357.5,36.6){1}
\put(380,40){\color{white}\circle*{10}} \put(377.5,36.6){2}
\put(400,40){\color{white}\circle*{10}} \put(397.5,36.6){1}
\put(420,40){\color{white}\circle*{10}} \put(417.5,36.6){0}

\put(340,60){\color{white}\circle*{10}} \put(337.5,56.6){0}
\put(360,60){\color{white}\circle*{10}} \put(357.5,56.6){2}
\put(380,60){\color{white}\circle*{10}} \put(377.5,56.6){0}
\put(400,60){\color{white}\circle*{10}} \put(397.5,56.6){0}
\put(420,60){\color{white}\circle*{10}} \put(417.5,56.6){0}

\end{picture}
\caption{Number of Bethe roots in the undeformed distinguished Q-functions for the four short multiplets that form the Konishi multiplet. As described in \cite{Marboe:2017dmb}, the four Q-systems are compatible due to certain roots being placed at $\spar=\frac{\ii}{2}\mathbb{Z}$ and due to symmetry transformations that suppress different Q-functions in $g$.}
\label{fig:YDroots}
\end{figure}

\paragraph{Counting Bethe roots.} One can use the asymptotics \eqref{Qasymp} to deduce the number of roots in $\dQ_{a,s}$, by subtracting the powers coming from the fusion factors. The pattern that one finds is that on the grading path where the operator in question is a HWS, the number of roots is the same in the undeformed and deformed theory. We will discuss the case where the operator is not a HWS in any grading in the undeformed theory in the example in section \ref{ex:DZX}. For states that are HWS in some grading in the undeformed theory, we can then simply use the counting rules from the undeformed theory \cite{Marboe:2017dmb} to find the number of roots in the Q-functions on the HWS grading path. The number of roots for operators from the four short multiplets forming the Konishi multiplet are depicted in figure \ref{fig:YDroots}. Knowing the number of roots on the grading path, one can deduce the number of roots in the remaining $\dQ_{a,s}$. One has to count powers in the QQ-relation \eqref{QQ3}, while also taking the fusion factors into account. Importantly, one should take into account whether the Q-functions have differing exponential factors or not. We give an examples of the counting procedure at the end of the section.

\subsection{Explicit solutions from zero-remainder conditions}
Knowing the number of roots in the distinguished Q-functions, we can write a precise ansatz for them in terms of a finite number of coefficients, i.e.
\be
q_{a,s}(\spar) &\propto& \spar^{M_{a,s}} + \sum_{j=0}^{M_{a,s}-1} c_{a,s,j} \, \spar^j \,, 
\ee
where $M_{a,s}$ is the number of Bethe roots in $q_{a,s}$. To determine the coefficients $c$, we choose a particular path through the Q-system from $\dQ_{0,0}$ to $\dQ_{4,4}$ (not necessarily the grading path!) and make an ansatz there. It is preferable to choose a path with as few roots as possible. This gives us a concrete ansatz for seven $\dQ$'s (and also $\dQ_{0,0}=\dQ_{4,4}=1$) in terms of a number of unknown $c$'s.

The remaining $\dQ$'s can then be generated form those on the chosen path through the QQ-relation \eqref{QQ3}, i.e.
\be \label{dQQ}
\dQ_{a,s}=\frac{\dQ_{a,s+1}^+\dQ_{a-1,s}^--\dQ_{a,s+1}^-\dQ_{a-1,s}^+}{\dQ_{a-1,s+1}}=\frac{\dQ_{a+1,s}^+\dQ_{a,s-1}^--\dQ_{a+1,s}^-\dQ_{a,s-1}^+}{\dQ_{a+1,s-1}} \,.
\ee
The polynomial part $q_{a,s}$ can be found as the quotient of this polynomial division, while the remainder gives us constraints on the coefficient $c$, as they have to vanish. An efficient strategy is to first generate all $\dQ_{a,s}$ and then impose the zero-remainder conditions simultaneously.

\subsection{Generating the full Q-system and the leading \texorpdfstring{$\bP\mu$}{Pmu}-system}
With the distinguished Q-functions at hand, it is rather straightforward to generate the remaining Q-functions. This was discussed in detail in \cite{Marboe:2017dmb}, and we use the same strategy

\begin{itemize}
\item First derive the functions $Q_{a|\emptyset} \equiv \bP_a$ from $\dQ_{a,0}$ and $Q_{\emptyset|i}\equiv \bQ_i$ from $\dQ_{0,i}$.
\item Then generate the 16 functions $Q_{a|i}$ through
\be \label{Qai}
Q_{a|i} &=& -\Psi\left( \bP_a \bQ_i \right)^+\,,
\ee
where $\Psi$ is the inverse of the difference operator $\nabla$, $\Psi\left(\nabla F(u)\right) = F(u) + \mathcal{P}$, and $\mathcal{P}$ is an $\ii$-periodic function, since any such function belongs to the kernel of $\nabla$.%
\item Generate the remaining Q-system from $\bP_a$, $\bQ_i$ and $Q_{a|i}$ through the determinant relations given in \cite{Gromov:2014caa}. In particular, we need the 36 functions $Q_{ab|ij}$ and the four functions $Q_{abc|1234}$.
\item Build $\mu_{ab}^{(0)}$ as
\be \label{mulead}
\mu_{ab}^{(0)}&=&\omega \, Q_{ab|12}^-\,,
\ee
where $\omega$ is a constant.
\item Then construct $\Pt_a^{(0)}=\mu_{ab}^{(0)}\bP^a_{(0)}$ and $\Pt^a_{(0)}=\mu^{ab}_{(0)}\bP_a^{(0)}$. 
Finally compare these expressions with the ansatz for $\Pt$ to fix remaining unknowns. 
\end{itemize}

\begin{shaded}
\noindent {\bf Example: $\Psi_{11}\FF_{11}$ - \grpath{2}{3}{3}{3}}\\
To find the precise boundary conditions for the solution corresponding to the submultiplet containing this operator, we need to look at the \grpath{2}{3}{3}{3} path in the leftmost diagram in figure \ref{fig:YDroots}. This path traces out the functions $\dQ_{0,1}$, $\dQ_{0,2}$, $\dQ_{1,2}$, $\dQ_{1,3}$, $\dQ_{2,3}$, $\dQ_{3,3}$ and $\dQ_{4,3}$. There is only a single Bethe root in $\dQ_{1,2}$, and the ansatz for these seven functions can thus be written as
\be\label{dQex}
&\dQ_{0,1}=\spar^2\,,\quad \dQ_{0,2}=(\spar^+\spar^-)^2\,, \quad \dQ_{1,2}=\spar^2 \left(\spar + c_{1,2,0}\right)\,, \\
&\dQ_{2,3}=\tx^{-\ii \spar}\,,\quad \dQ_{1,3}=\dQ_{3,3}=\dQ_{4,3}=1\,. \no
\ee
The number of roots in the other $\dQ_{a,s}$ can be found by power counting in the QQ-relations, and we find these numbers to be:

\begin{center}
\begin{picture}(150,100)

{\color{gray}
\linethickness{0.5mm}
\put(25,0){\line(1,0){100}}
\put(25,25){\line(1,0){100}}
\put(25,50){\line(1,0){100}}
\put(25,75){\line(1,0){100}}
\put(25,100){\line(1,0){100}}
\put(25,0){\line(0,1){100}}
\put(50,0){\color{black}\line(0,1){100}}
\put(75,0){\color{black}\line(0,1){100}}
\put(100,0){\color{black}\line(0,1){100}}
\put(125,0){\line(0,1){100}}
\put(50,0){\color{black}\line(1,0){50}}
\put(50,25){\color{black}\line(1,0){50}}
\put(50,50){\color{black}\line(1,0){50}}
\put(50,75){\color{black}\line(1,0){50}}
\put(50,100){\color{black}\line(1,0){50}}
}

\linethickness{1.3mm}
\put(25,0){\color{orange}\line(1,0){50}}
\put(75,0){\color{orange}\line(0,1){25}}
\put(75,25){\color{orange}\line(1,0){25}}
\put(100,25){\color{orange}\line(0,1){75}}
\put(100,100){\color{orange}\line(1,0){25}}

\thinlines



\linethickness{3mm}
\put(10,50){\color{twisthighlightcol}\line(1,0){130}}

{\color{shadecolor}
\put(25,0){\circle*{11}}
\put(50,0){\circle*{11}}
\put(75,0){\circle*{11}}
\put(100,0){\circle*{11}}
\put(125,0){\circle*{11}}

\put(25,25){\circle*{11}}
\put(50,25){\circle*{11}}
\put(75,25){\circle*{11}}
\put(100,25){\circle*{11}}
\put(125,25){\circle*{11}}

\put(25,50){\circle*{11}}
\put(50,50){\circle*{11}}
\put(75,50){\circle*{11}}
\put(100,50){\circle*{11}}
\put(125,50){\circle*{11}}

\put(25,75){\circle*{11}}
\put(50,75){\circle*{11}}
\put(75,75){\circle*{11}}
\put(100,75){\circle*{11}}
\put(125,75){\circle*{11}}

\put(25,100){\circle*{11}}
\put(50,100){\circle*{11}}
\put(75,100){\circle*{11}}
\put(100,100){\circle*{11}}
\put(125,100){\circle*{11}}
}

\put(22.3,-3.4){\color{gradrootcol}0}
\put(47.3,-3.4){\color{gradrootcol}0}
\put(72.3,-3.4){\color{gradrootcol}0}
\put(97.3,-3.4){\color{nongradrootcol}0}
\put(122.3,-2.7){\color{nongradrootcol}$\bullet$}

\put(22.3,22.3){\color{nongradrootcol}$\bullet$}
\put(47.3,21.6){\color{nongradrootcol}0}
\put(72.3,21.6){\color{gradrootcol}1}
\put(97.3,21.6){\color{gradrootcol}0}
\put(122.3,22.3){\color{nongradrootcol}$\bullet$}

\put(22.3,47.3){\color{nongradrootcol}$\bullet$}
\put(47.3,46.6){\color{nongradrootcol}2}
\put(72.3,46.6){\color{nongradrootcol}3}
\put(97.3,46.6){\color{gradrootcol}0}
\put(122.3,47.3){\color{nongradrootcol}$\bullet$}

\put(22.3,72.3){\color{nongradrootcol}$\bullet$}
\put(47.3,71.6){\color{nongradrootcol}4}
\put(72.3,71.6){\color{nongradrootcol}3}
\put(97.3,71.6){\color{gradrootcol}0}
\put(122.3,72.3){\color{nongradrootcol}$\bullet$}

\put(22.3,97.3){\color{nongradrootcol}$\bullet$}
\put(47.3,96.6){\color{nongradrootcol}4}
\put(72.3,96.6){\color{nongradrootcol}2}
\put(97.3,96.6){\color{gradrootcol}0}
\put(122.3,96.6){\color{gradrootcol}0}

\end{picture}
\end{center}
where the {\color{xcol}green} highlighting signals that the $\dQ$'s carry a twist-factor. The symbol {\color{nongradrootcol}$\bullet$} signals that the corresponding $\dQ$ vanishes at the leading order in $\g$.

Note that we could in fact have made a completely trivial ansatz on the path {\bf 3333}, but to illustrate the zero-remainder conditions, we stick with the starting point \eqref{dQex}.

Using \eqref{dQQ} to find $\dQ_{0,3}$, we get
\be
\dQ_{0,3}=q_{0,3} \propto \frac{\spar}{\spar+c_{1,2,0}} = 1 - \frac{c_{1,2,0}}{\spar+c_{1,2,0}}
\ee
from which it is obvious that we have to set $c_{1,2,0}=0$ to kill the remainder term. Going in the other direction, we can generate the other $\dQ$'s, e.g.
\be
\dQ_{2,2}=\tx^{-\ii \spar} \tx^{-\frac{1}{2}} \left((\tx-1)\spar^3-\frac{3\ii}{2}(\tx+1)\spar^2 - \frac{3}{4}(\tx-1) \spar + \frac{\ii}{8}(\tx+1) \right)
\ee
Notice that when sending $\tx\to1$, the function reduces to the well-known Konishi solution $\dQ_{2,2}\propto \spar^2-\frac{1}{12}$.

The full set of leading distinguished Q-functions are listed in the following table.
It is arranged in analogy to the $4\times 4$ diagrams. The lower left corner holds $\dQ_{0,0}$ while the upper right contains $\dQ_{4,4}$, and the grading line between them is indicated with the \textcolor{gradlin}{colored} frames. The exponential twist factors, being the same across each row, are relegated to the right. The highlighting indicates the momentum carrying Q-function $\dQ_{2,2}$. The presented normalization is chosen for brevity, where in general the leading $\spar$-power has unit coefficient.
{
\arrayrulecolor{gray}\setlength{\extrarowheight}{\distQTabRowHeight}
\scriptsize
\hspace{\distQTabShift}\begin{equation*}
\begin{array}{|c|c|c|c|c|cl}\hhline{|-|-|->{\arrgr}|-|-|>{\arrc}~~}
 \rhigh \grzero & \trivAnz{\big({\sparg^{[-2]}} \sparg {\sparg^{[2]}}\big)^{-2}} \left(\spar^4+\frac{2 \left(\tx^2+7 \tx+1\right) \spar^2}{(\tx-1)^2}+\frac{\tx^4-14 \tx^3+50 \tx^2-14 \tx+1}{(\tx-1)^4}\right) & \grcell{\spar^2-\frac{\tx^2-26 \tx+1}{12 (\tx-1)^2}} & \grc \grone & \grcell{\grc \grone} &  &  \\[\exLinX]\hhline{-|-|->{\arrgr}|>{\arrc}->{\arrgr}|-|>{\arrc}~~}
 \rhigh \grzero & \trivAnz{\big({\sparg^-} {\sparg^+}\big)^{-2}} \left(\spar^4+\frac{\left(\tx^2+22 \tx+1\right) \spar^2}{2 (\tx-1)^2}+\frac{\left(\tx^2-26 \tx+1\right)^2}{16 (\tx-1)^4}\right) & \grcell{\spar^3+\frac{6 \tx \spar}{(\tx-1)^2}} & \grcell{\grc \grone} & \grzero &  &  \\[\exLinX]\hhline{-|-|->{\arrgr}|>{\arrc}->{\arrgr}|>{\arrc}-~~}
 \grzero & \trivAnz{\sparg^{-2}} \left(\spar^2+\frac{6 \tx}{(\tx-1)^2}\right) & \grcell{\cellcolor{momcell}\spar^3-\frac{3 \ii (\tx+1) \spar^2}{2 (\tx-1)}-\frac{3 \spar}{4}+\frac{\ii (\tx+1)}{8 (\tx-1)}} & \grcell{\grc \grone} & \grzero &  & \rule{\exSpace}{0pt} \tx^{-\ii \spar} \\[\exLin]\hhline{-|->{\arrgr}|->{\arrgr}|>{\arrc}->{\arrgr}|>{\arrc}-~~}
 \grzero & \grcell{\grone} & \grc \trivAnz{\sparg^{2}} \spar & \grcell{\grc \grone} & \grzero &  &  \\[\exLin]\hhline{>{\arrgr}|-|->{\arrgr}|>{\arrc}->{\arrgr}|-|>{\arrc}-~~}
 \grcellFirstL{\grc \grone} & \grc \trivAnz{\sparg^{2}}  & \grcell{\grc \trivAnz{\big({\sparg^-} {\sparg^+}\big)^{2}} } & \grone & \grzero &  &  \\[\exLin]\hhline{>{\arrgr}|-|-|-|>{\arrc}-|-|~~}
\end{array}
\end{equation*}
}
\bigskip

\noindent \emph{Deriving $\bP_a$ and $\bQ_i$.}\\
Moving on to the rest of the Q-system, we use the relations between $Q_{a|\emp}, Q_{\emp|i}$ and the distinguished Q-functions to get $\bP_a$ and $\bQ_i$, as explained in \cite{Marboe:2017dmb}. Exemplified for $\bP_2 = Q_{2|\emp}$, we make the ansatz
\begin{equation}
Q_{2|\emp} = \tx^{-\ii \spar} \frac{c_2 \spar^2 + c_1 \spar + c_0}{\spar^2}
\end{equation}
in accordance with \eqref{eq:PAnz} and solve for the coefficients $c_k$ through the determinant relation
\begin{equation}
  \dQ_{2,0} = \det
  \begin{pmatrix}
    Q_{1|\emp}^+	&	Q_{1|\emp}^-	\\
    Q_{2|\emp}^+	&	Q_{2|\emp}^-	
  \end{pmatrix}
  \,.
\end{equation}
The solution is
\begin{align}
  c_0&=  -\frac{6}{(\tx-1) (\tx+1)}\,,  &  c_1&=  0 \,, &  c_2&=  -\frac{\tx-1}{\tx (\tx+1)}
  \,.
\end{align}
For the other relations, we refer to \cite{Marboe:2017dmb} as the procedure is entirely analogous. The leading $\bP_a$ and $\bQ_i$ are
{\small\allowdisplaybreaks
\begin{align}
 \bP_1 &= 0 \,, & \bQ_1 & = \frac{\ii \spar^2}{4} \,,\no
\\
 \bP_2 &= \tx^{-\ii \spar} \left(\frac{\tx-1}{\tx (\tx+1)}+\frac{6}{\spar^2 \left(\tx^2-1\right)}\right) \,, & \bQ_2 & = -\frac{\spar^3}{10} \,,\no
\\
 \bP_3 &= \tx^{\ii \spar} \left(-(\tx-1)^2-\frac{6 \tx}{\spar^2}\right) \,, & \bQ_3 & = 40 \ii \etaF2 \spar^3-40 \spar^2-20 \ii \spar+\frac{20}{3} \,,\no
\\
 \bP_4 &= \frac{6 \ii}{\spar^2} \,, & \bQ_4 & = 0 
 \,.
\end{align}
}
\bigskip

\noindent \emph{Generating $Q_{a|i}$.}\\
With $\bP_a$ and $\bQ_i$ we can now genererate the $Q_{a|i}$ through \eqref{Qai}. As an example we have
\begin{align}
  Q_{3|3}^- &= \tx^{\ii \spar} \bigg(-40 \spar^2 (\tx-1) \tx-20 \ii \spar \tx (\tx+5)+\frac{20}{3} (\tx-10) \tx \no\\
  &\quad\,
  +\etaF2 \Big(40 \ii \spar^3 (\tx-1) \tx-120 \spar^2 \tx+120 \ii \spar \tx+40 \tx\Big)
\bigg)
  \,,
\end{align}
where $\etaF2$ is a Hurwitz $\eta$-function, defined below in equation \eqref{eq:defEta}.  
An important detail is that $\Psi$ returns a constant for the arguments that do not contain an exponential twist factor, as the only allowed periodic function. These constants will be fixed together with the other coefficients when comparing the expressions for $\Pt$. For instance, we have
\begin{equation}
  Q_{4|1}^- = \frac{3 \ii}{2} \spar - c_{4|1}
  \,.
\end{equation}

\bigskip

\noindent \emph{Obtaining $Q_{ab|ij}, Q_{abc|1234}$ and $\mu^{(0)}$.}\\
The determinant relations given in \cite{Gromov:2014caa} are straightforward to apply and we omit these intermediate steps. Through equations \eqref{mulead} and \eqref{eq:muUpper} we then get $\mu^{(0)}$. This introduces the unknown coefficients $\omega$ and $\Pf(\mu)^{(0)}$.

\bigskip

\noindent \emph{Fixing coefficients through matching the expressions for $\Pt$.}\\
We now have all we need to define $\Pt_a^{(0)}  = \mu_{ab}^{(0)} \bP^b_{(0)}$ and similarily for $\Pt^a_{(0)}$ such that we can compare with the ansatz \eqref{eq:PAnz}.
 Let us fix a few coefficients in this way:
\begin{align*}
  &		&	\text{\underline{calculated \eqref{eq:PmuPt}}} \hspace{7ex} &\phantom{=} \hspace{13ex} \text{\underline{ansatz \eqref{eq:PAnz}}} \\
  & \bP_2 : &
  \tx^{-\ii \spar} \left(\frac{6}{\spar^2 \left(\tx^2-1\right)}+\frac{\tx-1}{\tx (\tx+1)}\right) &=
  \tx^{-\ii \spar} \left(\frac{\cd_{2,0}^{(0)}}{\spar^2}+\frac{\cd_{2,1}^{(0)}}{\spar}+\frac{\tx-1}{\tx (\tx+1)}\right) \,,
  \\[1.7ex]
  &\Pt_1 : &
  -\frac{\ii \spar^3 \tx \omega }{40 (\tx-1)^2} &=
  \spar^6 \cc_{1,4}^{(0)}+\spar^5 \cc_{1,3}^{(0)}+\spar^4 \cc_{1,2}^{(0)}-\frac{\ii \ado{1} \spar^3 \tx}{2 (\tx-1)^2} \,,
  \\[1.7ex]
  & \Pt_2 : &
  \frac{\spar^2 \omega  \tx^{-\ii \spar} \left((\spar+\ii)^2-(\spar-\ii)^2 \tx\right)}{40 (\tx-1)^2 (\tx+1)} &=
  \tx^{-\ii \spar} \left(\spar^2 \cd_{2,0}^{(0)} +  \sum_{k = 1}^{4} \spar^{k+2} \cc_{2,k}^{(0)} \right)
  .
\end{align*}
The equation system for these coefficients are solved by
\begin{gather}
  \cc_{2,1}^{(0)} = \frac{12\ii}{(\tx-1)^2} \,,	\hspace{5em}	\cc_{2,2}^{(0)} = -\frac{6}{\tx^2-1}  \,,	\hspace{5em} \cd_{2,0}^{(0)} = \frac{6}{\tx^2-1} \,,	\no
  \\
  \omega = 240 \,,	\hspace{5em}  \ado{1} = 12 \,,
\end{gather}
while all the others vanish. We see that we already obtain the 1-loop anomalous dimension.  Repeating this for all $\Pt^{(0)}$, we can fix all introduced coefficients.

As a check and an example of the difference equation \eqref{eq:muDiffEq}, we can look at
\begin{align}
  \nabla \mu_{12}^{(0)} &= -\bP_1^{(0)} \Pt_2^{(0)} + \bP_2^{(0)} \Pt_1^{(0)} = 
  \tx^{-\ii \spar} \left(-\frac{6 \ii \spar^3}{\tx^2-1}-\frac{36 \ii \spar \tx}{(\tx-1)^3 (\tx+1)}\right)
  \,,
\end{align}
which is adequately satisfied by the found expression 
\begin{equation}
  \mu_{12}^{(0)} = 
  \frac{\tx^{-\ii \spar}}{(\tx - 1)^2 (\tx+1)} \left(6 \ii \spar^3+ \frac{\tx}{(\tx-1)} \left( 18 \spar^2 -18 \ii \spar -6 \right) \right)
  \,.
\end{equation}
We \hyperref[sec:examplePertAlgo]{continue} this example in section \ref{sec:pertAlgo}, where we will look at the perturbative corrections.
\end{shaded}




\section{Perturbative corrections to the \texorpdfstring{$\bP\mu$}{Pmu}-system}\label{sec:pert}
In the perturbative solution of the QSC, we adapt the algorithm of \cite{Marboe:2018ugv}. It streamlines the redundant information in the QSC into a small set of steps that is carried out repeatedly order by order, only involving the quantities $\bP$ and $\mu$ in the $\bP \mu$-system. An overview of the steps is given in figure \ref{fig:pertAlgo} and further described below in section \ref{sec:pertAlgo}.

\paragraph{Difference equation on \texorpdfstring{$\mu$}{mu}.} The most central equation of the $\bP\mu$-system is the difference equation \eqref{eq:muDiffEq} for $\mu$, here repeated:
\begin{equation}
  \label{eq:muDiffEqII}
  \mu_{ab} - \mu_{ab}^{[2]} = - \bP_a \bP^c \mu_{bc}^{[1 \pm 1]} + \bP_b \bP^c \mu_{ac}^{[1 \pm 1]}
  ,
\end{equation}
which couples all six component functions of  $\mu$. It can be rephrased for perturbative calculations as an inhomogenous equation for each order,
\begin{equation}
  \label{eq:muDiffEqPert}
  \mu_{ab}^{(n)} - {\mu_{ab}^{(n)}}^{[2]} = - \bP_a^{(0)} \bP^c_{(0)} {\mu_{bc}^{(n)}}^{[1 \pm 1]} + \bP_b^{(0)} \bP^c_{(0)} {\mu_{ac}^{(n)}}^{[1 \pm 1]} + U_{ab}^{(n)}
  ,
\end{equation}
where all terms involving lower orders of $\mu$ are collected into the source term $U_{ab}^{(n)}$. This equation can be solved iteratively, order by order, by using the ansatz \eqref{eq:PAnz} for $\bP$ and exploiting the relationships and analytic structures of the involved quantities to fix the constants. Normally%
, all introduced constants are fixed at the end of each iterative step such that the only unknown parts inside $U_{ab}^{(n)}$ come from $\bP^{(n)}$. 
The solution to the difference equation \eqref{eq:muDiffEqPert} can be written as \cite{Marboe:2018ugv}
\begin{equation}
  \mu_{ab}^{(n)} =  \frac{1}{2} f_{ab| k} \Psi \left( f^{cd| k} U_{cd}^{(n)}\right)
  \label{eq:muDiffSol}
  ,
  \quad\quad f_{ab|k}\equiv Q^{(0)\,-}_{ab|\{12,13,14,23,24,34\}_k} \,.
\end{equation}

\paragraph{Basis of functions.} 
Importantly, the action of $\Psi$ on the functions appearing in the QSC closes such that the entire set of functions that appear are expressible through the basis
\begin{itemize}
  \item polynomials in $\spar$ and shifted inverse powers $\tfrac{1}{(\spar+\ii n)^m}$,
  \item Hurwitz $\eta$-functions\footnote{The $\eta$-functions are convergent for $a_i\geq 2$ while the special case $a = 1$ is defined as $\etaF{1} = \ii \psi(-\ii \spar)$, where $\psi$ is the digamma function. We leave the dependence on the spectral parameter implicit.} 
    \begin{equation}
      \etaF{a_1, \ldots, a_k} = \sum_{0\leq j_1 < \ldots < j_k}^{\infty} \frac{1}{(\spar + \ii j_1)^{a_1} \cdots (\spar + \ii j_k)^{a_k}}
      \,,
      \label{eq:defEta}
    \end{equation}
  \item $\ii$-periodic functions with at most constant $\spar$-asymptotics and poles only at $\ii \mathbb{Z}$, written in the basis
    \begin{equation}
      \Per{m} = \sum_{j=-\infty}^{\infty} \frac{1}{(\spar + \ii j)^m}
      \label{eq:defPer}
      \,,
    \end{equation}
  \item and overall exponential factors of $\tx_a^{\ii \spar}$.
\end{itemize}
All these functions form algebraic rings such that any expression can be written as quadrolinear combinations of them. %
This property ensures the closure of the $\Psi$-operation and allows for a fast and simple computer implementation, as discussed thorougly in \cite{Leurent:2013mr,Marboe:2014gma,Gromov:2015dfa}. 
Note that for the fully twisted QSC, one has to extend the above basis to include twisted $\eta$-functions, see e.g. \cite{Gromov:2015dfa}. These would arise from $\Psi$-actions like $\Psi\left( \frac{\tx^{\ii \spar}}{(\spar+\ii n)^m} \right)$, $\Psi\left( \tx^{\ii \spar} \etaF{A} \right)$, but such expressions never appear in our calculation. We do not have a proof of this property, but it strongly hints that these functions only appear for twists of the $\su(2,2)$ part of the symmetry. 
Conceptually, the perturbative computations in the $\tb$-deformed theory are thus very similar to those in the undeformed theory, and the $\Psi$-operation needs only a mild generalization. 
Whereas $\Psi$ maps a polynomial in $\spar$ to another polynomial of one degree higher, an overall $\tx^{\ii \spar}$-factor times such a polynomial is mapped to a product of the same exponential factor and a polynomial of the same degree. 

The functions $\Per{m}$ enter through the $\ii$-periodic ambiguity in the solution of equation \eqref{eq:muDiffSol}, i.e. 
\begin{equation}
\mathcal{P} =  \perphi{k}{0}{n} + \sum_{m = 1}^{\infty} \perphi{k}{m}{n} \, \Per{m}
  \label{eq:perAdded}
  ,
\end{equation}
where the coefficients $\perphi{k}{m}{n}$ are fixed later in the algorithm. In practice, the infinite sum in \eqref{eq:perAdded} is truncated rather soon.

\paragraph{Numbers.}
Practically, the fact that the coefficients in the QSC functions contain the twist instead of just being numbers as in the undeformed case is a computational challenge, as we will see. 
As in the undeformed case, the numbers that appear in the functions are the algebraic numbers arising when solving the zero-remainder conditions for the leading Q-system, and multiple zeta values (MZVs) $\zv{A}$ that arise in the power expansion of $\eta$-functions at $\spar = 0$, e.g.
\begin{equation}
  \etaF{2} =  \frac{1}{\spar^2} -\zv2 -2 \ii \zv3 \spar +\frac{6}{5} \zv2^2 \spar^2 +4 \ii \zv5 \spar^3 -\frac{8}{7} \zv2^3 \spar^4 - 6 \ii \zv7 \spar^5 + \CO(\spar^6)
  .
\end{equation}
The MZVs also appear in the corresponding expansion of the $\ii$-periodic functions $\Per{m}$.

\subsection{The perturbative algorithm}\label{sec:pertAlgo}
The algorithm for the perturbative calculation of the $\bP \mu$-system \cite{Marboe:2018ugv} consists of five steps, repeated at each order. We describe them here in chronological order, while a pictorial overview is given in figure \ref{fig:pertAlgo}.
\newcommand{\stepbox}[1]{\fcolorbox{step#1}{white}{Step #1}}
\begin{description}
  \item[\stepbox{1}] Define $\bP_a^{(n)}$ and $\bP^a_{(n)}$ through the ansatz \eqref{eq:PAnz}. This introduces a finite number of coefficients, including the perturbative corrections to the anomalous dimension $\ado{n}$, which the following steps aim to fix.

    It is convenient to impose equation \eqref{eq:PmuRels}, $\bP_a \bP^a = 0$,  at this point to already fix a few constants.
  \item[\stepbox{2}] Construct $\mu_{ab}^{(n)}$ through equation \eqref{eq:muDiffSol}. This automatically defines $\mu^{ab}_{(n)}$ through the relation \eqref{eq:muUpper} and also introduces a few more constants $\perphi{n}{k}{m}$ due to the $\ii$-periodic ambiguity. In our implementation, this is the most computationally expensive step.
  \item[\stepbox{3}] Impose the regularity conditions \eqref{eq:regConds} on $\mu_{ab}$. 
    This amounts to expanding the expressions $\mu_{ab}^{(n)} + {\mu_{ab}^{(n)}}^{[2]}$ and $\frac{\mu_{ab} - \mu_{ab}^{[2]}}{\sqrt{\spar^2 - 4 \g^2}}$ at $\spar = 0$ and imposing that all poles vanish. %
 This fixes many of the introduced constants. It is also where the MZVs first appear.
  \item[\stepbox{4}] Define $\Pt_a^{(n)}$ and $\Pt^a_{(n)}$ through equation \eqref{eq:PmuPt}, $\Pt_a = \mu_{ab} \bP^b$ and ${\Pt^a} = \mu^{ab} \bP_b$.
  \item[\stepbox{5}] Match the expressions defined in step 4 with the ansatz \eqref{eq:PAnz}. This again requires power expanding at $\spar = 0$, and introduces more MVZs. This normally fixes the last introduced constants such that all quantities at order $\g^{2n}$ are fixed and can be used as input in the calculation of the next order.
\end{description}
%

\newcommand{\steptext}[1]{\textcolor{step#1}{step #1}}
\begin{figure}
  \centering
  \includegraphics{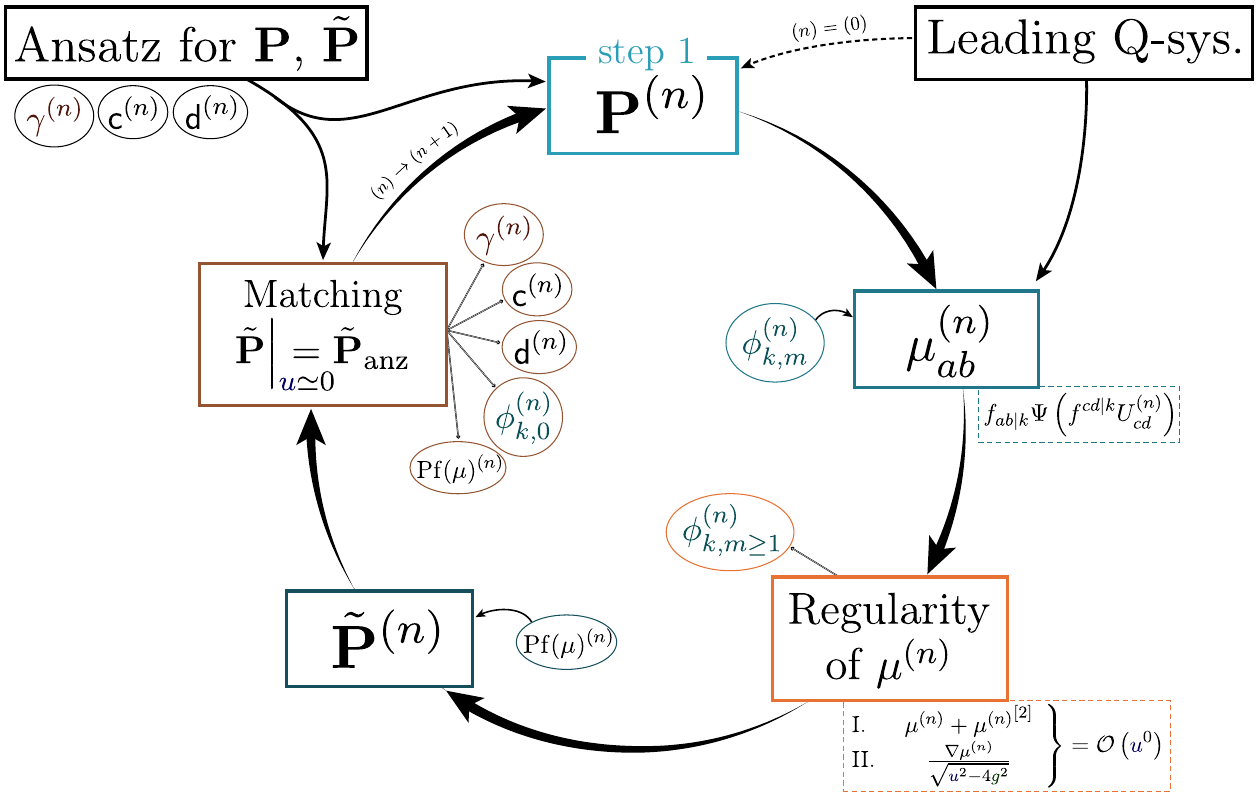}
  \caption{An overview of the five steps in the perturbative algorithm. \steptext{1}, \steptext{2} and \steptext{4} introduce constants while they are fixed in \steptext{3} and \steptext{5}.}
  \label{fig:pertAlgo}
\end{figure}

\begin{shaded}
\newcommand{\stepboxg}[1]{\fcolorbox{step#1}{shadecolor}{Step #1}}
\noindent {\bf Example: $\Psi_{11}\FF_{11}$ - \grpath{2}{3}{3}{3}}
\label{sec:examplePertAlgo}

\noindent 
We return to our 
example to illustrate the perturbative algorithm. 
Having all quantities at leading order already, we go through the algorithm step by step at the subleading order.\bigskip

\noindent\stepboxg{1}\\
We define $\bP$ through the ansatz, substituting all coefficients we already fixed at leading order:
\begin{align}
  \bP_1^{(1)} &=  -\frac{6 \ii \tx}{\spar^3 (\tx-1)^2} \,,	&  \bP^1_{(1)}  & = -30  \,, \no\\
  \bP_2^{(1)} &=  \tx^{-\ii \spar} \left(\frac{12}{\spar^4 \left(\tx^2-1\right)}+\frac{12 \ii}{\spar^3 (\tx-1)^2}+ \frac{\cd_{2,0}^{(1)}}{\spar^2}+\frac{\cd_{2,1}^{(1)}}{\spar}\right) \,,	&  \bP^2_{(1)}  & = \frac{3 \ii \tx^{\ii \spar+1} (\tx+1)}{\spar (\tx-1)}  \,, \no\\
  \bP_3^{(1)} &=  \tx^{\ii \spar} \left(-\frac{12 \tx}{\spar^4}+\frac{12 \ii (\tx+1) \tx}{\spar^3 (\tx-1)}+\frac{\cd_{3,0}^{(1)}}{\spar^2}+\frac{\cd_{3,1}^{(1)}}{\spar}\right) \,,	&  \bP^3_{(1)}  & = -\frac{3 \ii \tx^{-\ii \spar}}{\spar (\tx-1)^2}  \,, \no\\
  \bP_4^{(1)} &=  \frac{12 \ii}{\spar^4}+\frac{66 \ii}{\spar^2} \,,	&  \bP^4_{(1)}  & = -\frac{6 \tx}{\spar (\tx-1)^2}  \,.
\end{align}
The condition $\bP_a\bP^a = 0$ immediately fixes two of the four $\cd$-coefficients such that only $\cd_{2,0}^{(1)}$ and $\cd_{3,1}^{(1)}$ remain.\bigskip

\noindent\stepboxg{2}\\
\newcommand{\polu}[1]{\text{pol}_{\spar}^{#1}}
We move on to calculate $\mu_{ab}^{(1)}$. These functions become bulky already at subleading order so we will only sketch the procedure.

Introducing the notation $\polu{k}(\text{coeff.s})$ for polynomials in $\spar$ of degree $k$ that contain specified, undetermined, coefficients, we can display the structure of the $\Psi$-operation in step 2. It acts on one of the terms in equation \eqref{eq:muDiffSol} as
\begin{gather*}
  f^{cd|1} U_{cd}^{(1)} = \polu{4}\left(\cd_{2,0}^{(1)}, \cd_{3,1}^{(1)}\right) + \frac{1440\ii}{\spar} 
  +  \polu{5}\left(\cd_{2,0}^{(1)}, \cd_{3,1}^{(1)}\right) \etaF{2}
  \\
  \Big\downarrow \Psi 
  \\
  \polu{5}\left(\cd_{2,0}^{(1)}, \cd_{3,1}^{(1)}\right) - 48\ii\left(18+(\tx^2-1)\cd_{3,1}^{(1)} \right) \etaF{1}
  + \polu{6}\left(\cd_{2,0}^{(1)}, \cd_{3,1}^{(1)}\right)  \etaF{2}
  + \perphi{1}{0}{1} + \sum_{m=1}^{3} \perphi{1}{m}{1} \Per{m}
  \,.
\end{gather*}
In accordance with equation \eqref{eq:muDiffSol}, this is then multiplied by $f_{ab|1}$ and summed with the other five terms to yield the full $\mu_{ab}^{(1)}$. \bigskip

\noindent\stepboxg{3}\\
We impose the regularity conditions for $\mu_{ab}^{(1)}$ at $\spar=0$, first by expanding $\mu_{ab}^{(1)} + {\mu_{ab}^{(1)}}^{[2]}$, here for $ab = 14$,
\begin{align}
  \mu_{14}^{(1)} + {\mu_{14}^{(1)}}^{[2]} &\propto 
  -\frac{\perphi{1}{3}{1}}{\spar^3} -\frac{\perphi{1}{2}{1}-200 \ii \perphi{2}{0}{1}}{\spar^2}   \label{eq:exRegCond1}\\
   &  \quad\, +\frac{ \ii 24 \left(\tx^2-1\right) \cd_{2,0}^{(1)}- \perphi{1}{1}{1}+3 \perphi{1}{3}{1}-600 \perphi{2}{0}{1}+432 \ii}{\spar } +\CO\big( \spar^0 \big) \notag
\end{align}
and demanding that all negative powers vanish. For example, we immediately see that $\perphi{1}{3}{1} = 0$ while the other relations between the coefficients are collected for all six $\mu_{ab}^{(1)}$. 

Secondly, we do the same for the second regularity condition, again here for $\mu_{14}^{(1)}$:
{\footnotesize
\begin{align*}
  \left( \frac{\nabla \mu_{14}}{\sqrt{\spar^2 - 4\g^2}} \right)^{(1)} \!\!\!\!&\propto
  \frac{\tx (3 \perphi{1}{3}{1}-200 \perphi{2}{0}{1})}{\spar^3} + \frac{3 \tx \left(-8 \left(\tx^2-1\right) \cd_{2,0}^{(1)}+\perphi{1}{2}{1}-200 \ii \perphi{2}{0}{1}+336\right)}{\spar^2}   \\
  &  \quad -\frac{3 \ii \left(24 \tx^3 \cd_{2,0}^{(1)}+\tx (-24 \cd_{2,0}^{(1)}+\ii \perphi{1}{1}{1}+400 \ii \perphi{2}{0}{1}+432)+20 \ii (\cd_{3,1}^{(1)}+3 \ii)+60 \tx^2\right)}{\spar} + \CO\big( \spar^0 \big)
  \\
  &\subeq -\frac{200 \tx \perphi{2}{0}{1}}{\spar^3} -\frac{24 \tx \left(\left(\tx^2-1\right) \cd_{2,0}^{(1)}-42\right)}{\spar^2}   
   + \frac{60 \left(\cd_{3,1}^{(1)}-10 \tx \perphi{2}{0}{1}-3 \ii \tx^2+3 \ii\right)}{\spar} + \CO\big( \spar^0 \big)
  \,.
\end{align*}
}%
\noindent We have substituted the solutions from equation \eqref{eq:exRegCond1} in the second step. Most remaining coefficients are again very easy to identify and are, still at this loop order, very simple rational expressions in $\tx$. A few coefficients survive until step 5.

Generically, we would expect MZVs to appear in this step but that is first when the $\eta$-functions multiply negative powers of $\spar$, which doesn't happen at this order.

\bigskip

\noindent\stepboxg{4}\\
Next, we calculate $\mu^{ab}_{(1)}$ and $\Pt^{(1)}$ from relations \eqref{eq:muUpper} and \eqref{eq:PmuPt}. The smallest examples are
{\small
\begin{align*}
  \Pt_1^{(1)} &= 
  \frac{\tx}{(\tx-1)^2} \bigg(-\frac{\ii \spar^3  (\perphi{1}{0}{1}-1440)}{40 }-72 \etaF1 \spar^3 +36 \Per{1} \spar^3 +36 \spar^2 +18 \ii \spar  \bigg)
  \,,
  \\
  \Pt^4_{(1)} &=
  \spar^3 \left(-\frac{1}{240} \perphi{1}{0}{1}-\frac{1}{6} \ii \Pf(\mu)^{(1)}+6\right)+\frac{\spar \left(-\tx \perphi{1}{0}{1}-40 \ii \Pf(\mu)^{(1)} \tx+120 \tx^2+120\right)}{40 (\tx-1)^2}  \\
  &  \quad\, +\etaF1 \left(12 \ii \spar^3+\frac{72 \ii \spar \tx}{(\tx-1)^2}\right)+\Per{1} \left(-6 \ii \spar^3-\frac{36 \ii \spar \tx}{(\tx-1)^2}\right)-6 \ii \spar^2+\frac{6 \tx}{\spar (\tx-1)^2}-\frac{36 \ii \tx}{(\tx-1)^2}
  \,.
\end{align*}
}%
\noindent Note that the Pfaffian in the definition of $\mu^{ab}$ is introduced as another constant to fix, with its own $\g$-expansion.
\bigskip

\noindent\stepboxg{5}\\
In the final step, we match the obtained expressions for $\Pt$ with the ansatz \eqref{eq:PAnz} (with $\zh \to 1/\zh$). We expand the expressions from step 4 around $\spar = 0$ to the relevant order
\begin{align*}
  \Pt_1^{(1)} &= 
  \spar^3 \left(\frac{72 \ii \zv1 \tx}{(\tx-1)^2}-\frac{\ii \tx (\perphi{1}{0}{1}-1440)}{40 (\tx-1)^2}\right)+\frac{18 \ii \spar \tx}{(\tx-1)^2}
  + \CO\big( \spar^5 \big) \,,
  \\
  \Pt^4_{(1)} &=
  \frac{6 \tx}{\spar (\tx-1)^2}+\spar \left(\frac{72 \zv1 \tx}{(\tx-1)^2}+\frac{-\tx \perphi{1}{0}{1}-40 \ii \Pf(\mu)^{(1)} \tx+120 \tx^2+120}{40 (\tx-1)^2}\right)
  + \CO\big( \spar^4 \big) \,,
\end{align*}
and compare it with 
\begin{align*}
  \Pt_1^{(1)} &\stackrel{\text{\tiny \eqref{eq:PAnz}}}{=}
  \spar^4 (\cc_{1,2}^{(1)}-6 \cc_{1,4}^{(0)})+\spar^3 \left(-5 \cc_{1,3}^{(0)}-\frac{\ii \left(\ado{1}^2+2 \ado{2}\right) \tx}{4 (\tx-1)^2}\right)-4 \spar^2 \cc_{1,2}^{(0)}+\frac{3 \ii \ado{1} \spar \tx}{2 (\tx-1)^2}
  \\
  &\subeq 
  \spar^4 \cc_{1,2}^{(1)}-\frac{\ii \left(2 \ado{2}+144\right) \spar^3 \tx}{4 (\tx-1)^2}+\frac{18 \ii \spar \tx}{(\tx-1)^2}
  \,,
  \\
  \Pt^4_{(1)} &\stackrel{\text{\tiny \eqref{eq:PAnz}}}{=} 
  \spar^2 (\cc^{4,2}_{(1)}-4 \cc^{4,4}_{(0)})+\spar \left(-3 \cc^{4,3}_{(0)}-\frac{\ado{2} \tx}{2 (\tx-1)^2}\right)-2 \cc^{4,2}_{(0)}+\frac{\ado{1} \tx}{2 \spar (\tx-1)^2}
  \\
  &\subeq
  \spar^2 \cc^{4,2}_{(1)}+\spar \left(3-\frac{\ado{2} \tx}{2 (\tx-1)^2}\right)+\frac{6 \tx}{\spar (\tx-1)^2}
  \,,
\end{align*}
where again we have substituted all coefficients known from previous steps in the second equalities. We have here truncated the ansatz in a way consistent with this loop order being our final aim.

Here we see the MZVs entering, although the anomalous dimension is still a simple integer at this loop order. The non-zero coefficients that explicitly appeared in this example are fixed to
\begin{gather*}
  \perphi{1}{0}{1} =  960 \left(3 \zv1+2\right)  \,, \hspace{5em}
  \Pf(\mu)^{(1)} =  66 \ii \,,   \hspace{5em}   \ado{2} =  -48
  \,.
\end{gather*}

The full results of our perturbative calculations for this solution are shown in the \hyperref[sec:exampleResults]{conclusion} of this example in section \ref{sec:results}.
\end{shaded}

\subsection{Results, performance and challenges}\label{sec:results}
We have applied a \emph{Mathematica}-implementation of the described algorithm to the examples in table \ref{tab:examples}. The success varies significantly, however, depending on the operator in question. For the simplest cases, we have reached 7- and 8-loop results on a standard laptop while for the most challenging one only the 2-loop anomalous dimension could be fixed within reasonable time. In this section, we discuss the general features and challenges, while we present the individual calculations in section \ref{sec:ex}.

As expected, the results for the anomalous dimensions contain MZVs, while the dependence on the twist comes in the form $\cos (n \tb)$, with $n$ being integer. They all agree with former results where such were known and they all reduce to the known result for the Konishi anomalous dimension in the undeformed theory in the limit $\tb \to 0$. 

\begin{shaded}
  \label{sec:exampleResults}
\noindent {\bf Example:  $\Psi_{11} \FF_{11}$ - \grpath{2}{3}{3}{3}}

\noindent 
For this example, we were able to complete eight perturbative loops with our implementation. The result for the anomalous dimension is:
{\small
  \allowdisplaybreaks
\begin{align}
\ado{1} &= 12 \label{PsiFres}\\[1mm]
\ado{2} &= -48 \no\\[1mm]
\ado{3} &= -12 \big(\tc{1}-29\big)\no \\[1mm]
\ado{4} &= -192 \zv3 \big(\tc{1}-4\big)+348 \tc{1}-1440 \zv5-2844 \no\\[1mm]
\ado{5} &= 96 \zv3 \big(29 \tc{1}+43\big)+2880 \zv5 \big(\tc{1}-4\big)-7380 \tc{1}-5184 \zv3^2+30240 \zv7+22548 
	\no\\[1mm]
\ado{6} &= 6912 \zv3^2 \big(\tc{1}-4\big)-44832 \zv3 \tc{1}+192 \zv3 \tc{2}+288 \zv5 \big(539-149 \tc{1}\big) \no\\
  &\quad\, +336 \zv7 \big(334-109 \tc{1}\big)+136428 \tc{1}-156 \tc{2} \no\\
    &\quad\, +155520 \zv5 \zv3-218016 \zv3-489888 \zv9-143952 
   \no \\[1mm]
\ado{7} &= 576 \zv3^2 \big(-17 \tc{1}+3 \tc{2}-718\big)+597984 \zv3 \tc{1}-7824 \zv3 \tc{2} \no\\
  &\quad\, +576 \zv5 \zv3 \big(863-149 \tc{1}\big)-48 \zv5 \big(-13586 \tc{1}+95 \tc{2}+18267\big) \no\\
  &\quad\, +15120 \zv7 \big(30 \tc{1}-113\big)+451584 \zv9 \tc{1}-2315988 \tc{1}+9984 \tc{2} \no\\
  &\quad\, +124416 \zv3^3-1935360 \zv7 \zv3+4639920 \zv3-993600 \zv5^2-1287072 \zv9+7318080 \zv{11}+170964 
 \no \\[1mm]
 \ado{8} &= -\frac{684288}{5} \Zv+27648 \zv3^3 \big(5 \tc{1}+67\big)-1085184 \zv3^2 \tc{1}-71136 \zv3^2 \tc{2}-5217888 \zv3 \tc{1} \no \\
 &  \quad\,  +248112 \zv3 \tc{2}-1728 \zv5 \zv3 \big(-233 \tc{1}+35 \tc{2}+2718\big)-4032 \zv7 \zv3 \big(1573-28 \tc{1}\big) \no\\
 &  \quad\,  -11520 \zv5^2 \big(11 \tc{1}-218\big)-9759936 \zv5 \tc{1}+188400 \zv5 \tc{2}-8016816 \zv7 \tc{1} \no\\
 &  \quad\,  +92856 \zv7 \tc{2}-4287840 \zv9 \tc{1}-5575680 \zv{11} \tc{1}+36845004 \tc{1}-384636 \tc{2} \no\\
 &  \quad\,  -3255552 \zv5 \zv3^2+9091296 \zv3^2+23224320 \zv9 \zv3-78527184 \zv3-10106928 \zv5+22256640 \zv5 \zv7 \no \\
 &  \quad\,  +29792664 \zv7+13615584 \zv9+\frac{93807648 \zv{11}}{5}-106007616 \zv{13}+17947824 \no
 \,,
\end{align}
}%
with the shorthand notation
\begin{equation}
  \tc{k} = \cos (k \tb)
  .
\end{equation}
At the 8th loop, we here introduced the single-valued MZV \cite{Brown:2013gia,Schnetz:2013hqa} 
\begin{equation}
  \Zv = -\zv{3,5,3} + \zv{3} \zv{3,5}
  .
\end{equation}
As a nice check, this result reduces to the known 8-loop result for the untwisted Konishi multiplet \cite{Leurent:2013mr}, which is given below in \eqref{ZZbres}.

\end{shaded}
\paragraph{Computational challenges.} How far we have been able to push the calculations is highly solution dependent. The main complication is the difficulty of dealing with the symbolical expressions involving twists and algebraic expressions arising in the solution of the zero-remainder conditions at the leading order.

The appearance of a square root in a solution poses a practical complication as \emph{Mathematica} has more difficulties simplifying such expressions. Much of this is by design, in order to avoid any assumptions of branch cuts, but even in simple cases (such as $\sqrt{6}$), there are significant slowdowns. Our attempts to remedy this have been to first manipulate all expressions such that the square root appears in the numerators and not the denominators, whenever it is possible. Secondly, it may be beneficial at certain points in the algorithm to replace the square root with a placeholder variable that squares to the square root argument. In the end, after various timing tests, we only used this in an iterative solver for the regularity conditions in step 3. It is still possible though that going back and forward in between the placeholder and the explicit square root would improve performance in other places too.

Although the $\tb$-deformation is arguably the mildest of all $\gamma$-deformations and only depends on a single parameter, it seems it is enough to seriously bog down the perturbative algorithm. Rapidly growing rational expressions of the variable $\tx$ can start to accumulate at each order, in particular in the case where all the $\tx_a$ are different. In fact, even the leading Q-system can be rather complicated (as can be seen in the example for $\ZZ \Psi_{11}^2$ in appendix \ref{sec:largeQSys}).

\paragraph{Performance.}
\begin{figure}
  \centering
  \includegraphics{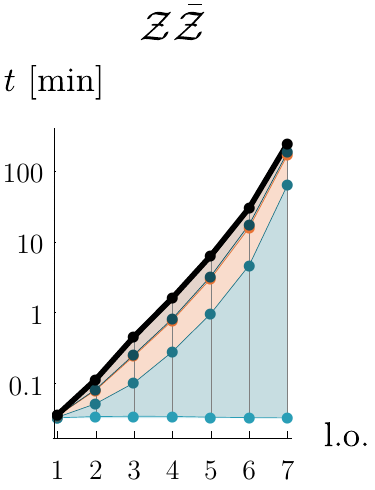}\hspace{-5pt}
  \includegraphics{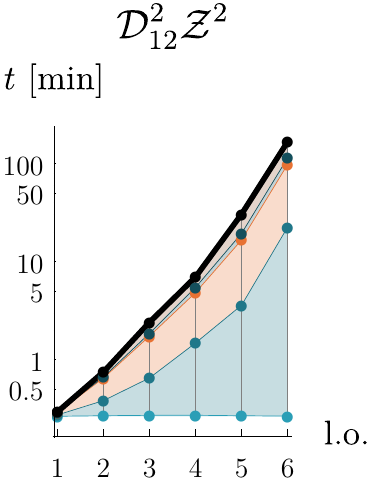}\hspace{-5pt}
  \includegraphics{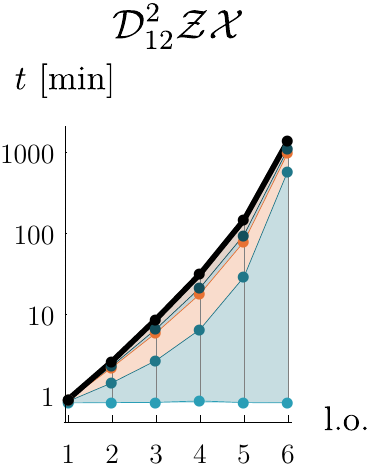}\hspace{-5pt}
  \includegraphics{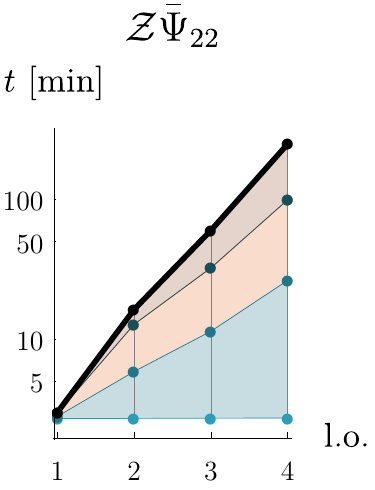}\\
  \caption{The scaling of computation times for four example operators: the $\ZZ\ZZb$ which has no twist, the simple $\DD_{12}^2 \ZZ^2$, the $\DD_{12}^2 \ZZ \XX$ which contains $\sqrt{6}$ and $\ZZ\bar{\Psi}_{22}$ which contains the more complicated square root $\sqrt{\tx \left(2 \tx^2+5 \tx+2\right)}$. The total computation time is plotted as the thick black line in accordance with the logarithmic time scale on the $y$-axis. The shaded regions below that line indicate the fractional computation times (in percent) spent at each step in the algorithm and are hence independent of the scale on the axis. The step coloration follows section \ref{sec:pertAlgo} as in \textcolor{step1}{step 1} (which is almost instant), \textcolor{step2}{step 2}, \textcolor{step3}{step 3}, \textcolor{step4}{step 4} and \textcolor{step5}{step 5}. Note how the computation times increase significantly with the presence of the square roots. }
  \label{fig:compTimes}
\end{figure}
As discussed, the computation times vary a lot depending on the solution. Overall, the computation time seems to scale roughly exponentially with the loop order, illustrated for four examples in figure \ref{fig:compTimes}. The costliest step is by far the construction of $\mu_{ab}^{(n)}$ in step 2, which also has the worst scaling. It is followed by the regularity conditions in step 3. 
The total times for the illustrated calculations at $\CO(\g^{8})$ were 1.6 min for $\ZZ\ZZb$, 7 min for $\DD_{12}^2 \ZZ^2$, 31 min for  $\DD_{12}^2 \ZZ \XX$ and 248 min for $\ZZ\bar{\Psi}_{22}$ which shows what an impact the twist and the square root expressions have.

The memory required to store the $\bP\mu$-system and replacement rules for the constants is once again solution dependent. Typically, the computations have a roughly exponential memory scaling with the loop orders. The required memory for the $\Psi_{11}\FF_{11}$ solution is
\begin{center}
  {%
    \setlength{\extrarowheight}{3pt}
    \begin{tabular}{l|*{7}{c}}
    $\g$-order	& 2	& 4	& 6	& 8	& 10	& 12	& 14	\\\hline
    mem. (Mb)	& .1	& .2	& .7	& 2.4	& 8.2	& 30.7	& 125.8
  \end{tabular}
  ~.
}
\end{center}
The memory usage for the more complicated solutions is much higher, where for instance the solution for the $\ZZ \Psi_{11}^2$ solution requires about 30 Mb already at $\CO(\g^{2})$.

\section{Examples}\label{sec:ex}
In this section we analyze the solutions for the submultiplets given in table \ref{tab:examples}, except for $\Psi_{11}\FF_{11}$ which was treated along the way.

\subsection{The \texorpdfstring{$\su(4)$}{su(4)} operator \texorpdfstring{$\ZZ\ZZb$}{ZZbar} - HWS in \texorpdfstring{\grpath{2}{2}{2}{2}}{2222}}
\begin{shaded}
\noindent Following the procedure of section \ref{sec:lead}, this operator (and any other operator from the Konishi multiplet with $n_{\fff_1}=n_{\fff_2}=n_{\fff_3}$) corresponds to the same boundary as the Konishi multiplet in the undeformed theory. Consequently, the solution is exactly the same, and it has been treated extensively in \cite{Marboe:2014gma,Marboe:2018ugv}, so we simply state the results to compare its simplicity to the other Konishi solutions in the $\tb$-deformation.

The boundary conditions for the solution can be summed up as
\begin{center}
\begin{picture}(300,105)

{\color{gray}
\linethickness{0.5mm}
\put(25,0){\line(1,0){100}}
\put(25,25){\line(1,0){100}}
\put(25,50){\line(1,0){100}}
\put(25,75){\line(1,0){100}}
\put(25,100){\line(1,0){100}}
\put(25,0){\line(0,1){100}}
\put(50,0){\color{black}\line(0,1){100}}
\put(75,0){\color{black}\line(0,1){100}}
\put(100,0){\color{black}\line(0,1){100}}
\put(125,0){\line(0,1){100}}
\put(50,0){\color{black}\line(1,0){50}}
\put(50,25){\color{black}\line(1,0){50}}
\put(50,50){\color{black}\line(1,0){50}}
\put(50,75){\color{black}\line(1,0){50}}
\put(50,100){\color{black}\line(1,0){50}}
}

\linethickness{1.3mm}
\put(25,0){\color{orange}\line(1,0){50}}
\put(75,0){\color{orange}\line(0,1){100}}
\put(75,100){\color{orange}\line(1,0){50}}


\linethickness{3mm}
\put(10,50){\color{betacol!30}\line(1,0){0}}

{\color{shadecolor}
\put(25,0){\circle*{11}}
\put(50,0){\circle*{11}}
\put(75,0){\circle*{11}}
\put(100,0){\circle*{11}}
\put(125,0){\circle*{11}}

\put(25,25){\circle*{11}}
\put(50,25){\circle*{11}}
\put(75,25){\circle*{11}}
\put(100,25){\circle*{11}}
\put(125,25){\circle*{11}}

\put(25,50){\circle*{11}}
\put(50,50){\circle*{11}}
\put(75,50){\circle*{11}}
\put(100,50){\circle*{11}}
\put(125,50){\circle*{11}}

\put(25,75){\circle*{11}}
\put(50,75){\circle*{11}}
\put(75,75){\circle*{11}}
\put(100,75){\circle*{11}}
\put(125,75){\circle*{11}}

\put(25,100){\circle*{11}}
\put(50,100){\circle*{11}}
\put(75,100){\circle*{11}}
\put(100,100){\circle*{11}}
\put(125,100){\circle*{11}}
}


\put(22.3,-3.4){\color{rootcol1}0}
\put(47.3,-3.4){\color{rootcol1}0}
\put(72.3,-3.4){\color{rootcol1}0}
\put(97.3,-3.4){\color{rootcol2}0}
\put(122.3,-2.7){\color{rootcol2}$\bullet$}

\put(22.3,22.3){\color{rootcol2}$\bullet$}
\put(47.3,21.6){\color{rootcol2}0}
\put(72.3,21.6){\color{rootcol1}1}
\put(97.3,21.6){\color{rootcol2}0}
\put(122.3,22.3){\color{rootcol2}$\bullet$}

\put(22.3,47.3){\color{rootcol2}$\bullet$}
\put(47.3,46.6){\color{rootcol2}0}
\put(72.3,46.6){\color{rootcol1} 2}
\put(97.3,46.6){\color{rootcol2}0}
\put(122.3,47.3){\color{rootcol2}$\bullet$}

\put(22.3,72.3){\color{rootcol2}$\bullet$}
\put(47.3,71.6){\color{rootcol2}0}
\put(72.3,71.6){\color{rootcol1}1}
\put(97.3,71.6){\color{rootcol2}0}
\put(122.3,72.3){\color{rootcol2}$\bullet$}

\put(22.3,97.3){\color{rootcol2}$\bullet$}
\put(47.3,96.6){\color{rootcol2}0}
\put(72.3,96.6){\color{rootcol1}0}
\put(97.3,96.6){\color{rootcol1}0}
\put(122.3,96.6){\color{rootcol1}0}

\normalsize

\put(200,85){$\ns=[0,0|1,1,1,1|0,0]$}
\put(200,60){$\hl_a=\{3,2,1,0\}+\Lambda$}
\put(200,35){$\hn_i=\{-2,-3,2,1\}-\Lambda$}
\put(200,10){$\tx_a=\{1,1,1,1\}$}

\end{picture}
\end{center}

\noindent where the diagram shows the number of roots in the distinguished Q-functions $\dQ_{a,s}$. Note that it is possible to choose a path from $\dQ_{0,0}$ to $\dQ_{4,4}$ with no Bethe roots, so the solution is in fact trivial. The distinguished Q-functions are
{
\arrayrulecolor{gray}\setlength{\extrarowheight}{\distQTabRowHeight}
\scriptsize
\hspace{\distQTabShift}\begin{equation*}
\begin{array}{|c|c|c|c|c|cl}\hhline{|-|->{\arrgr}|-|-|-|>{\arrc}~~}
 \grzero & \grcell{\trivAnz{\big(\sparg {\sparg^{[-2]}} {\sparg^{[2]}}\big)^{-2}} } & \grc \grone & \grc \grone & \grcell{\grc \grone} &  &  \\[\exLin]\hhline{-|->{\arrgr}|>{\arrc}->{\arrgr}|-|-|>{\arrc}~~}
 \grzero & \grcell{\trivAnz{\big({\sparg^-} {\sparg^+}\big)^{-2}} } & \grcell{\grc \spar} & \grone & \grzero &  &  \\[\exLin]\hhline{-|->{\arrgr}|>{\arrc}->{\arrgr}|>{\arrc}-|-~~}
 \grzero & \grcell{\trivAnz{\sparg^{-2}} } & \grcell{\cellcolor{momcell}\spar^2-\frac{\grone}{12}} & \grone & \grzero &  &  \\[\exLin]\hhline{-|->{\arrgr}|>{\arrc}->{\arrgr}|>{\arrc}-|-~~}
 \grzero & \grcell{\grone} & \grcell{\grc \trivAnz{\sparg^{2}} \spar} & \grone & \grzero &  &  \\[\exLin]\hhline{>{\arrgr}|-|->{\arrgr}|>{\arrc}->{\arrgr}|>{\arrc}-|-~~}
 \grcellFirstL{\grc \grone} & \grc \trivAnz{\sparg^{2}}  & \grcell{\grc \trivAnz{\big({\sparg^-} {\sparg^+}\big)^{2}} } & \grone & \grzero &  &  \\[\exLin]\hhline{>{\arrgr}|-|-|-|>{\arrc}-|-|~~}
\end{array}
\end{equation*}
}
from which one finds
{\footnotesize
\begin{align}
  \bP_1^{(0)} &= 0 \,,   &    \bQ_1^{(0)} &= \frac{\ii \spar^2}{12} \,,  \no \\
  \bP_2^{(0)} &= \frac{1}{\spar^2} \,,   &    \bQ_2^{(0)} &= -\frac{\spar^3}{20} \,,  \no \\
  \bP_3^{(0)} &= \frac{6 \ii}{\spar} \,,   &    \bQ_3^{(0)} &= -720 \etaF2 \spar^3-720 \ii \spar^2+360 \spar+120 i \,,  \no \\
  \bP_4^{(0)} &= -12 \,,   &    \bQ_4^{(0)} &= 0 \,.
\end{align}}
\noindent Going through the prescribed procedure, we can reproduce the first eight loop corrections to the anomalous dimensions:
{\allowdisplaybreaks\begin{align}
\ado{1} &= 12 \label{ZZbres}\\[1mm]
\ado{2} &= -48 \no\\[1mm]
\ado{3} &= 336 \no\\[1mm]
\ado{4} &= 576 \zv3-1440 \zv5-2496\no \\[1mm]
\ado{5} &= -5184 \zv3^2+6912 \zv3-8640 \zv5+30240 \zv7+15168 \no\\[1mm]
\ado{6} &= -20736 \zv3^2+155520 \zv5 \zv3-262656 \zv3+112320 \zv5+75600 \zv7-489888 \zv9-7680 \no\\[1mm]
\ado{7} &= 124416 \zv3^3-421632 \zv3^2+411264 \zv5 \zv3-1935360 \zv7 \zv3+5230080 \zv3 \no\\
  &\quad\,
  -993600 \zv5^2-229248 \zv5-1254960 \zv7-835488 \zv9+7318080 \zv{11}-2135040 \no\\[1mm]
\ado{8} &= -\frac{684288}{5} \Zv\no \\
&\quad\,
  +1990656 \zv3^3-3255552 \zv5 \zv3^2+7934976 \zv3^2-4354560 \zv5 \zv3-6229440 \zv7 \zv3 \no\\
    &\quad\,
    +23224320 \zv9 \zv3-83496960 \zv3+2384640 \zv5^2-19678464 \zv5+22256640 \zv5 \zv7 \no\\
      &\quad\,
      +21868704 \zv7+9327744 \zv9 
      +\frac{65929248 \zv{11}}{5}-106007616 \zv{13}+54408192\,.\no
\end{align}}
This result was first found in \cite{Leurent:2013mr} and has been extended to 11 loops in \cite{Marboe:2018ugv}, so we mainly include it for reference.


\end{shaded}

\subsection{The \texorpdfstring{$\sl(2)$}{sl(2)} operator \texorpdfstring{$\DD_{12}^2 \ZZ^2$}{D2Z2} - HWS in \texorpdfstring{\grpath{1}{1}{3}{3}}{1133}}

\begin{shaded}
\noindent For the $\sl(2)$ Konishi, the boundary conditions are
\begin{center}
\begin{picture}(300,105)

\linethickness{3.5mm}
\put(10,25){\color{betacol!30}\line(1,0){130}}

{\color{gray}
\linethickness{0.5mm}
\put(25,0){\line(1,0){100}}
\put(25,25){\line(1,0){100}}
\put(25,50){\line(1,0){100}}
\put(25,75){\line(1,0){100}}
\put(25,100){\line(1,0){100}}
\put(25,0){\line(0,1){100}}
\put(50,0){\color{black}\line(0,1){100}}
\put(75,0){\color{black}\line(0,1){100}}
\put(100,0){\color{black}\line(0,1){100}}
\put(125,0){\line(0,1){100}}
\put(50,0){\color{black}\line(1,0){50}}
\put(50,25){\color{black}\line(1,0){50}}
\put(50,50){\color{black}\line(1,0){50}}
\put(50,75){\color{black}\line(1,0){50}}
\put(50,100){\color{black}\line(1,0){50}}
}

\linethickness{1.3mm}
\put(25,0){\color{orange}\line(1,0){25}}
\put(50,0){\color{orange}\line(0,1){50}}
\put(50,50){\color{orange}\line(1,0){50}}
\put(100,50){\color{orange}\line(0,1){50}}
\put(100,100){\color{orange}\line(1,0){25}}

{\color{shadecolor}
\put(25,0){\circle*{11}}
\put(50,0){\circle*{11}}
\put(75,0){\circle*{11}}
\put(100,0){\circle*{11}}
\put(125,0){\circle*{11}}

\put(25,25){\circle*{11}}
\put(50,25){\circle*{11}}
\put(75,25){\circle*{11}}
\put(100,25){\circle*{11}}
\put(125,25){\circle*{11}}

\put(25,50){\circle*{11}}
\put(50,50){\circle*{11}}
\put(75,50){\circle*{11}}
\put(100,50){\circle*{11}}
\put(125,50){\circle*{11}}

\put(25,75){\circle*{11}}
\put(50,75){\circle*{11}}
\put(75,75){\circle*{11}}
\put(100,75){\circle*{11}}
\put(125,75){\circle*{11}}

\put(25,100){\circle*{11}}
\put(50,100){\circle*{11}}
\put(75,100){\circle*{11}}
\put(100,100){\circle*{11}}
\put(125,100){\circle*{11}}
}

\put(22.3,-3.4){\color{rootcol1}0}
\put(47.3,-3.4){\color{rootcol1}0}
\put(72.3,-3.4){\color{rootcol2}2}
\put(97.3,-3.4){\color{rootcol2}4}
\put(122.3,-2.7){\color{rootcol2}$\bullet$}

\put(22.3,21.6){\color{rootcol2}0}
\put(47.3,21.6){\color{rootcol1}0}
\put(72.3,21.6){\color{rootcol2}2}
\put(97.3,21.6){\color{rootcol2}2}
\put(122.3,22.3){\color{rootcol2}$\bullet$}

\put(22.3,46.6){\color{rootcol2}0}
\put(47.3,46.6){\color{rootcol1}0}
\put(72.3,46.6){\color{rootcol1}2}
\put(97.3,46.6){\color{rootcol1}0}
\put(122.3,47.3){\color{rootcol2}$\bullet$}

\put(22.3,72.3){\color{rootcol2}$\bullet$}
\put(47.3,71.6){\color{rootcol2}0}
\put(72.3,71.6){\color{rootcol2}1}
\put(97.3,71.6){\color{rootcol1}0}
\put(122.3,72.3){\color{rootcol2}$\bullet$}

\put(22.3,97.3){\color{rootcol2}$\bullet$}
\put(47.3,96.6){\color{rootcol2}0}
\put(72.3,96.6){\color{rootcol2}0}
\put(97.3,96.6){\color{rootcol1}0}
\put(122.3,96.6){\color{rootcol1}0}

\normalsize

\put(200,85){$\ns= 	[0,  2|2,  2,  0,  0|2,  0]$}
\put(200,60){$\hl_a=\{2, 2, 3, 2\}+\Lambda$}
\put(200,35){$\hn_i=\{-2, -5, 0, -1\}-\Lambda$}
\put(200,10){$\tx_a=\left\{\tx^2, \tx^{-2}, 1, 1\right\} $}

\end{picture}
\end{center}
Again, we see that there is a path without any Bethe roots, so the $\dQ$'s can again be trivially generated and are rational in the twist. The $\dQ$'s are
{
\arrayrulecolor{gray}\setlength{\extrarowheight}{\distQTabRowHeight}
\scriptsize
\hspace{\distQTabShift}\begin{equation*}
\begin{array}{|c|c|c|c|c|cl}\hhline{|-|-|->{\arrgr}|-|-|>{\arrc}~~}
 \grzero & \trivAnz{\big({\sparg^{[-2]}} \sparg {\sparg^{[2]}}\big)^{-2}}  & \grcell{\grone} & \grc \grone & \grcell{\grc \grone} &  &  \\[\exLin]\hhline{-|-|->{\arrgr}|>{\arrc}->{\arrgr}|-|>{\arrc}~~}
 \grzero & \trivAnz{\big({\sparg^-} {\sparg^+}\big)^{-2}}  & \grcell{\spar} & \grcell{\grc \grone} & \grzero &  &  \\[\exLin]\hhline{->{\arrgr}|-|->{\arrgr}|>{\arrc}->{\arrgr}|>{\arrc}-~~}
 \grcellL{\trivAnz{\big({\sparg^-} {\sparg^+}\big)^{-2}} } & \grc \trivAnz{\sparg^{-2}}  & \cellcolor{momcell}\spar^2-\frac{\grone}{12} & \grcell{\grc \grone} & \grzero &  &  \\[\exLin]\hhline{->{\arrgr}|>{\arrc}->{\arrgr}|-|-|>{\arrc}-~~}
 \grcellL{\rhigh \trivAnz{\sparg^{-2}} } & \grcell{\grc \grone} & \trivAnz{\sparg^{2}} \left(\spar^2+\frac{\ii \left(\tx^2+1\right) \spar}{\tx^2-1}-\frac{\grone}{3}\right) & \spar^2-\frac{\ii \left(\tx^2+1\right) \spar}{\tx^2-1}-\frac{\tx^4-26 \tx^2+1}{4 \left(\tx^2-1\right)^2} & \grzero &  & \rule{\exSpace}{0pt} \tx^{2 \ii \spar} \\[\exLinX]\hhline{>{\arrgr}|->{\arrgr}|>{\arrc}->{\arrgr}|>{\arrc}-|-|-~~}
 \grcellFirstL{\rhigh \grc \grone} & \grcell{\grc \trivAnz{\sparg^{2}} } & \trivAnz{\big({\sparg^-} {\sparg^+}\big)^{2}} \left(\spar^2-\frac{\tx^4-26 \tx^2+1}{12 \left(\tx^2-1\right)^2}\right) & \spar^4+\frac{2 \left(\tx^4+7 \tx^2+1\right) \spar^2}{\left(\tx^2-1\right)^2}+\frac{\tx^8-14 \tx^6+50 \tx^4-14 \tx^2+1}{\left(\tx^2-1\right)^4} & \grzero &  &  \\[\exLinX]\hhline{>{\arrgr}|-|-|>{\arrc}-|-|-|~~}
\end{array}
\end{equation*}
}
from which we get
{\footnotesize\allowdisplaybreaks
\begin{align*}
  \bP_1^{(0)} &= -\frac{\tx^{2+2 \ii \spar}}{\spar^2 \left(\tx^2-1\right)^2 \left(\tx^2+1\right)} \,,   &    \bQ_1^{(0)} &= \frac{\ii \spar^2}{6} \,,   \\
  \bP_2^{(0)} &= \frac{\tx^{2-2 \ii \spar}}{\spar^2 \left(\tx^2-1\right)} \,,   &    \bQ_2^{(0)} &= \frac{\spar^5 \left(\tx^2-1\right)^2}{20 \tx^2}+\frac{3 \spar^3}{10} \,,   \\
  \bP_3^{(0)} &= 0 \,,   &    \bQ_3^{(0)} &= \text{see below}\,,   \\
  \bP_4^{(0)} &= \frac{4 \ii}{\spar^2} \,,   &    \bQ_4^{(0)} &= 0 \,,  
\end{align*}
\begin{align*}
  \bQ_3^{(0)} & = \etaF2 \left(\frac{30 \spar^5 \left(\tx^2-1\right)^2}{\tx^2}+180 \spar^3\right) + \frac{30 \ii \spar^4 \left(\tx^2-1\right)^2}{\tx^2}-\frac{15 \spar^3 \left(\tx^2-1\right)^2}{\tx^2} \\
  & \quad\, -\frac{5 \ii \spar^2 \left(\tx^4-38 \tx^2+1\right)}{\tx^2}-90 \spar-30 \ii
  .
\end{align*}
}
With the dependence on the twist being rational and rather simple, we have been able to run this example through seven loop orders in a few hours, yielding the result
{\allowdisplaybreaks
\begin{align}
\ado{1} &= 12 \label{DZres} \\[1mm]
\ado{2} &= -48 \no\\[1mm]
\ado{3} &= -12 \big(\tc{2}-29\big)\no \\[1mm]
\ado{4} &= -192 \zv3 \big(\tc{2}-4\big)-1440 \zv5+348 \tc{2}-2844 \no\\[1mm]
\ado{5} &= 96 \zv3 (29 \tc{2}+43)+2880 \zv5 (\tc{2}-4)-5184 \zv3^2+30240 \zv7 \no\\
  &\quad\, -7380 \tc{2}+22548\no\\[1mm]
\ado{6} &= 6912 \zv3^2 \big(\tc{2}-4\big)-44832 \zv3 \tc{2}+192 \zv3 \tc{4} \no\\
  &\quad\, +288 \zv5 \big(539-149 \tc{2}\big)+336 \zv7 \big(334-109 \tc{2}\big)+155520 \zv5 \zv3-218016 \zv3 \no\\
    &\quad\, -489888 \zv9+136428 \tc{2}-156 \tc{4}-143952\no \\[1mm]
\ado{7} &= 576 \zv3^2 \big(-17 \tc{2}+3 \tc{4}-718\big)+597984 \zv3 \tc{2}-7824 \zv3 \tc{4}\no \\
  &\quad\, +576 \zv5 \zv3 \big(863-149 \tc{2}\big) +652128 \zv5 \tc{2}-4560 \zv5 \tc{4}\no \\
    &\quad\, +453600 \zv7 \tc{2}+451584 \zv9 \tc{2}+124416 \zv3^3-1935360 \zv7 \zv3+4639920 \zv3  \no\\
      &\quad\, -993600 \zv5^2 -876816 \zv5-1708560 \zv7-1287072 \zv9+7318080 \zv{11} \no\\
      &\quad\, -2315988 \tc{2}+9984 \tc{4}+170964 \,.\no
\end{align}}
Again, this result nicely reduces to \eqref{ZZbres} in the limit $\beta\to0$, and the first four orders are in agreement with the known result \cite{deLeeuw:2010ed}.


\end{shaded}

\subsection{The operator \texorpdfstring{$\DD_{12}^2 \ZZ \XX$}{D2ZX} - \texorpdfstring{$R$}{R}-symmetry descendant in the undeformed theory}\label{ex:DZX}
\begin{shaded}

\noindent
This is an example of an operator that is not a HWS in any grading in the undeformed theory. It is obtained by acting on the $\sl(2)$ Konishi operator with content $\DD_{12}^2\ZZ^2$ with the $R$-symmetry generator $\fff_3^\dagger\fff_2$. This symmetry is broken in the $\tb$-deformed theory, and thus these two types of operators are no longer in the same multiplet.

As the action of the $R$-symmetry does not correspond to a fermionic duality transformation, we keep the grading of the $\sl(2)$ operator, \grpath{1}{1}{3}{3}, and then the charges and boundary conditions for such a state should be
\begin{center}
\begin{picture}(300,105)

\linethickness{3.5mm}
\put(10,50){\color{betacol!30}\line(1,0){130}}

{\color{gray}
\linethickness{0.5mm}
\put(25,0){\line(1,0){100}}
\put(25,25){\line(1,0){100}}
\put(25,50){\line(1,0){100}}
\put(25,75){\line(1,0){100}}
\put(25,100){\line(1,0){100}}
\put(25,0){\line(0,1){100}}
\put(50,0){\line(0,1){100}}
\put(75,0){\color{black}\line(0,1){100}}
\put(100,0){\line(0,1){100}}
\put(125,0){\line(0,1){100}}
\put(50,0){\color{black}\line(1,0){50}}
\put(25,25){\color{black}\line(1,0){75}}
\put(25,50){\color{black}\line(1,0){100}}
\put(50,75){\color{black}\line(1,0){75}}
\put(50,100){\color{black}\line(1,0){50}}
\put(100,0){\color{black}\line(0,1){25}}
\put(25,25){\color{black}\line(0,1){25}}
\put(125,50){\color{black}\line(0,1){25}}
\put(50,75){\color{black}\line(0,1){25}}
}

\linethickness{1.3mm}
\put(25,0){\color{orange}\line(1,0){25}}
\put(50,0){\color{orange}\line(0,1){50}}
\put(50,50){\color{orange}\line(1,0){50}}
\put(100,50){\color{orange}\line(0,1){50}}
\put(100,100){\color{orange}\line(1,0){25}}

{\color{shadecolor}
\put(25,0){\circle*{11}}
\put(50,0){\circle*{11}}
\put(75,0){\circle*{11}}
\put(100,0){\circle*{11}}
\put(125,0){\circle*{11}}

\put(25,25){\circle*{11}}
\put(50,25){\circle*{11}}
\put(75,25){\circle*{11}}
\put(100,25){\circle*{11}}
\put(125,25){\circle*{11}}

\put(25,50){\circle*{11}}
\put(50,50){\circle*{11}}
\put(75,50){\circle*{11}}
\put(100,50){\circle*{11}}
\put(125,50){\circle*{11}}

\put(25,75){\circle*{11}}
\put(50,75){\circle*{11}}
\put(75,75){\circle*{11}}
\put(100,75){\circle*{11}}
\put(125,75){\circle*{11}}

\put(25,100){\circle*{11}}
\put(50,100){\circle*{11}}
\put(75,100){\circle*{11}}
\put(100,100){\circle*{11}}
\put(125,100){\circle*{11}}
}

\put(22.3,-3.4){\color{rootcol1}0}
\put(47.3,-3.4){\color{rootcol1}0}
\put(72.3,-3.4){\color{rootcol2}1}
\put(97.3,-3.4){\color{rootcol2}2}
\put(122.3,-2.7){\color{rootcol2}$\bullet$}

\put(22.3,22.3){\color{rootcol2}$\bullet$}
\put(47.3,21.6){\color{rootcol1}0}
\put(72.3,21.6){\color{rootcol2}2}
\put(97.3,21.6){\color{rootcol2}2}
\put(122.3,22.3){\color{rootcol2}$\bullet$}

\put(22.3,47.3){\color{rootcol2}$\bullet$}
\put(47.3,46.6){\color{rootcol1}1}
\put(72.3,46.6){\color{rootcol1}3}
\put(97.3,46.6){\color{rootcol1}1}
\put(122.3,47.3){\color{rootcol2}$\bullet$}

\put(22.3,72.3){\color{rootcol2}$\bullet$}
\put(47.3,71.6){\color{rootcol2}2}
\put(72.3,71.6){\color{rootcol2}2}
\put(97.3,71.6){\color{rootcol1}0}
\put(122.3,72.3){\color{rootcol2}$\bullet$}

\put(22.3,97.3){\color{rootcol2}$\bullet$}
\put(47.3,96.6){\color{rootcol2}2}
\put(72.3,96.6){\color{rootcol2}1}
\put(97.3,96.6){\color{rootcol1}0}
\put(122.3,96.6){\color{rootcol1}0}

\normalsize

\put(200,85){$\ns= 	[0,  2|2,  1,  1,  0|2,  0]$}
\put(200,60){$\hl_a=\{3, 1, 1, 2\}+\Lambda$}
\put(200,35){$\hn_i=\{-2, -4, 1, -1\}-\Lambda$}
\put(200,10){$\tx_a=\left\{1, \tx^{-1}, \tx, 1\right\} $}

\end{picture}
\end{center}

\noindent Loosely, one can think of the action of the $R$-symmetry as deforming the Young diagram as
\begin{center}
\begin{picture}(250,70)
\setlength{\unitlength}{0.6mm}
\linethickness{0.5mm}\color{black}
\put(12,36){\line(1,0){12}}
\put(12,30){\line(1,0){12}}
\put(12,12){\line(1,0){12}}
\put(12,18){\line(1,0){12}}
\put(12,24){\line(1,0){12}}
\put(12,12){\line(0,1){24}}
\put(18,12){\line(0,1){24}}
\put(24,12){\line(0,1){24}}
\linethickness{1mm}\color{orange}
\put(6,12){\line(1,0){6}}
\put(12,12){\line(0,1){12}}
\put(12,24){\line(1,0){12}}
\put(24,24){\line(0,1){12}}
\put(24,36){\line(1,0){6}}

\put(62,22){\color{black}\huge$\rightarrow$}
\put(63,32){\color{black}$\fff_3^\dagger\fff_2$}

\linethickness{0.5mm}\color{black}
\put(112,12){\line(1,0){12}}
\put(106,18){\line(1,0){18}}
\put(106,24){\line(1,0){24}}
\put(112,30){\line(1,0){18}}
\put(112,36){\line(1,0){12}}
\put(106,18){\line(0,1){6}}
\put(112,12){\line(0,1){12}}\put(112,30){\line(0,1){6}}
\put(118,12){\line(0,1){24}}
\put(124,12){\line(0,1){6}}\put(124,24){\line(0,1){12}}
\put(130,24){\line(0,1){6}}
\linethickness{1mm}\color{orange}
\put(106,12){\line(1,0){6}}
\put(112,12){\line(0,1){12}}
\put(112,24){\line(1,0){12}}
\put(124,24){\line(0,1){12}}
\put(124,36){\line(1,0){6}}
\end{picture}
\end{center}
since then the rule for determining the number of Bethe roots by counting boxes in the diagram \cite{Marboe:2017dmb} can be applied to get the number of roots on the \grpath{1}{1}{3}{3} path.

This time, the lowest number of Bethe roots on a path from $\dQ_{0,0}$ to $\dQ_{4,4}$ is five, and the solution of the corresponding zero-remainder conditions in fact gives rise to two solutions that both reduce to the untwisted Konishi solution in the limit $\tb\to0$. Correspondingly, there must be two operators of this type in the Konishi multiplet. One solution is
{
\arrayrulecolor{gray}\setlength{\extrarowheight}{\distQTabRowHeight}
\scriptsize
\hspace{\distQTabShift}\begin{equation*}
\begin{array}{|c|c|c|c|c|cl}\hhline{|-|-|->{\arrgr}|-|-|>{\arrc}~~}
 \grzero & \trivAnz{\big({\sparg^{[-2]}} \sparg {\sparg^{[2]}}\big)^{-2}} \left(\spar^2-\frac{2 \ii \sqrt{6\tx} \spar}{\tx-1}+\frac{\tx^2-6 \tx+1}{(\tx-1)^2}\right) & \grcell{\spar-\frac{\ii \sqrt{\frac{3}{2}\tx}}{\tx-1}} & \grc \grone & \grcell{\grc \grone} &  &  \\[\exLin]\hhline{-|-|->{\arrgr}|>{\arrc}->{\arrgr}|-|>{\arrc}~~}
 \grzero & \trivAnz{\big({\sparg^-} {\sparg^+}\big)^{-2}} \left(\spar^2-\frac{2 \ii \sqrt{6\tx} \spar}{\tx-1}+\frac{\tx^2-26 \tx+1}{4 (\tx-1)^2}\right) & \grcell{\spar^2-\frac{\ii \sqrt{6\tx}}{\tx-1}} & \grcell{\grc \grone} & \grzero &  &  \\[\exLin]\hhline{->{\arrgr}|-|->{\arrgr}|>{\arrc}->{\arrgr}|>{\arrc}-~~}
 \grcellL{\grzero} & \grc \trivAnz{\sparg^{-2}} \left(\spar-\frac{\ii \sqrt{6\tx}}{\tx-1}\right) & \cellcolor{momcell}\spar^3-\frac{3 \ii (\tx+1) \spar^2}{2 (\tx-1)}-\frac{3 \spar}{4}+\frac{\ii (\tx+1)}{8 (\tx-1)} & \grcell{\grc \spar+\frac{\ii \sqrt{6\tx}}{\tx-1}} & \grzero &  & \rule{\exSpace}{0pt} \tx^{-\ii \spar} \\[\exLin]\hhline{->{\arrgr}|>{\arrc}->{\arrgr}|-|-|>{\arrc}-~~}
 \grcellL{\grzero} & \grcell{\grc \grone} & \trivAnz{\sparg^{2}} \left(\spar^2+\frac{\ii \sqrt{6\tx} \spar}{\tx-1}\right) & \spar^2+\frac{2 \ii \sqrt{6\tx} \spar}{\tx-1}+\frac{\tx^2-26 \tx+1}{4 (\tx-1)^2} & \grzero &  &  \\[\exLin]\hhline{>{\arrgr}|->{\arrgr}|>{\arrc}->{\arrgr}|>{\arrc}-|-|-~~}
 \grcellFirstL{\grc \grone} & \grcell{\grc \trivAnz{\sparg^{2}} } & \trivAnz{\big({\sparg^-} {\sparg^+}\big)^{2}} \left(\spar+\frac{\ii \sqrt{\frac{3}{2}\tx}}{\tx-1}\right) & \spar^2+\frac{2 \ii \sqrt{6\tx} \spar}{\tx-1}+\frac{\tx^2-6 \tx+1}{(\tx-1)^2} & \grzero &  &  \\[\exLin]\hhline{>{\arrgr}|-|-|>{\arrc}-|-|-|~~}
\end{array}
\end{equation*}
}
%
\noindent while the second one is given by the replacement $\sqrt{6}\to-\sqrt{6}$. 
For the first solution, the above distinguished Q-functions lead to the single-indexed Q-functions
{\footnotesize
\begin{align*}
  \bP_1^{(0)} &= 0 \,,   &    \bQ_1^{(0)} &= \frac{\ii \spar^2}{6} \,,  \no \\
  \bP_2^{(0)} &= \tx^{-\ii \spar} \left(\frac{1}{\spar (\tx+1)}-\frac{\ii \sqrt{6} \sqrt{\tx}}{\spar^2 \left(\tx^2-1\right)}\right) \,,   &    \bQ_2^{(0)} &= \frac{2 \sqrt{\frac{2}{3}} \spar^3 \sqrt{\tx}}{5 (\tx-1)}-\frac{2 \ii \spar^4}{15} \,, \no  \\
  \bP_3^{(0)} &= \left(\frac{\tx-1}{\spar}-\frac{\ii \sqrt{6} \sqrt{\tx}}{\spar^2}\right) \tx^{\ii \spar} \,,   &    \bQ_3^{(0)} &= \text{see below} \,, \no  \\
  \bP_4^{(0)} &= \frac{6 \ii}{\spar^2} \,,   &    \bQ_4^{(0)} &= 0 \,,  
\end{align*}
\begin{align}
  \bQ_3^{(0)} &= \etaF2 \left(\frac{45 \sqrt{6} \spar^3 (\tx-1)}{\sqrt{\tx}}-\frac{45 \ii \spar^4 (\tx-1)^2}{\tx}\right) + \frac{45 \spar^3 (\tx-1)^2}{\tx}+\frac{45 \ii \spar^2 (\tx-1) \left(\tx+2 \sqrt{6} \sqrt{\tx}-1\right)}{2 \tx} \no \\
  &  \quad\, -\frac{15 \spar (\tx-1) \left(\tx+3 \sqrt{6} \sqrt{\tx}-1\right)}{2 \tx}-\frac{15 \ii \sqrt{\frac{3}{2}} (\tx-1)}{\sqrt{\tx}} 
  \,.
\end{align}
}
Again, $\bP_a^{(0)}$ and $\bQ_i^{(0)}$ for the second solution is obtained by the replacement $\sqrt{6}\to-\sqrt{6}$.

The appearance of the $\sqrt{6}$ slows down the perturbative calculation somewhat such that reaching the 7-loop anomalous dimension with our code requires about 25 hours on a standard laptop. The anomalous dimension is the same for both solutions and is given by
{\allowdisplaybreaks
\begin{align}
\ado{1} &= 12 \label{DZXres}\\[1mm]
\ado{2} &= -48 \no\\[1mm]
\ado{3} &= -12\big(\tc1-29\big)\no \\[1mm]
\ado{4} &= -2832 + 336 \tc1 -192 \zv3 \big(\tc1-4\big) -1440 \zv5 \no\\[1mm]
\ado{5} &= 22128 -6960 \tc1 + 2304 \zv3 \big(\tc1+2\big) -5184 \zv3^2\no  \\
  &\quad\, + 480 \zv5 \big(7 \tc1-25\big) + 30240 \zv7 \no\\[1mm]
\ado{6} &= -133890 + 126360 \tc1 -150 \tc{2} -225696 \zv3 -37152 \zv3 \tc1  \no\\
  &\quad\, + 192 \zv3 \tc{2} + 1728 \zv3^2 \big(5 \tc1-17\big) + 192 \zv5 \big(791-206 \tc1\big)  \no\\
  &\quad\, + 155520 \zv3 \zv5 + 3024 \zv7 \big(41-16 \tc1\big) -489888 \zv9 \no\\[1mm]
\ado{7} &= -32844 -2111472 \tc1 + 9276 \tc{2} + 4727760 \zv3 + 509568 \zv3 \tc1  \no\\
  &\quad\, -7248 \zv3 \tc{2} + 576 \zv3^2 \big(-14 \tc1+3 \tc{2}-721\big) + 124416 \zv3^3   \no\\
  &\quad\, -48 \zv5 \big(-10934 \tc1+95 \tc{2}+15615\big) + 1152 \zv3 \zv5 \big(484-127 \tc1\big)  \no\\
  &\quad\, -993600 \zv5^2 + 5040 \zv7 \big(92 \tc1-341\big) -1935360 \zv3 \zv7 -1498752 \zv9  \no\\
  &\quad\, + 663264 \zv9 \tc1 + 7318080 \zv{11}\,. \no
\end{align}}
\vspace{-5mm}
\end{shaded}

\subsection{The operator \texorpdfstring{$\ZZ \Psib_{22}$}{ZPsi} - HWS in \texorpdfstring{\grpath{1}{2}{2}{2}}{1222}}
\begin{shaded}
\noindent This operator corresponds to the boundary conditions
\begin{center}
\begin{picture}(300,105)

\linethickness{3.5mm}
\put(10,50){\color{betacol!30}\line(1,0){130}}

{\color{gray}
\linethickness{0.5mm}
\put(25,0){\line(1,0){100}}
\put(25,25){\line(1,0){100}}
\put(25,50){\line(1,0){100}}
\put(25,75){\line(1,0){100}}
\put(25,100){\line(1,0){100}}
\put(25,0){\line(0,1){100}}
\put(50,0){\color{black}\line(0,1){100}}
\put(75,0){\color{black}\line(0,1){100}}
\put(100,0){\color{black}\line(0,1){100}}
\put(125,0){\line(0,1){100}}
\put(50,0){\color{black}\line(1,0){50}}
\put(50,25){\color{black}\line(1,0){50}}
\put(50,50){\color{black}\line(1,0){50}}
\put(50,75){\color{black}\line(1,0){50}}
\put(50,100){\color{black}\line(1,0){50}}
}

\linethickness{1.3mm}
\put(25,0){\color{orange}\line(1,0){25}}
\put(50,0){\color{orange}\line(0,1){25}}
\put(50,25){\color{orange}\line(1,0){25}}
\put(75,25){\color{orange}\line(0,1){75}}
\put(75,100){\color{orange}\line(1,0){50}}

{\color{shadecolor}
\put(25,0){\circle*{11}}
\put(50,0){\circle*{11}}
\put(75,0){\circle*{11}}
\put(100,0){\circle*{11}}
\put(125,0){\circle*{11}}

\put(25,25){\circle*{11}}
\put(50,25){\circle*{11}}
\put(75,25){\circle*{11}}
\put(100,25){\circle*{11}}
\put(125,25){\circle*{11}}

\put(25,50){\circle*{11}}
\put(50,50){\circle*{11}}
\put(75,50){\circle*{11}}
\put(100,50){\circle*{11}}
\put(125,50){\circle*{11}}

\put(25,75){\circle*{11}}
\put(50,75){\circle*{11}}
\put(75,75){\circle*{11}}
\put(100,75){\circle*{11}}
\put(125,75){\circle*{11}}

\put(25,100){\circle*{11}}
\put(50,100){\circle*{11}}
\put(75,100){\circle*{11}}
\put(100,100){\circle*{11}}
\put(125,100){\circle*{11}}
}

\put(22.3,-3.4){\color{rootcol1}0}
\put(47.3,-3.4){\color{rootcol1}0}
\put(72.3,-3.4){\color{rootcol2}0}
\put(97.3,-3.4){\color{rootcol2}2}
\put(122.3,-2.7){\color{rootcol2}$\bullet$}

\put(22.3,22.3){\color{rootcol2}$\bullet$}
\put(47.3,21.6){\color{rootcol1}0}
\put(72.3,21.6){\color{rootcol1}1}
\put(97.3,21.6){\color{rootcol2}2}
\put(122.3,22.3){\color{rootcol2}$\bullet$}

\put(22.3,47.3){\color{rootcol2}$\bullet$}
\put(47.3,46.6){\color{rootcol2}1}
\put(72.3,46.6){\color{rootcol1}2}
\put(97.3,46.6){\color{rootcol2}1}
\put(122.3,47.3){\color{rootcol2}$\bullet$}

\put(22.3,72.3){\color{rootcol2}$\bullet$}
\put(47.3,71.6){\color{rootcol2}2}
\put(72.3,71.6){\color{rootcol1}1}
\put(97.3,71.6){\color{rootcol2}0}
\put(122.3,72.3){\color{rootcol2}$\bullet$}

\put(22.3,97.3){\color{rootcol2}$\bullet$}
\put(47.3,96.6){\color{rootcol2}2}
\put(72.3,96.6){\color{rootcol1}0}
\put(97.3,96.6){\color{rootcol1}0}
\put(122.3,96.6){\color{rootcol1}0}

\normalsize

\put(200,85){$\ns= [0,  1|2,  1,  1,  1|0,  0]$}
\put(200,60){$\hl_a=\{3, 1, 1, 2\}+\Lambda$}
\put(200,35){$\hn_i=\{-2, -3, 0, -1\}-\Lambda$}
\put(200,10){$\tx_a=\left\{1, \tx^{-1}, \tx, 1\right\}  $}

\end{picture}
\end{center}
The solution of the distinguished Q-system contains the square root, 
\begin{equation*}
\sqrt{\tx \left(2 \tx^2+5 \tx+ 2\right)} \equiv \koSqrt 
\;,
\end{equation*}
which significantly slows down the computer calculations, as discussed in section \ref{sec:results}. 
The distinguished Q-functions, using this notation, are
{
\arrayrulecolor{gray}\setlength{\extrarowheight}{\distQTabRowHeight}
\scriptsize
\hspace{\distQTabShift}\begin{equation*}
\begin{array}{|c|c|c|c|c|cl}\hhline{|-|->{\arrgr}|-|-|-|>{\arrc}~~}
 \rhigh \grzero & \grcell{\trivAnz{\big({\sparg^{[-2]}} \sparg {\sparg^{[2]}}\big)^{-2}} \left(\spar^2+\frac{\tx^2+4 \tx+2 \koSqrt+1}{(\tx-1)^2}\right)} & \grc \grone & \grc \grone & \grcell{\grc \grone} &  &  \\[\exLinX]\hhline{-|->{\arrgr}|>{\arrc}->{\arrgr}|-|-|>{\arrc}~~}
 \rhigh \grzero & \grcell{\trivAnz{\big({\sparg^-} {\sparg^+}\big)^{-2}} \left(\spar^2+\frac{\tx^2+10 \tx+4 \koSqrt+1}{4 (\tx-1)^2}\right)} & \grcell{\grc \spar} & \grone & \grzero &  &  \\[\exLinX]\hhline{-|->{\arrgr}|>{\arrc}->{\arrgr}|>{\arrc}-|-~~}
 \rhigh \grzero & \grcell{\trivAnz{\sparg^{-2}} \left(\spar+\frac{\ii (\koSqrt+\tx)}{\tx^2-1}\right)} & \grcell{\cellcolor{momcell}\spar^2-\frac{\ii \left(\tx^2+\tx-\koSqrt+1\right) \spar}{\tx^2-1}-\frac{\tx^2+4 \tx-2 \koSqrt+1}{4 (\tx-1)^2}} & \spar+\frac{\ii (\koSqrt+\tx)}{\tx^2-1} & \grzero &  & \rule{\exSpace}{0pt} \tx^{-\ii \spar} \\[\exLinX]\hhline{->{\arrgr}|->{\arrgr}|>{\arrc}->{\arrgr}|>{\arrc}-|-~~}
 \grcellL{\grzero} & \grc \grone & \grcell{\grc \trivAnz{\sparg^{2}} \spar} & \spar^2+\frac{\tx^2+10 \tx+4 \koSqrt+1}{4 (\tx-1)^2} & \grzero &  &  \\[\exLin]\hhline{>{\arrgr}|->{\arrgr}|>{\arrc}->{\arrgr}|-|>{\arrc}-|-~~}
 \grcellFirstL{\grc \grone} & \grcell{\grc \trivAnz{\sparg^{2}} } & \trivAnz{\big({\sparg^-} {\sparg^+}\big)^{2}}  & \spar^2+\frac{\tx^2+4 \tx+2 \koSqrt+1}{(\tx-1)^2} & \grzero &  &  \\[\exLin]\hhline{>{\arrgr}|-|-|>{\arrc}-|-|-|~~}
\end{array}
\end{equation*}
}
while the leading $\bP$ and $\bQ$ read
{\small\allowdisplaybreaks
\begin{align*}
  \bP_1^{(0)} &= 0 \,,   &    \bQ_1^{(0)} &= \frac{\ii \spar^2}{2} \,,   \\
  \bP_2^{(0)} &= \tx^{-\ii \spar} \left(\frac{1}{\spar (\tx+1)}+\frac{\ii(\koSqrt+ \tx)}{\spar^2 (\tx-1) (\tx+1)^2}\right) \,,   &    \bQ_2^{(0)} &= -\frac{\ii \spar^3}{6} \,,   \\
  \bP_3^{(0)} &= \tx^{\ii \spar} \left(\frac{\tx-1}{\spar}-\frac{\ii(\koSqrt+ \tx)}{\spar^2 (\tx+1)}\right) \,,   &    \bQ_3^{(0)} &= \text{see below}\,,   \\
  \bP_4^{(0)} &= -\frac{2}{\spar^2} \,,   &    \bQ_4^{(0)} &= 0 \,,
\end{align*}
\begin{align}
  \bQ_3^{(0)} &=  -\frac{12 \etaF2 \spar^3 \left(-3 \ii \koSqrt+\tx^2+7 \tx+1\right)}{\tx}-\frac{12 \ii \spar^2 \left(-3 \ii \koSqrt+\tx^2+7 \tx+1\right)}{\tx} \no \\
  & \quad\: +\frac{6 \spar \left(-3 \ii \koSqrt+\tx^2+7 \tx+1\right)}{\tx}+\frac{6 \koSqrt}{\tx}+18 \ii 
  .
\end{align}
}

Despite the challenge of working with such expressions, the anomalous dimension up to five loops is nevertheless accessible within a few hours on a standard laptop:
\newcommand{\wilDelta}{\textcolor{xcol}{\Delta}}
{\allowdisplaybreaks
\begin{align}
\ado{1} &= 6 + 6 \wilDelta \label{ZPsibres}  \\[1mm]
\ado{2} &= -\frac{3}{\wilDelta } -15 -21 \wilDelta  -9 \wilDelta ^2 \no\\[1mm]
\ado{3} &= -\frac{3}{4 \wilDelta ^3} + \frac{153}{4 \wilDelta} + 114 + \frac{495 \wilDelta }{4} + 54 \wilDelta ^2 + \frac{27 \wilDelta ^3}{4} \no\\[1mm]
\ado{4} &= -\frac{3}{8 \wilDelta ^5} + \frac{33}{2 \wilDelta ^3} -\frac{1701}{4 \wilDelta } -1230 -\frac{2427 \wilDelta }{2} -180 \wilDelta ^2 + \frac{2997 \wilDelta ^3}{8} + 162 \wilDelta ^4 \no \\
  &\quad\, + \left(-243 \wilDelta ^4-405 \wilDelta ^3+234 \wilDelta ^2+702 \wilDelta +297 -\frac{9}{\wilDelta }\right) \zv3 + \left(-360 \wilDelta ^2-720 \wilDelta -360\right) \zv5 \no\\[1mm]
\ado{5} &= -\frac{15}{64 \wilDelta ^7} + \frac{375}{32 \wilDelta ^5} -\frac{16725}{64 \wilDelta ^3} + \frac{76605}{16 \wilDelta } + 14244 + \frac{982455 \wilDelta }{64} \no \\
  &\quad\, + 1440 \wilDelta ^2 -\frac{331425 \wilDelta ^3}{32} -8100 \wilDelta ^4 -\frac{124659 \wilDelta ^5}{64} \no \\
  &\quad\, + \left(\frac{3645 \wilDelta ^5}{2}+6156 \wilDelta ^4+6129 \wilDelta ^3-576 \wilDelta ^2-4266 \wilDelta -2124 -\frac{225}{\wilDelta } -\frac{9}{2 \wilDelta ^3}\right) \zv3  \no \\
  &\quad\, + \left(3240 \wilDelta ^4+4860 \wilDelta ^3-4320 \wilDelta ^2-9720 \wilDelta -3240 +\frac{540}{\wilDelta }\right) \zv5 \no \\
    &\quad\, + \left(-648 \wilDelta ^3-1944 \wilDelta ^2-1944 \wilDelta -648\right) \zv3^2 + \left(7560 \wilDelta ^2+15120 \wilDelta +7560\right) \zv7\no
\end{align}}
where
\begin{equation}
  \wilDelta \equiv \frac{\sqrt{5 + 4 \cos \tb}}{3}
  .
\end{equation}
As we will now discuss, this solution is, up to the replacement $\tb\to2\tb$, the same for the two examples $\XX\YY\ZZ$ and $\ZZ^2 \XX^2$.
\end{shaded}

\subsection{The operator \texorpdfstring{$\XX \YY \ZZ$}{XYZ} - HWS in \texorpdfstring{\grpath{0}{2}{2}{2}}{0222}}

\begin{shaded}
\noindent 
This operator has the same solution as $\ZZ \Psib_{22}$ in the example above, with the only difference that $\tx \to \tx^2$. The grading, oscillator content, twist factors and the shifted quantum numbers are
\begin{center}
\begin{picture}(300,105)

\linethickness{3.5mm}
\put(10,50){\color{betacol!30}\line(1,0){130}}

{\color{gray}
\linethickness{0.5mm}
\put(25,0){\line(1,0){100}}
\put(25,25){\line(1,0){100}}
\put(25,50){\line(1,0){100}}
\put(25,75){\line(1,0){100}}
\put(25,100){\line(1,0){100}}
\put(25,0){\line(0,1){100}}
\put(50,0){\line(0,1){25}}
\put(75,0){\line(0,1){25}}
\put(100,0){\line(0,1){25}}
\put(25,25){\color{black}\line(0,1){75}}
\put(50,25){\color{black}\line(0,1){75}}
\put(75,25){\color{black}\line(0,1){75}}
\put(100,25){\color{black}\line(0,1){75}}
\put(125,0){\line(0,1){100}}
\put(25,25){\color{black}\line(1,0){75}}
\put(25,50){\color{black}\line(1,0){75}}
\put(25,75){\color{black}\line(1,0){75}}
\put(25,100){\color{black}\line(1,0){75}}
}

\linethickness{1.3mm}
\put(25,0){\color{orange}\line(0,1){25}}
\put(25,25){\color{orange}\line(1,0){50}}
\put(75,25){\color{orange}\line(0,1){75}}
\put(75,100){\color{orange}\line(1,0){50}}

{\color{shadecolor}
\put(25,0){\circle*{11}}
\put(50,0){\circle*{11}}
\put(75,0){\circle*{11}}
\put(100,0){\circle*{11}}
\put(125,0){\circle*{11}}

\put(25,25){\circle*{11}}
\put(50,25){\circle*{11}}
\put(75,25){\circle*{11}}
\put(100,25){\circle*{11}}
\put(125,25){\circle*{11}}

\put(25,50){\circle*{11}}
\put(50,50){\circle*{11}}
\put(75,50){\circle*{11}}
\put(100,50){\circle*{11}}
\put(125,50){\circle*{11}}

\put(25,75){\circle*{11}}
\put(50,75){\circle*{11}}
\put(75,75){\circle*{11}}
\put(100,75){\circle*{11}}
\put(125,75){\circle*{11}}

\put(25,100){\circle*{11}}
\put(50,100){\circle*{11}}
\put(75,100){\circle*{11}}
\put(100,100){\circle*{11}}
\put(125,100){\circle*{11}}
}

\put(22.3,-3.4){\color{rootcol1}0}
\put(47.3,-2.7){\color{rootcol2}$\bullet$}
\put(72.3,-2.7){\color{rootcol2}$\bullet$}
\put(97.3,-2.7){\color{rootcol2}$\bullet$}
\put(122.3,-2.7){\color{rootcol2}$\bullet$}

\put(22.3,21.6){\color{rootcol1}0}
\put(47.3,21.6){\color{rootcol1}0}
\put(72.3,21.6){\color{rootcol1}0}
\put(97.3,21.6){\color{rootcol2}2}
\put(122.3,22.3){\color{rootcol2}$\bullet$}

\put(22.3,46.6){\color{rootcol2}2}
\put(47.3,46.6){\color{rootcol2}2}
\put(72.3,46.6){\color{rootcol1}2}
\put(97.3,46.6){\color{rootcol2}1}
\put(122.3,47.3){\color{rootcol2}$\bullet$}

\put(22.3,71.6){\color{rootcol2}4}
\put(47.3,71.6){\color{rootcol2}4}
\put(72.3,71.6){\color{rootcol1}1}
\put(97.3,71.6){\color{rootcol2}0}
\put(122.3,72.3){\color{rootcol2}$\bullet$}

\put(22.3,96.6){\color{rootcol2}4}
\put(47.3,96.6){\color{rootcol2}5}
\put(72.3,96.6){\color{rootcol1}0}
\put(97.3,96.6){\color{rootcol1}0}
\put(122.3,96.6){\color{rootcol1}0}

\normalsize

\put(200,85){$\ns=[0,  0|3,  1,  1,  1|0,  0] $}
\put(200,60){$\hl_a=\{3, 1, 1, 2\}+\Lambda$}
\put(200,35){$\hn_i=\{-2, -3, 0, -1\}-\Lambda$}
\put(200,10){$\tx_a=\left\{1, \tx^{-2}, \tx^2, 1\right\} $}

\end{picture}
\end{center}

\noindent Though $\XX \YY \ZZ$ has length $3$, it still has shifted weights identical to the operator $\ZZ\Psib_{22}$ treated above, and we thus set the modified length used in the ansatz \eqref{eq:PAnz} to $\Lanz = 2$.

Our procedure for finding the leading solution leads to the distinguished Q-functions \hyperref[tab:XYZDistQ]{given} in appendix \ref{sec:largeQSys} due to their large size. They differ from the distinguished Q-functions of $\ZZ \Psib_{22}$, but the difference only lies in which Q-functions are suppressed in $\g$. By using the freedom to choose $\cA_a$ and $\cB_i$, $\dQ_{1,0} \propto \tfrac{1}{\spar^3}$ can be traded for $\dQ_{0,1} \propto \spar^2$ through a redefinition of the asymptotic normalization. With the rescaling
\begin{align*}
  \cA_1 &\to \g^2 \cA_1 \,,	&	\cA^1 \to \frac{1}{\g^2} \cA^1 \,, \\
  \cB_1 &\to \frac{1}{\g^2} \cB_1 \,,	&	\cB^1 \to \g^2 \cB^1 \,, 
\end{align*}
which obviously respects \eqref{AABB},
we can bring the entire set of distinguished Q-functions into the ones for $\ZZ \Psib_{22}$, again with the substitution $\tx \to \tx^2$. 
Naturally, both the anomalous dimension and $\bP$ and $\bQ$ are the same as for $\ZZ \Psib_{22}$, with the mentioned change of power for the twist factor $\tx$, corresponding to $\tb\to2\tb$.
\end{shaded}

\subsection{The \texorpdfstring{$\su(2)$}{su(2)} operator \texorpdfstring{$\ZZ^2 \XX^2$}{Z2X2} - HWS in \texorpdfstring{\grpath{0}{2}{2}{4}}{0224}}
\begin{shaded}
\noindent 
We have already discussed that the $\su(2)$ Konishi operator $\ZZ^2\XX^2$ is in the same submultiplet as $\ZZ\XX\YY$, so they should correspond to the same solution to the QSC. 
The boundary conditions for this solution are 
\begin{center}
\begin{picture}(300,105)

\linethickness{3.5mm}
\put(10,50){\color{betacol!30}\line(1,0){130}}

{\color{gray}
\linethickness{0.5mm}
\put(25,0){\line(1,0){100}}
\put(25,25){\line(1,0){100}}
\put(25,50){\line(1,0){100}}
\put(25,75){\line(1,0){100}}
\put(25,100){\line(1,0){100}}
\put(25,0){\line(0,1){100}}
\put(50,0){\line(0,1){100}}
\put(75,0){\line(0,1){100}}
\put(100,0){\line(0,1){100}}
\put(125,0){\line(0,1){100}}
\put(25,25){\color{black}\line(0,1){50}}
\put(50,25){\color{black}\line(0,1){50}}
\put(75,25){\color{black}\line(0,1){50}}
\put(100,25){\color{black}\line(0,1){50}}
\put(125,25){\color{black}\line(0,1){50}}
\put(25,25){\color{black}\line(1,0){100}}
\put(25,50){\color{black}\line(1,0){100}}
\put(25,75){\color{black}\line(1,0){100}}
}

\linethickness{1.3mm}
\put(25,0){\color{orange}\line(0,1){25}}
\put(25,25){\color{orange}\line(1,0){50}}
\put(75,25){\color{orange}\line(0,1){50}}
\put(75,75){\color{orange}\line(1,0){50}}
\put(125,75){\color{orange}\line(0,1){25}}

{\color{shadecolor}
\put(25,0){\circle*{11}}
\put(50,0){\circle*{11}}
\put(75,0){\circle*{11}}
\put(100,0){\circle*{11}}
\put(125,0){\circle*{11}}

\put(25,25){\circle*{11}}
\put(50,25){\circle*{11}}
\put(75,25){\circle*{11}}
\put(100,25){\circle*{11}}
\put(125,25){\circle*{11}}

\put(25,50){\circle*{11}}
\put(50,50){\circle*{11}}
\put(75,50){\circle*{11}}
\put(100,50){\circle*{11}}
\put(125,50){\circle*{11}}

\put(25,75){\circle*{11}}
\put(50,75){\circle*{11}}
\put(75,75){\circle*{11}}
\put(100,75){\circle*{11}}
\put(125,75){\circle*{11}}

\put(25,100){\circle*{11}}
\put(50,100){\circle*{11}}
\put(75,100){\circle*{11}}
\put(100,100){\circle*{11}}
\put(125,100){\circle*{11}}
}

\put(22.3,-3.4){\color{rootcol1}0}
\put(47.3,-2.7){\color{rootcol2}$\bullet$}
\put(72.3,-2.7){\color{rootcol2}$\bullet$}
\put(97.3,-2.7){\color{rootcol2}$\bullet$}
\put(122.3,-2.7){\color{rootcol2}$\bullet$}

\put(22.3,21.6){\color{rootcol1}0}
\put(47.3,21.6){\color{rootcol1}0}
\put(72.3,21.6){\color{rootcol1}0}
\put(97.3,21.6){\color{rootcol2}4}
\put(122.3,21.6){\color{rootcol2}4}

\put(22.3,46.6){\color{rootcol2}2}
\put(47.3,46.6){\color{rootcol2}2}
\put(72.3,46.6){\color{rootcol1}2}
\put(97.3,46.6){\color{rootcol2}2}
\put(122.3,46.6){\color{rootcol2}2}

\put(22.3,71.6){\color{rootcol2}4}
\put(47.3,71.6){\color{rootcol2}4}
\put(72.3,71.6){\color{rootcol1}0}
\put(97.3,71.6){\color{rootcol1}0}
\put(122.3,71.6){\color{rootcol1}0}

\put(22.3,97.3){\color{rootcol2}$\bullet$}
\put(47.3,97.3){\color{rootcol2}$\bullet$}
\put(72.3,97.3){\color{rootcol2}$\bullet$}
\put(97.3,97.3){\color{rootcol2}$\bullet$}
\put(122.3,97.3){\color{rootcol1}0}

\normalsize

\put(200,85){$\ns=[0,  0|4,  2,  2,  0|0,  0] $}
\put(200,60){$\hl_a=\{4, 2, 2, 3\}+\Lambda$}
\put(200,35){$\hn_i=\{-3, -4, -1, -2\}-\Lambda$}
\put(200,10){$\tx_a=\left\{1, \tx^{-2}, \tx^2, 1\right\} $}

\end{picture}
\end{center}

The solution for the distinguished Q-functions are displayed in appendix \ref{sec:largeQSys}. Again they differ from those of $\ZZ \Psib_{22}$ and $\ZZ\XX\YY$, but can be brought into the same form. The normalization then requires the rescalings
\begin{align*}
  \cA_1 &\to \g^2 \cA_1 \,,	&	\cA^1 \to \frac{1}{\g^2} \cA^1 \,,	\\
  \cA_4 &\to \frac{1}{\g^2} \cA_4 \,,	&	\cA^4 \to \g^2 \cA^4 \,, 	\\
  \cB_1 &\to \frac{1}{\g^2} \cB_1 \,,	&	\cB^1 \to \g^2 \cB^1 \,, 	\\
  \cB_4 &\to \g^2 \cB_4 \,,	&	\cB^4 \to \frac{1}{\g^2} \cB^4 \,. 
\end{align*}
The powers of $\spar$ still do not match after such a rescaling but can be made to do so by a gauge transformation \eqref{gauge} with $\Lambda = 1$. In the ansatz, we need to set $\Lanz= 2$ and $\deltaL=1$ for it to match that of $\ZZ\Psib_{22}$. Performing these manipulations it is evident that the $\bP, \bQ$ and the anomalous dimensions are the same as for $\ZZ \Psib_{22}$, with the change $\tx \to \tx^{2}$.

We can thus conclude that the 5-loop anomalous dimension of the $\su(2)$ Konishi operator is given by \eqref{ZPsibres} with the replacement $\tb\to2\tb$, which is in agreement with the known 4-loop result \cite{Fiamberti:2008sm,Ahn:2010yv}.

\end{shaded}

\subsection{The operator \texorpdfstring{$\ZZ \Psi_{11}^2$}{ZPsi} - HWS in \texorpdfstring{\grpath{0}{2}{3}{3}}{0233}}

\begin{shaded}
\noindent 
This example has the property that all twist factors are different: 
\begin{center}
\begin{picture}(300,105)

\linethickness{3.5mm}
\put(10,50){\color{betacol!42}\line(1,0){130}}
\put(10,25){\color{betacol!25}\line(1,0){130}}

{\color{gray}
\linethickness{0.5mm}
\put(25,0){\line(1,0){100}}
\put(25,25){\line(1,0){100}}
\put(25,50){\line(1,0){100}}
\put(25,75){\line(1,0){100}}
\put(25,100){\line(1,0){100}}
\put(25,0){\line(0,1){100}}
\put(50,0){\line(0,1){25}}
\put(75,0){\line(0,1){25}}
\put(100,0){\line(0,1){25}}
\put(25,25){\color{black}\line(0,1){75}}
\put(50,25){\color{black}\line(0,1){75}}
\put(75,25){\color{black}\line(0,1){75}}
\put(100,25){\color{black}\line(0,1){75}}
\put(125,0){\line(0,1){100}}
\put(25,25){\color{black}\line(1,0){75}}
\put(25,50){\color{black}\line(1,0){75}}
\put(25,75){\color{black}\line(1,0){75}}
\put(25,100){\color{black}\line(1,0){75}}
}

\linethickness{1.3mm}
\put(25,0){\color{orange}\line(0,1){25}}
\put(25,25){\color{orange}\line(1,0){50}}
\put(50,25){\color{orange}\line(1,0){25}}
\put(75,25){\color{orange}\line(0,1){25}}
\put(75,50){\color{orange}\line(1,0){25}}
\put(100,50){\color{orange}\line(0,1){50}}
\put(100,100){\color{orange}\line(1,0){25}}

{\color{shadecolor}
\put(25,0){\circle*{11}}
\put(50,0){\circle*{11}}
\put(75,0){\circle*{11}}
\put(100,0){\circle*{11}}
\put(125,0){\circle*{11}}

\put(25,25){\circle*{11}}
\put(50,25){\circle*{11}}
\put(75,25){\circle*{11}}
\put(100,25){\circle*{11}}
\put(125,25){\circle*{11}}

\put(25,50){\circle*{11}}
\put(50,50){\circle*{11}}
\put(75,50){\circle*{11}}
\put(100,50){\circle*{11}}
\put(125,50){\circle*{11}}

\put(25,75){\circle*{11}}
\put(50,75){\circle*{11}}
\put(75,75){\circle*{11}}
\put(100,75){\circle*{11}}
\put(125,75){\circle*{11}}

\put(25,100){\circle*{11}}
\put(50,100){\circle*{11}}
\put(75,100){\circle*{11}}
\put(100,100){\circle*{11}}
\put(125,100){\circle*{11}}
}

\put(22.3,-3.4){\color{rootcol1}0}
\put(47.3,-3.4){\color{rootcol2}0}
\put(72.3,-3.4){\color{rootcol2}0}
\put(97.3,-3.4){\color{rootcol2}4}
\put(122.3,-2.7){\color{rootcol2}$\bullet$}

\put(22.3,21.6){\color{rootcol1}0}
\put(47.3,21.6){\color{rootcol1}0}
\put(72.3,21.6){\color{rootcol1}0}
\put(97.3,21.6){\color{rootcol2}1}
\put(122.3,22.3){\color{rootcol2}$\bullet$}

\put(22.3,46.6){\color{rootcol2}2}
\put(47.3,46.6){\color{rootcol2}2}
\put(72.3,46.6){\color{rootcol1}2}
\put(97.3,46.6){\color{rootcol1}0}
\put(122.3,47.3){\color{rootcol2}$\bullet$}

\put(22.3,71.6){\color{rootcol2}5}
\put(47.3,71.6){\color{rootcol2}5}
\put(72.3,71.6){\color{rootcol2}2}
\put(97.3,71.6){\color{rootcol1}0}
\put(122.3,72.3){\color{rootcol2}$\bullet$}

\put(22.3,96.6){\color{rootcol2}5}
\put(47.3,96.6){\color{rootcol2}6}
\put(72.3,96.6){\color{rootcol2}1}
\put(97.3,96.6){\color{rootcol1}0}
\put(122.3,96.6){\color{rootcol1}0}

\normalsize

\put(200,85){$\ns=[0,  0|3,  1,  0,  0|2,  0]$}
\put(200,60){$\hl_a=\{3, 1, 0, 3\}+\Lambda$}
\put(200,35){$\hn_i=\{-3, -4, 0, -2\}-\Lambda$}
\put(200,10){$\tx_a=\left\{\tx, \tx^{-3}, \tx^2, 1\right\} $}

\end{picture}
\end{center}
The fact that the twist factors are distinct does not make any difference conceptually and, in principle, this operator should be treatable with the presented algorithm just as were the previous examples. In practice, however, the twist dependent expressions grow very rapidly making computer calculations very slow. Already the leading order Q-system is very bulky and is put in appendix \ref{sec:largeQSys}. 

The leading $\bP_a$ are
{\footnotesize\allowdisplaybreaks
\begin{align}
  \bP_1^{(0)} &= -\frac{\tx^{2+\ii \spar}}{\spar^3 (\tx-1)^3 \left(\tx^3+\tx^2+\tx+1\right)} \,,  \no\\
  \bP_2^{(0)} &= \tx^{-3 \ii \spar} \bigg(-\frac{\left(\tx^8-\tx^7+\tx^5+\tx^3-\tx+1\right) \tx^2}{\spar^3 \left(\tx^2+\tx+1\right) \left(\tx^4-1\right)^2 \left(\tx^4+\tx^3+\tx^2+\tx+1\right)} \no \\
  &  \quad\, +\frac{\ii \left(\tx^4+2 \tx^3+4 \tx^2+2 \tx+1\right) \tx}{\spar^2 (\tx-1) (\tx-\ii) (\tx+\ii) (\tx+1) \left(\tx^4+\tx^3+\tx^2+\tx+1\right)}  
  +\frac{\tx^2+\tx+1}{\spar \left(\tx^4+\tx^3+\tx^2+\tx+1\right)}\bigg) \,,  \no\\
  \bP_3^{(0)} &= \tx^{2 \ii \spar} \bigg(\frac{\ii \tx \left(\tx^8+6 \tx^7+12 \tx^6+16 \tx^5+14 \tx^4+16 \tx^3+12 \tx^2+6 \tx+1\right)}{\spar^3 \left(\tx^6+\tx^5-\tx-1\right)} \no \\
  &  \quad\, +\frac{\tx \left(3 \tx^6+5 \tx^5+5 \tx^4+4 \tx^3+5 \tx^2+5 \tx+3\right)}{\spar^2 \left(\tx^4+\tx^3+\tx^2+\tx+1\right)}  
  -\frac{\ii \tx \left(3 \tx^6+2 \tx^5-\tx^4+\tx^2-2 \tx-3\right)}{\spar \left(\tx^4+\tx^3+\tx^2+\tx+1\right)}-\left(\tx^2-1\right)^2\bigg) \,,\no  \\
  \bP_4^{(0)} &= \frac{4 \ii}{\spar^3} \,, 
\intertext{\normalsize while the leading $\bQ$ are}
  \bQ_1^{(0)} &= -\frac{1}{3} \ii \spar^3 (\tx-1) \,,  \no\\
  \bQ_2^{(0)} &= \frac{1}{8} \spar^4 (\tx-1) \,, \no \\
  \bQ_3^{(0)} &= -\frac{6 \etaF3 \spar^4 (\tx-1)^2 \left(\tx^2+\tx+1\right)}{\tx^3}+\etaF2 \left(\frac{12 \spar^3 \left(\tx^2+\tx+1\right)^2}{\tx^3}+\frac{18 \ii \spar^4 \left(\tx^4+\tx^3-\tx-1\right)}{\tx^3}\right)  \no\\
  &  \quad\, -\frac{18 \spar^3 \left(\tx^4+\tx^3-\tx-1\right)}{\tx^3}+\frac{18 \ii \spar^2 \left(\tx^3+2 \tx^2+2 \tx+1\right)}{\tx^3}-\frac{6 \spar \left(2 \tx^3+3 \tx^2+3 \tx+1\right)}{\tx^3}-\frac{6 \ii \left(\tx^2+\tx+1\right)}{\tx^2} \,,  \no\\
  \bQ_4^{(0)} &= 0\,. 
\end{align}
}

Notice that for this operator, the length $L=3$ and the modified length $\Lanz=3$ are in correspondence, since no operator with $L=2$ can give rise to the same boundary conditions.

We have so far not found an efficient way of dealing with the large $\tx$-expressions beyond the first correction in the perturbative algorithm, which fixes the 2-loop anomalous dimension. The fully simplified expressions of the objects at this first step require the same amount of memory as does step 6 for the simpler operators, and the memory scaling is much worse. Our results for this operator are thus limited to the 2-loop result
\begin{align}
\ado{1} &= 4 (\cos (\tb )+2) \\[1mm]
\ado{2} &= -4 (5 \cos (\tb )+7) \no
.
\end{align}
\end{shaded}

\section{Conclusion}\label{sec:con}

In this paper, we discussed how to find explicit solutions to the twisted QSC in one of the simplest possible cases, the $\tb$-deformation. We considered several operators that in the undeformed theory belong to the Konishi multiplet, and we were able to produce a range of new results. Our main results are summed up in table \ref{tab:res}.
\begin{table}[h!]
\centering
\begin{tabular}{|c|c|c|} \hline
Operator & Loop order & Equation \\\hline\hline
$\Psi_{11}\FF_{11}$ & 8 loops & \eqref{PsiFres} \\\hline
$\DD_{12}^2\ZZ^2$ & 7 loops & \eqref{DZres} \\\hline
$\DD_{12}^2\ZZ\XX$  & 7 loops & \eqref{DZXres} \\\hline
$\ZZ\bar\Psi_{22}$ & \multirow{2}{*}{5 loops} & \eqref{ZPsibres} \\
$\ZZ^2\XX^2$, $\ZZ\XX\YY$ &  & $\quad\quad\;\,\eqref{ZPsibres}|_{\tb\to2\tb}$ \\\hline
\end{tabular}
\caption{Our main results and where to find them in the paper. The operator $\DD_{12}^2\ZZ^2$ is the $\sl(2)$ Konishi operator, %
while the operator $\ZZ^2\XX^2$ is the $\su(2)$ Konishi operator.} \label{tab:res}
\end{table}

Though we only scratched the surface by considering operators from the Konishi multiplet, the strategy is general and should be generalizable to the remaining spectrum. 
It would be interesting to have a classification of all submultiplets of the Konishi multiplet in the $\tb$-deformed theory. Furthermore, it would be very interesting to study operators that in the undeformed theory belong to the $L=2$ BMN vacuum multiplet, in particular 1-magnon states, e.g. $\ZZ\XX$. Our preliminary studies\footnote{Done in collaboration with Kasper E. Vardinghus.} indicate that it might be necessary to expand the QSC functions in odd powers of $\g$ for such states, similar to the special cases found in \cite{Marboe:2018ugv}.  We may return to this question in future work, but also encourage others to attack it.

As we saw, it quickly becomes technically challenging to handle calculations with twist variables. The ultimate challenge would be to use the procedure to construct perturbative corrections to Q-operators \cite{Bazhanov:2010ts,Bazhanov:2010jq,Frassek:2010ga,Frassek:2011aa,Kazakov:2010iu}. A procedure to explicitly construct the 1-loop Q-operators was given in \cite{Frassek:2017bfz}, but the technicality of constructing perturbative corrections in the fully twisted theory will probably require a courageous computational effort. However, this could lead to new results about perturbative corrections to eigenstates, the dilatation operator, and perhaps about the still mysterious integrable model that underlies the AdS/CFT spectral problem.

\acknowledgments

We thank Kasper E. Vardinghus for collaboration on related topics. 
We thank Dmytro Volin and Matthias Wilhelm for very useful discussions. 
CM would like to thank the Niels Bohr Institute for kind hospitality during the final stages of this work. 
The work of CM was supported by the grant {\it Exact Results in Gauge and String Theories} from the Knut and Alice Wallenberg foundation, while the work of EW was supported by the ERC advanced grant No 341222.

\appendix
\section{Larger 1-loop Q-system examples}
\label{sec:largeQSys}
The leading Q-system for the operators $\ZZ \Psi_{11}^2$, $\ZZ\XX\YY$ and $\ZZ^2\XX^2$ are presented here in landscape mode due to their size.

\begin{landscape}
  \begin{shaded}
    \textbf{Leading distinguished Q-functions for $\ZZ \Psi_{11}^2$}
{
\arrayrulecolor{gray}\setlength{\extrarowheight}{3ex}
\scriptsize
\hspace{\distQTabShift}\begin{equation*}
  \begin{array}{|m{28em}|m{21em}|c|m{7em}|c|cl}\hhline{-|-|->{\arrgr}|-|-|>{\arrc}~~}
 $\trivAnz{\big({\sparg^{[-3]}} {\sparg^-} {\sparg^+} {\sparg^{[3]}}\big)^{-3}} \bigg(\spar^5-\frac{\ii \left(3 \tx^4+\tx^3+10 \tx^2+\tx+3\right) \spar^4}{2 \left(\tx^4+\tx^3-\tx-1\right)}
   $\newline$
   +\frac{3 \left(3 \tx^8+8 \tx^7+15 \tx^6+28 \tx^5+32 \tx^4+28 \tx^3+15 \tx^2+8 \tx+3\right) \spar^3}{2 \left(\tx^4+\tx^3-\tx-1\right)^2} 
   $\newline$
   -\frac{3 \ii \left(9 \tx^{10}+18 \tx^9+12 \tx^8+12 \tx^7+23 \tx^6+20 \tx^5+23 \tx^4+12 \tx^3+12 \tx^2+18 \tx+9\right) \spar^2}{4 \left(\tx^2-1\right)^3 \left(\tx^2+\tx+1\right)^2} 
   $\newline$
   +\frac{9 \left(9 \tx^{10}+12 \tx^9+30 \tx^8-10 \tx^7+65 \tx^6+44 \tx^5+65 \tx^4-10 \tx^3+30 \tx^2+12 \tx+9\right) \spar}{16 (\tx-1)^4 (\tx+1)^2 \left(\tx^2+\tx+1\right)^2} 
 $\newline$
 -\frac{3 \ii \left(81 \tx^{12}+54 \tx^{11}-63 \tx^{10}-252 \tx^9-273 \tx^8+118 \tx^7-98 \tx^6\right)}{32 (\tx-1)^5 (\tx+1)^3 \left(\tx^2+\tx+1\right)^2}
 $\newline$
 -\frac{3 \ii\left(118 \tx^5-273 \tx^4-252 \tx^3-63 \tx^2+54 \tx+81\right)}{32 (\tx-1)^5 (\tx+1)^3 \left(\tx^2+\tx+1\right)^2}\bigg) 
 $
 &
 $\trivAnz{\big({\sparg^{[-2]}} \sparg {\sparg^{[2]}}\big)^{-3}} \bigg(\spar^6 
 -\frac{3 \ii \left(\tx^2+1\right)^2 \spar^5}{4 \left(\tx^4+\tx^3-\tx-1\right)} 
 $\newline$
 +\frac{3 \left(3 \tx^8+8 \tx^7+16 \tx^6+24 \tx^5+30 \tx^4+24 \tx^3+16 \tx^2+8 \tx+3\right) \spar^4}{4 \left(\tx^4+\tx^3-\tx-1\right)^2}
 $\newline$
 -\frac{\ii \left(3 \tx^8-11 \tx^6-\tx^5-18 \tx^4-\tx^3-11 \tx^2+3\right) \spar^3}{2 (\tx-1)^3 (\tx+1) \left(\tx^2+\tx+1\right)^2} 
 $\newline$
 +\frac{\left(3 \tx^8+9 \tx^7+22 \tx^6+32 \tx^5+30 \tx^4+32 \tx^3+22 \tx^2+9 \tx+3\right) \spar^2}{2 \left(\tx^4+\tx^3-\tx-1\right)^2} 
 $\newline$
 -\frac{\ii \left(3 \tx^8+3 \tx^7-4 \tx^6-17 \tx^5-42 \tx^4-17 \tx^3-4 \tx^2+3 \tx+3\right) \spar}{4 (\tx-1)^3 (\tx+1) \left(\tx^2+\tx+1\right)^2} 
 $\newline$
 -\frac{-\tx^6+\tx^4+6 \tx^3+\tx^2-1}{4 \left(\tx^3-1\right)^2}\bigg) 
 $
 & 
 \grcell{\spar+\frac{\ii \tx \left(\tx^2+4 \tx+1\right)}{2 \left(\tx^2-1\right) \left(\tx^2+\tx+1\right)}} 
 & 
 \multicolumn{1}{c|}{\grc \grone}
 & \grcell{\grc \grone} &  &  \\[4\exLin]\hhline{-|-|->{\arrgr}|>{\arrc}->{\arrgr}|-|>{\arrc}~~}
 $
 \trivAnz{\big({\sparg^{[-2]}} \sparg {\sparg^{[2]}}\big)^{-3}} \bigg(\spar^5-\frac{\ii \left(\tx^2+1\right)^2 \spar^4}{\tx^4+\tx^3-\tx-1}
 $\newline$
 +\frac{2 \left(\tx^8+3 \tx^7+7 \tx^6+13 \tx^5+17 \tx^4+13 \tx^3+7 \tx^2+3 \tx+1\right) \spar^3}{\left(\tx^4+\tx^3-\tx-1\right)^2}
 $\newline$
 -\frac{2 \ii \left(\tx^{10}+2 \tx^9-2 \tx^8-5 \tx^7-6 \tx^6-10 \tx^5-6 \tx^4-5 \tx^3-2 \tx^2+2 \tx+1\right) \spar^2}{\left(\tx^2-1\right)^3 \left(\tx^2+\tx+1\right)^2}
 $\newline$
 +\frac{\left(\tx^{10}+2 \tx^9+6 \tx^8-6 \tx^7+7 \tx^6+16 \tx^5+7 \tx^4-6 \tx^3+6 \tx^2+2 \tx+1\right) \spar}{(\tx-1)^4 (\tx+1)^2 \left(\tx^2+\tx+1\right)^2}
 $\newline$
 -\frac{\ii \left(\tx^{12}+\tx^{11}-2 \tx^{10}-9 \tx^9-11 \tx^8+8 \tx^7+8 \tx^5-11 \tx^4-9 \tx^3-2 \tx^2+\tx+1\right)}{(\tx-1)^5 (\tx+1)^3 \left(\tx^2+\tx+1\right)^2}\bigg) 
 $
 &
 $
 \trivAnz{\big({\sparg^-} {\sparg^+}\big)^{-3}} \bigg(\spar^5-\frac{\ii \left(\tx^4-\tx^3-2 \tx^2-\tx+1\right) \spar^4}{2 \left(\tx^4+\tx^3-\tx-1\right)}
 $\newline$
 +\frac{\left(\tx^8+4 \tx^7+13 \tx^6+24 \tx^5+36 \tx^4+24 \tx^3+13 \tx^2+4 \tx+1\right) \spar^3}{2 \left(\tx^4+\tx^3-\tx-1\right)^2}
 $\newline$
 -\frac{\ii \left(\tx^8-17 \tx^6-2 \tx^5-36 \tx^4-2 \tx^3-17 \tx^2+1\right) \spar^2}{4 (\tx-1)^3 (\tx+1) \left(\tx^2+\tx+1\right)^2}
 $\newline$
 +\frac{\left(\tx^8+6 \tx^7+25 \tx^6+18 \tx^5-52 \tx^4+18 \tx^3+25 \tx^2+6 \tx+1\right) \spar}{16 \left(\tx^4+\tx^3-\tx-1\right)^2}
 $\newline$
 -\frac{\ii \left(\tx^8+2 \tx^7-\tx^6-36 \tx^5-76 \tx^4-36 \tx^3-\tx^2+2 \tx+1\right)}{32 (\tx-1)^3 (\tx+1) \left(\tx^2+\tx+1\right)^2}\bigg) 
 $
 &
 \grcell{\spar^2+\frac{\ii \tx \left(\tx^2+4 \tx+1\right) \spar}{\tx^4+\tx^3-\tx-1}} 
 &
 \grcell{\grc \grone} 
 &
 \grzero &  &  \\[2\exLin]\hhline{-|->{\arrgr}|->{\arrgr}|>{\arrc}->{\arrgr}|>{\arrc}-~~}
 $
 \trivAnz{\big({\sparg^-} {\sparg^+}\big)^{-3}} \left(\spar^2+\frac{\ii \left(\tx^3+\tx^2+\tx+1\right) \spar}{\tx^3-1}-\frac{\tx^6+2 \tx^5+6 \tx^4-10 \tx^3+6 \tx^2+2 \tx+1}{4 \left(\tx^3-1\right)^2}\right) 
 $
 &
 \grcell{\trivAnz{\sparg^{-3}} \left(\spar^2+\frac{\ii \tx (\tx+1) \spar}{\tx^3-1}-\frac{\tx^2}{\left(\tx^2+\tx+1\right)^2}\right)} 
 &
 \cellcolor{momcell}\spar^2-\frac{\ii \left(\tx^2-1\right) \spar}{\tx^2+\tx+1}+\frac{-\tx^2+\tx-1}{4 \left(\tx^2+\tx+1\right)} 
 &
 \grcell{\grc \grone} 
 &
 \grzero &  & \rule{\exSpace}{0pt} \tx^{-2 \ii \spar} \\[3\exLin]\hhline{>{\arrgr}|-|->{\arrgr}|>{\arrc}->{\arrgr}|-|>{\arrc}-~~}
 \grcellFirstL{\grc \trivAnz{\sparg^{-3}} } 
 &
 \multicolumn{1}{c|}{\grc \grone}
 &
 \grcell{\grc \trivAnz{\sparg^{3}} } 
 &
 \multicolumn{1}{c|}{\spar-\frac{\ii (\tx+1)}{2 (\tx-1)}}
 &
 \grzero &  & \rule{\exSpace}{0pt} \tx^{\ii \spar} \\[3\exLin]\hhline{>{\arrgr}|>{\arrc}->{\arrgr}|-|-|>{\arrc}-|-~~}
 \grcellFirstLR{\grc \grone} 
 & 
 \multicolumn{1}{c|}{\trivAnz{\sparg^{3}}}
 & 
 \trivAnz{\big({\sparg^-} {\sparg^+}\big)^{3}}  
 &
 $
 \spar^4+\frac{2 \ii (\tx+1) \spar^3}{\tx-1}
 $\newline$
 +\frac{6 \tx \spar^2}{(\tx-1)^2}
 $\newline$
 +\frac{2 \ii (\tx+1) \spar}{\tx-1}
 $\newline$
 -\frac{\tx^2+1}{(\tx-1)^2} 
 $
 & \grzero &  &  \\[3\exLin]\grline{1-1}\cline{2-5}
\end{array}
\end{equation*}
}

\end{shaded}

\begin{shaded}\label{tab:XYZDistQ}
  \textbf{Leading distinguished Q-functions for $\XX\YY\ZZ$}
{%
\arrayrulecolor{gray}\setlength{\extrarowheight}{\distQTabRowHeight}
\scriptsize
\hspace{\distQTabShift}\begin{equation*}
  \begin{array}{|m{23em}|m{14em}|c|c|c|cl}\hhline{|-|->{\arrgr}|-|-|-|>{\arrc}~~}
 $
  \trivAnz{\big({\sparg^{[-3]}} {\sparg^-} {\sparg^+} {\sparg^{[3]}}\big)^{-3}} \bigg(\spar^4+\frac{3 \left(3 \tx^4+8 \tx^2+2 \ii \koSqrt \tx+3\right) \spar^2}{2 \left(\tx^2-1\right)^2}
  $\newline$
  +\frac{3 \left(9 \tx^4-2 \tx^2+9\right) \left(3 \left(\tx^2+1\right)^2+4 \ii \koSqrt \tx\right)}{16 \left(\tx^2-1\right)^4}\bigg) 
  $
  & \multicolumn{1}{m{14em}!{\color{orange}\vrule}}{
  $
  \trivAnz{\big({\sparg^{[-2]}} \sparg {\sparg^{[2]}}\big)^{-3}} \bigg(\spar^{[-2]} \spar^{[2]} \spar^3
  $\newline$
 +\frac{\spar^{[-2]} \spar^{[2]} \left(\tx^4+4 \tx^2+2 \ii \koSqrt \tx+1\right) \spar}{\left(\tx^2-1\right)^2}\bigg)
 $
 } & \grc \grone & \grc \grone & \grcell{\grc \grone} &  &  \\[\exLin]\hhline{-|->{\arrgr}|>{\arrc}->{\arrgr}|-|-|>{\arrc}~~} 
  $
  \trivAnz{\big({\sparg^{[-2]}} \sparg {\sparg^{[2]}}\big)^{-3}} \bigg(\spar^4+\frac{2 \left(\tx^4+3 \tx^2+\ii \koSqrt \tx+1\right) \spar^2}{\left(\tx^2-1\right)^2}
  $\newline$
  +\frac{\tx^8+2 \tx^6+2 \tx^2+2 \ii \koSqrt \left(\tx^5-\tx^3+\tx\right)+1}{\left(\tx^2-1\right)^4}\bigg) 
  $
  & 
  \multicolumn{1}{m{14em}!{\color{orange}\vrule}}{$
  \trivAnz{\big({\sparg^-} {\sparg^+}\big)^{-3}} \bigg( \spar^{-} \spar^{+} \spar^2
  $\newline$
 +\frac{\spar^{-} \spar^{+} \left(\tx^4+10 \tx^2+4 \ii \koSqrt \tx+1\right)}{4 \left(\tx^2-1\right)^2}\bigg)
 $} & \grcell{\grc \spar} & \grone & \grzero &  &  \\[\exLin]\hhline{-|->{\arrgr}|>{\arrc}->{\arrgr}|>{\arrc}-|-~~}
 $
 \trivAnz{\big({\sparg^-} {\sparg^+}\big)^{-3}} \bigg(\spar^2+\frac{\ii \left(\tx^4+3 \tx^2+\ii \koSqrt \tx+1\right) \spar}{\tx^4-1}
 $\newline$
 -\frac{\tx^4+2 \ii \koSqrt \tx+1}{4 \left(\tx^2-1\right)^2}\bigg) 
 $
 & \grcell{\trivAnz{\sparg^{-3}} \left(\spar^2-\frac{\spar (\koSqrt-\ii \tx) \tx}{\tx^4-1}\right)} & \grcell{\cellcolor{momcell}\spar^2-\frac{\ii \left(\tx^4+\tx^2-\ii \koSqrt \tx+1\right) \spar}{\tx^4-1}-\frac{\tx^4+4 \tx^2-2 \ii \koSqrt \tx+1}{4 \left(\tx^2-1\right)^2}} & \spar+\frac{\ii \tx (\ii \koSqrt+\tx)}{\tx^4-1} & \grzero &  & \rule{\exSpace}{0pt} \tx^{-2 \ii \spar} \\[\exLin]\hhline{>{\arrgr}|-|->{\arrgr}|>{\arrc}->{\arrgr}|>{\arrc}-|-~~}
 \grcellFirstL{\grc \trivAnz{\sparg^{-3}} } 
 & 
 \multicolumn{1}{c|}{\grc \grone}
 & \grcell{\grc \trivAnz{\sparg^{3}} } & \spar^2+\frac{\tx^4+10 \tx^2+4 \ii \koSqrt \tx+1}{4 \left(\tx^2-1\right)^2} & \grzero &  &  \\[\exLin]\hhline{>{\arrgr}|>{\arrc}->{\arrgr}|-|-|>{\arrc}-|-~~}
 \grcellFirstLR{\grc \grone} 
 & 
 \multicolumn{1}{c|}{\grzero} & \multicolumn{1}{c|}{\grzero} & \grzero & \grzero &  &  \\[\exLin]\hhline{>{\arrgr}|-|>{\arrc}-|-|-|-|~~} 
\end{array}
\end{equation*}
}
\end{shaded}
\vspace{-1.2cm}
\begin{shaded}\label{tab:ZZXXDistQ}
  \hspace{-1.5cm}\textcolor{white}{\rule{1.1\linewidth}{2pt}}\\[1mm]
  \textbf{Leading distinguished Q-functions for $\ZZ^2 \XX^2$}
{
\arrayrulecolor{gray}\setlength{\extrarowheight}{\distQTabRowHeight}
\scriptsize
\hspace{\distQTabShift}\begin{equation*}
  \begin{array}{|m{19em}|m{13em}|m{11em}|m{12em}|m{14em}|cl}\hhline{|-|-|-|->{\arrgr}|-|>{\arrc}~~}
    \multicolumn{1}{|c|}{\grzero} & \multicolumn{1}{c|}{\grzero} & \multicolumn{1}{c|}{\grzero} & \grcell{\grzero} & \grcell{\grc \grone} &  &  \\[\exLin]\hhline{-|->{\arrgr}|-|->{\arrgr}|>{\arrc}->{\arrgr}|>{\arrc}~~}
    $
 \trivAnz{\big({\sparg^{[-2]}} \sparg {\sparg^{[2]}}\big)^{-4}} \bigg(\spar^4+\frac{2 \left(\tx^4+3 \tx^2+\ii \koSqrt \tx+1\right) \spar^2}{\left(\tx^2-1\right)^2} 
 $\newline$
 +\frac{\tx^8+2 \tx^6+2 \tx^2+2 \ii \koSqrt \left(\tx^5-\tx^3+\tx\right)+1}{\left(\tx^2-1\right)^4}\bigg)
 $
 &
 \multicolumn{1}{m{13em}!{\color{orange}\vrule}}{
 $
 \trivAnz{\big({\sparg^-} {\sparg^+}\big)^{-4}} \bigg( \spar^- \spar^+ \spar^2
 $\newline$
 +\frac{\spar^- \spar^+ \left(\tx^4+10 \tx^2+4 \ii \koSqrt \tx+1\right)}{16 \left(\tx^2-1\right)^2}\bigg)
 $} 
 & \multicolumn{1}{c|}{\grc \grone} & \multicolumn{1}{c|}{\grc \grone} & \grcell{\grc \grone} &  &  \\[\exLin]\hhline{-|->{\arrgr}|>{\arrc}->{\arrgr}|-|-|>{\arrc}~~}
 $
 \trivAnz{\big({\sparg^-} {\sparg^+}\big)^{-4}} \bigg(\spar^2+\frac{\ii \left(\tx^4+3 \tx^2+\ii \koSqrt \tx+1\right) \spar}{\tx^4-1}
 $\newline$
 -\frac{\tx^4+2 \ii \koSqrt \tx+1}{4 \left(\tx^2-1\right)^2}\bigg)
 $& \grcell{\trivAnz{\sparg^{-4}} \left(\spar^2-\frac{\spar (\koSqrt-\ii \tx) \tx}{\tx^4-1}\right)} & \multicolumn{1}{m{11em}!{\color{orange}\vrule}}{\cellcolor{momcell}
 $
 \spar^2-\frac{\ii \left(\tx^4+\tx^2-\ii \koSqrt \tx+1\right) \spar}{\tx^4-1}
 $\newline$
 -\frac{\tx^4+4 \tx^2-2 \ii \koSqrt \tx+1}{4 \left(\tx^2-1\right)^2}
 $} & 
 \multicolumn{1}{c|}{\spar^2-\frac{\spar (\koSqrt-\ii \tx) \tx}{\tx^4-1}} 
 & 
 $
 \spar^2 +\frac{\ii \left(\tx^4+3 \tx^2+\ii \koSqrt \tx+1\right) \spar}{\tx^4-1}
 $\newline$
 -\frac{\tx^4+2 \ii \koSqrt \tx+1}{4 \left(\tx^2-1\right)^2} 
 $
 &  & \rule{\exSpace}{0pt} \tx^{-2 \ii \spar} \\[\exLin]\hhline{>{\arrgr}|-|->{\arrgr}|>{\arrc}->{\arrgr}|>{\arrc}-|-~~}
 \grcellFirstL{\grc \trivAnz{\sparg^{-4}} } & \multicolumn{1}{c|}{\grc \grone} & \grcell{\grc \trivAnz{\sparg^{4}} } & 
 $
 \spar^4+\frac{\left(\tx^4+4 \tx^2+2 \ii \koSqrt \tx+1\right) \spar^2}{2 \left(\tx^2-1\right)^2}
 $\newline$
 +\frac{\tx^4+10 \tx^2+4 \ii \koSqrt \tx+1}{16 \left(\tx^2-1\right)^2} 
 $
 &
 $
 \spar^4+\frac{2 \left(\tx^4+3 \tx^2+\ii \koSqrt \tx+1\right) \spar^2}{\left(\tx^2-1\right)^2}
 $\newline$
 +\frac{\tx^8+2 \tx^6+2 \tx^2+2 \ii \koSqrt \left(\tx^5-\tx^3+\tx\right)+1}{\left(\tx^2-1\right)^4} 
 $
 &  &  \\[\exLin]\hhline{>{\arrgr}|>{\arrc}->{\arrgr}|-|-|>{\arrc}-|-~~}
 \grcellFirstLR{\grc \grone} & \multicolumn{1}{c|}{\grzero} & \multicolumn{1}{c|}{\grzero} & \multicolumn{1}{c|}{\grzero} & \multicolumn{1}{c|}{\grzero} &  &  \\[\exLin]\hhline{>{\arrgr}|-|>{\arrc}-|-|-|-|~~}
\end{array}
\end{equation*}
}
\end{shaded}

\end{landscape}






\newpage

\addcontentsline{toc}{section}{References}

\bibliographystyle{elsarticle-num}
\bibliography{bibliography}








\end{document}